\documentclass[11pt,fleqn]{article}

\usepackage[left=1in,right=1in,top=1in,bottom=1in]{geometry}

\usepackage{amsfonts,amsmath,amssymb,amsthm,array,booktabs,chngcntr,dsfont,enumerate,graphicx}
\usepackage{mathrsfs,mathtools,MnSymbol,multirow,rotating,setspace,tabu,upgreek,ushort}

\usepackage[english]{babel}
\usepackage[longnamesfirst,round]{natbib}
\usepackage[colorlinks=true,citecolor=black,linkcolor=black,urlcolor=blue]{hyperref}

\usepackage[inline]{enumitem}

\usepackage[affil-it]{authblk}
\usepackage{color,soul}

\usepackage{algorithm,algpseudocode}

\renewcommand{\algorithmicrequire}{\textbf{Input:}\ }
\renewcommand{\algorithmicensure}{\textbf{Output:}\ }

\usepackage{array,booktabs,multirow,rotating,tabu}
\makeatletter
\newcommand{\thickhline}{%
	\noalign {\ifnum 0=`}\fi \hrule height 1.5pt
	\futurelet \reserved@a \@xhline
}
\newcolumntype{"}{@{\hskip\tabcolsep\vrule width 1pt\hskip\tabcolsep}}
\makeatother

\usepackage{tabularx}

\newcommand\clearrow{\global\let\rowmac\relax}
\clearrow

\newtheorem{assump}{Assumption}

\newtheorem{lemma}{Lemma}
\newtheorem{alemma}{Lemma}[section]

\newtheorem{propo}{Proposition}
\newtheorem{theorem}{Theorem}

\newcommand{\lowmathcal}[1]{{\operatorname{\text{\usefont{U}{BOONDOX-cal}{m}{n}#1}}}}

\newcommand{\E}{\mathbb{E}}
\newcommand{\V}{\mathbb{V}}

\newcommand{\Prob}{\mathbb{P}}

\newtheorem{innercustomgeneric}{\customgenericname}
\providecommand{\customgenericname}{}
\newcommand{\newcustomtheorem}[2]{%
  \newenvironment{#1}[1]
  {%
   \renewcommand\customgenericname{#2}%
   \renewcommand\theinnercustomgeneric{##1}%
   \innercustomgeneric
  }
  {\endinnercustomgeneric}
}

\newcustomtheorem{customassump}{Assumption}

\makeatletter
\newcommand*\bigcdot{{\mathpalette\bigcdot@{.75}}}
\newcommand*\bigcdot@[2]{\mathbin{\vcenter{\hbox{\scalebox{#2}{$\m@th#1\bullet$}}}}}
\makeatother

\usepackage{accents}

\begin{document}

\baselineskip=1.1\baselineskip

\title{Counterfactual Revenues in Ascending Auctions:\\ Estimation and Inference}

\author{Federico Zincenko\thanks{Address: 730 N.\ 14th St., HLH 525J, Lincoln, NE 68588, U.S.A. Phone: (402)\,472-2319.

E-mail: \href{mailto:fzincenko2@unl.edu}{fzincenko2@unl.edu}.
}
}
\affil{College of Business (Department of Economics)\\ University of Nebraska--Lincoln}
\date{\today}
\maketitle

\begin{abstract}  
\noindent 
I propose estimators for the seller’s expected revenue function and the set of optimal reserve prices in an open ascending auction under the symmetric IPV paradigm with observable auction-specific heterogeneity. These estimators are constructed using transaction payments, which reveal the second-highest valuations, and auction-specific covariates.
The proposed estimator of the expected revenue function is piecewise linear, with coefficients that can be computed without requiring numerical optimization or density estimation, while the estimator of the set of optimal reserve prices takes the form of a finite union of intervals whose endpoints can also be computed without either procedure.

I establish weak convergence of the centered and scaled estimator of the expected revenue function to a Gaussian process at the parametric rate, which also implies uniform consistency. I then construct confidence intervals and uniform confidence bands for the expected revenue, as well as (conservative) confidence regions for the set of optimal reserve prices. Critical values are obtained using the smooth bootstrap method, whose asymptotic validity I establish, thereby validating the proposed inference procedures. Monte Carlo experiments and an empirical application to timber auction data are also provided.
\end{abstract}

\newpage


\section{Introduction}

Ascending auctions are widely used to allocate resources. In these auctions, the reserve price is a central policy instrument because it directly affects the seller’s expected revenue. Therefore, choosing reserve prices requires reliable estimators of the expected revenue function and the set of optimal reserve prices.

This paper proposes estimators of the seller’s expected revenue, as a function of the reserve price, and of the set of optimal reserve prices in an ascending auction. I consider a symmetric private-values environment in which bidders’ valuations are independent and identically distributed. The winning bidder pays the maximum of the reserve price and the second-highest valuation, provided that the highest valuation exceeds the reserve price. To incorporate auction-specific covariates, I follow the widely used semi-nonparametric approach in the empirical auction literature. In particular, the distribution of private values is left unrestricted, while auction heterogeneity enters valuations through a log-linear specification, which allows for the standard homogenization procedure; see, e.g., \cite{hhs16}, \cite{res08}, and \cite{liuluo17}.

In this setting, the parameters of interest are the expected revenue function for an object with covariates normalized to zero, without loss of generality, and the corresponding set of optimal reserve prices. For each reserve price $r$, I characterize the expected revenue function $\pi(r)$ as a smooth nonlinear functional of the distribution of the winning bid, denoted by $F_W$. Specifically, $\pi(r)=\Pi(F_W,r)$, where the closed-form expression for $\Pi$ is provided in Eq.\ (\ref{eqa:clfunc}) of the Appendix.

This characterization suggests a simple three-step estimation procedure. First, I homogenize bids by regressing the logarithm of the winning bids on auction-specific covariates and take the resulting residuals (without subtracting the constant term). Second, I estimate the empirical c.d.f.\ $\hat F_U$ of these residuals. Third, using $\hat F_W(w)=\hat F_U(\log w)$ as the estimator of the c.d.f.\ of the winning bid, I estimate the expected revenue function by plugging $\hat F_W$ into the functional $\Pi$, so that
$ \hat \pi(r)=\Pi(\hat F_W,r) $. The resulting estimator is a piecewise linear function, with intercepts and slopes that can be computed directly without numerical optimization or density estimation.

An important feature of this environment is that estimating the expected revenue only requires an estimate of the c.d.f.\ of the winning bid $F_W$: neither its density nor the bidding function needs to be estimated. This contrasts with first-price auctions, where estimation of expected revenue typically requires estimating both objects \cite[]{zin24}. The reason is that, in ascending auctions with private values, the ascending process reveals bidders’ valuations through the order statistics, so the relevant information is contained in the distribution of the winning bid.

To establish weak convergence of $\hat \pi$ at the parametric rate, I first show that $\sqrt{L}(\hat F_W-F_W)$ converges weakly to a tight Gaussian process, where $L$ denotes the number of auctions in the sample. The main difficulty is that the first-step regression contributes to the limiting distribution, since its coefficients are estimated at the same parametric rate as the empirical c.d.f.\footnote{In particular, a super-consistent first-step estimator is not generally available because $f_W$ vanishes smoothly at the endpoints of its support if $n\geq3$, even when the p.d.f.\ of private values is bounded away from zero.}
To handle this issue, I rely on the arguments in \citet[Section 3.13.4.1]{vw23}. The functional delta method then implies that
$
\sqrt{L}(\hat \pi-\pi)
=
\sqrt{L}[ \Pi(\hat F_W,\cdot)-\Pi(F_W,\cdot)]
$
converges weakly to a tight Gaussian process $\mathbb G_\pi$, for which I provide an explicit covariance function without requiring its estimation. I also note that, unlike in first-price auction settings, the first-step coefficients cannot be estimated at a super-consistent rate, as discussed in the next section.

To construct confidence intervals and uniform-in-$r$ confidence bands, the distribution of $\mathbb{G}_\pi$ must be approximated. To this end, I employ the smooth bootstrap procedure of \cite{ko94boot} and \cite{neu09}, combined with the bootstrap delta method and adapted to account for the first-step OLS estimation error. A key feature of this procedure is that the bootstrap residuals for the first-stage regression are generated from a kernel-based estimate of their c.d.f., while the covariates are held fixed and are not bootstrapped. The procedure requires only one tuning parameter: the bandwidth used to construct the smoothed residual distribution.
Once the distribution of $\mathbb G_\pi$ is approximated, critical values for the confidence intervals and uniform confidence bands can be obtained by standard procedures, without the need to estimate the covariance function by nonparametric methods.

Using the estimator of the expected revenue function, I also construct a set estimator for the set of optimal reserve prices. Importantly, I do not impose restrictions that guarantee a unique optimum, such as the pseudo-concavity condition on the profit function imposed in \citet[Assumption 3]{ht03inf}, despite the otherwise weak restrictions imposed in that paper on bidding behavior and the auction environment.
Instead, I allow the set of optimal reserve prices to contain multiple values and estimate it using the criterion-function approach of \cite{cht07}. Because $\hat\pi$ is piecewise linear, the resulting estimator is a finite union of intervals whose endpoints can be computed directly, without numerical optimization. Relying on the weak convergence of $\hat\pi$, I also provide formulas that yield consistent-in-level, though possibly conservative, confidence regions for the set of optimal reserve prices. I further discuss the difficulties of obtaining exact asymptotic coverage when the optimal set is a singleton.

The approach of this paper is policy-oriented rather than focused on recovering the underlying primitives. The proposed estimators are designed to predict expected revenue under counterfactual reserve prices and to guide the choice of the reserve price. Accordingly, the paper contributes to the empirical literature on expected revenue and optimal reserve prices in ascending auctions within the symmetric IPV paradigm, dating back to the seminal contribution of \cite{paa97}. Following \cite{ht03inf}, one strand of this literature adopts an incomplete-model approach that imposes weak restrictions on the bidding behavior, resulting in a partially identified set of optimal reserve price(s) and motivating alternative criteria for selecting a reserve price from this set \cite[see, e.g.,][]{aryal13,junp24}.

In contrast, this paper imposes more structure on bidding behavior while still allowing the (true) set of optimal reserve prices to contain multiple elements. The closest related approach is \cite{gim17}, who uses quantile regression to estimate the private-value distribution and expected revenue and provides a point estimator of the optimal reserve price. This paper instead uses an OLS-based homogenization procedure to accommodate auction-level covariates, constructs uniform confidence bands for expected revenue, and provides set estimation and inference for potentially nonunique optimal reserve prices. Since the elements of the estimated set are expected to generate the same revenue, the choice among them may be guided by considerations outside the model. E.g., the smallest optimal reserve price may be preferred to encourage bidder participation.

The rest of this paper is organized as follows.
Section \ref{sec:frame} describes the auction model and presents the assumptions, allowing for observed auction-specific covariates.
Section \ref{sec:estier} introduces the estimator of the expected revenue together with its convenient computational procedure.
Section \ref{sec:asinf} presents the main asymptotic results, establishing weak convergence of the expected revenue function estimator and its bootstrap approximation, and also provides the formulas for constructing valid confidence intervals and bands.
Section \ref{subsec:orpresti} presents the set estimator for the set of optimal reserve prices, as well as the suggested inference procedures, and discusses the challenges of conducting inference on this set.
Section \ref{sec:mc} reports the results of Monte Carlo experiments, while Section \ref{sec:ei} provides an empirical application using the timber auction data of \cite{lp08}.
Section \ref{sec:con} concludes.
The proofs of the lemmas, propositions, and theorems stated in the main text are collected in the Appendix, whereas the proofs of the auxiliary lemmas used therein are provided in the supplementary material.

\vspace{.01in}

\paragraph{Notation.}

The following notation is used throughout the paper.
For $a,b \in \mathbb{R}$, $a \wedge b$ and $a \vee b$ denote $\min\{a,b\}$ and $\max\{ a,b \}$, respectively.
The interior and closure of a set $S$ are denoted by $\mathrm{int}(S)$ and $\mathrm{cl}(S)$, respectively.
For a real-valued function $\phi$ defined on a set $S$, we write $\|  \phi \|_{S,p}  : = ( \int_{S} |\phi (t)|^p dt )^{1/p}$ for $1 \leq p < \infty$ and $\|  \phi \|_{S,\infty} : = \sup\{ | \phi(t) | : t \in S  \}$. 
Moreover, $\ell^\infty(S)$ and $\mathscr{C}(S)$ denote the spaces of bounded and continuous real-valued functions on $S$, respectively, and we write $\mathscr{C}_b ( S) : = \ell^\infty ( S ) \cap\mathscr{C}( S)$ for the space of bounded continuous real-valued functions on $S$.
Unless otherwise stated, function spaces are endowed with the sup-norm when regarded as normed spaces.
Any element of $\mathbb{R}^m$, $m \in \mathbb{N}$, is regarded as a column vector and the superscript $^\tau$ denotes transpose. For $v \in \mathbb{R}^m$, with a slight abuse of notation, $\|v\|_2$ and $\|v\|_\infty$ denote the Euclidean and sup norms, respectively. Vectors separated by commas are interpreted as vertically concatenated; e.g., for two (column) vectors $v_1$ and $v_2$, we will simply write $v = ( v_1 ,  v_2)$ for $v = (v_1^\tau \,  v_2^\tau)^\tau$.

For a random vector $Y$, $\mu_Y$, $\mathscr{S}_Y$, and $\Sigma_Y$ denote its mean, support, and variance-covariance matrix, respectively; if $Y$ is a scalar random variable, $\sigma_Y^2$ denotes its variance. The notation $F_Y$ is used for the joint c.d.f.\ of $Y$ and $f_Y$ denotes its density, while $P_{Y}$ is used for the corresponding probability measure. For a sample $\{Y_1,\ldots,Y_m\}$, $\bar{Y}$ denotes the sample average, and $Y_{(j:m)}$ denotes the $j$th order statistic. We use $\leadsto$ and $\overset{p}{\rightarrow}$ to denote weak convergence and convergence in probability, respectively. The usual stochastic order symbols $o_p$ and $O_p$ are also used. Finally, “w.p.1” and “w.p.a.1” abbreviate “with probability one” and “with probability approaching one,” respectively.

\section{Framework}	\label{sec:frame}

This section is divided into two subsections. In the first, I present the auction model and derive an expression for the expected revenue as a function of the reserve price. In the second, I introduce the assumptions on the data that allow for observed auction-specific heterogeneity.

\subsection{Auction model, expected revenue and optimal reserve prices}	\label{sec:modrev}

An indivisible object is sold through an open ascending auction to $n \geq 2$ potential bidders. The seller sets a reserve price $r \geq 0$ prior to the auction: if the highest bid is below $r$, the seller retains the object.
Both $n$ and $r$ are assumed to be common knowledge.

Within the IPV paradigm, each bidder $i  \in \{ 1,\dots, n\} $ has a private value $V_i \geq 0$. We assume that $ V_1, \dots, V_n $ are independent and identically distributed (i.i.d.), each following an absolutely continuous c.d.f.\ $F_V$ that is supported on $[ \ushort{v} ,   \bar{v} ] \subset \mathbb{R}_{\geq0}$ and admits a continuous p.d.f.\ on this interval. The seller is risk-neutral and attaches a value $v_s \in [  \ushort{v} , \bar{v} )$ to the auctioned object: the seller's utility equals the auction payment $W$ if a sale occurs, and equals $v_s$ otherwise.
We abstract from discrete bid increments and restrict attention to the following equilibrium outcome: if the highest valuation exceeds the reserve price $r$, the bidder with the highest valuation wins the object and pays $W = \max \{ r , V_{(n-1 : n )} \}$, i.e., the maximum of $r$ and the second-highest valuation; otherwise, the seller retains the object.

In this setting, the seller's expected revenue from setting a reserve price $r \geq 0$ becomes
\begin{equation}
\pi (r) :  =  F_V (r )^{n}   v_s +  n  [ 1 - F_V(r) ] F_V(r  )^{n-1} r   +  n  \int_{r}^{\infty}  v [ 1 - F_V(v  ) ]   d F_V (v )^{n-1}   ,
\label{eq:exprev}
\end{equation}
The first term on the right-hand side of Eq.\ (\ref{eq:exprev}) represents the benefits of retaining the object, while the remaining terms represent the ex-ante expected payment received from bidders when the object is sold: see, e.g., \cite{kri10} and \cite{hmz10} for further discussion. From these references, we also know that the set of optimal reserve prices
\begin{equation*}
\mathcal{R}  \ : = \   \underset{r \geq 0}{\arg\max} \ \pi (r) 
\end{equation*}
is a nonempty and compact subset of $( v_s , \bar{v} )$, and it may contain more than one element because we do not impose additional restrictions on $\pi$: see Section \ref{sec:mc} below for a concrete example.

\subsection{Data and auction-specific heterogeneity}		\label{sec:dgp}

The econometrician knows $(n , v_s)$ and observes data from a random sample
\begin{equation*}
\mathcal{S}_L  : = \{ ( \tilde{W}_{1}, X_1 ) , \dots , ( \tilde{W}_L, X_L )   \} ,
\end{equation*}
where $ \tilde{W}_{l}$ denotes the transaction payment in the $l$th auction, $X_l \in \mathbb{R}^d$ is a vector of covariates describing the auctioned object, and $L$ is the number of auctions in the sample. It is also assumed that each of these auctions has $n$ bidders and a nonbinding reserve price.

The next assumption formally characterizes the data-generating process, specifying, among other aspects, how the covariates $X_l$ affect the private values. These conditions are maintained throughout the remainder of the paper.

\begin{assump}		\label{assump:dgp}
Private values are given by $\tilde{V}_{jl} : = \exp \left( X_l^\tau \beta \right)     V_{jl} $, where $\beta \in \mathbb{R}^d$ is a parameter and $\left\{  ( X_l , V_{1l},\dots, V_{nl}  )   : l =1,\dots, L \right \}$ are independent draws from the distribution of a random vector $ ( X , V_{1},\dots, V_{n}  ) \in \mathbb{R}^{d+n}$ satisfying the following conditions:
\begin{enumerate}

\item $F_X$ is supported on a compact set and its variance-covariance matrix $\Sigma_X$ is positive definite.

\item $X$ and $( V_{1} , \dots, V_{n})$ are independent.

\item Private values $\{  V_{1} , \dots, V_{n} \}$ are i.i.d.\ with c.d.f.\ given by $F_V$, as described in the previous subsection: $F_V$ is continuous on $\mathbb{R}$ and is supported on $[ \ushort{v} ,   \bar{v} ] \subset \mathbb{R}_{\geq0}$. In addition, it admits a p.d.f.\ $f_V$ that is Lipschitz continuous on $[ \ushort{v} ,   \bar{v} ] $.

\end{enumerate}
Moreover, each payment $\tilde{W}_l$ is the $(n-1)$th order statistic of $\{ \tilde{V}_{1l} , \dots , \tilde{V}_{nl} \}$: $\tilde{W}_l = \tilde{V}_{(n-1:n),l}$.
\end{assump}

This assumption introduces a random vector of observable covariates $X$ to characterize the auctioned object and imposes regularity conditions required for the asymptotic analysis below. The specification  $\tilde{V}_{jl}  = \exp ( X_l^\tau \beta ) V_{jl} $ ensures that the expected revenue function $\pi$ in Eq.\ (\ref{eq:exprev}) corresponds to that of a baseline object with characteristics normalized to $X=0$.
Hence, under this normalization and in the absence of a binding reserve price, the “homogenized” payment becomes $W  =V_{(n-1:n)}$ and its c.d.f.\ is given by $F_W (w) : = \Prob ( W \leq w ) = \Prob ( \tilde{W} \leq w  \mid X = 0 ) $, where $\tilde{W} : = \tilde{V}_{(n-1:n)} = \exp ( X^\tau \beta )  W$.

In words, under Assumption \ref{assump:dgp}, $V_{jl}$ can be interpreted as the valuation that bidder $j$ would have assigned to the auctioned object in auction $l$ had its characteristics been $X_l = 0$. This normalization is widely used in the empirical auction literature and provides a convenient benchmark to avoid the curse of dimensionality.
Note also that the normalization $X=0$ is without loss of generality, since the covariates can be redefined as $X-x$ when the object of interest has characteristics described by $x$.
Empirical examples of normalization procedures in the literature include \cite{hhs16} for timber auctions, \cite{leekim14} for used-car auctions, and \cite{res08} for eBay iPod Mini auctions.

We assume $d\geq1$ as the absence of covariates entails a different (simpler) bootstrap structure. The number of bidders $n$ is constant across auctions and known to bidders and the econometrician. The vector of covariates $X$ may be discrete or continuous, provided its variance-covariance matrix $\Sigma_X$ is nonsingular, which rules out having an intercept in $X$.
No strong restrictions are imposed on $F_V$; its p.d.f.\ $f_V$ may not be bounded away from zero and is only assumed Lipschitz continuous on its compact support. These conditions are weaker than the regularity assumptions often maintained in the empirical first-price auction literature.

Finally, I remark that, under Assumption \ref{assump:dgp}, super-consistent estimation of $\beta$ is generally ruled out, with the  exception of the case where $n=2$ and $f_V(\ushort{v})>0$.
Indeed, even if $f_V$ is bounded away from zero, the density of $W$ vanishes at the upper endpoint of its support and, provided $n\geq 3$, also at the lower endpoint: this rules out boundary-based procedures such as that of \cite{kt93}.
Consequently, estimating $\beta$ by OLS in a first stage is both natural and efficient. The resulting first-stage estimation error is explicitly incorporated into the asymptotic analysis developed below.

\section{Estimation of expected revenue: computational aspects}		\label{sec:estier}

In this section, for a given seller's valuation $v_s \in [ \ushort{v}  , \bar{v} ) $, I construct the estimator of the expected-revenue function $\pi$ and describe a convenient computational procedure. Before proceeding, recall that, under Assumption \ref{assump:dgp}, $\pi (r)$ is the expected revenue from an auction in which the exogenous covariates of the auctioned object are set to zero, i.e., $X = 0$.

The construction of the proposed estimator begins with a useful characterization of $F_V$ in terms of $F_W$, the c.d.f.\ of the $(n-1)$th order statistic $W= V_{(n-1 : n)}$. From \citet[Theorem 1]{ah02ecma}, for any $v \in \mathbb{R}$, we have
\begin{equation}
\label{eq:Fpoly}
F_V ( v )   =  
\left\{
\begin{array}{ll}
0  &  \text{if} \  v \leq \ushort{v}  , \\
\lowmathcal{p}^{-1} [  F_W (v) ] & \text{if} \  \ushort{v} < v  < \bar{v}  ,  \\
1  & \text{otherwise} ,
\end{array}
\right.
\end{equation}
where $\lowmathcal{p} ( t ) : =  n t^{n-1} - (n-1)t^n $ is a polynomial on $\mathbb{R}$ and, for $y \in [0, 1]$, $\lowmathcal{p}^{-1} ( y )$ denotes the unique root of the polynomial $\lowmathcal{p} ( \cdot) -  y$ lying in the interval $[0, 1 ]$. For our purposes, the usefulness of this result lies in the fact that the root of this polynomial can be computed at very low computational cost.
Moreover, it can be easily shown that
\begin{equation}
\label{eq:FW}
{F}_W (w )  \, =  \, \Prob \left( \tilde{W} \exp\left( - X^\tau \beta  \right) \leq w   \right) 
\, =  \,  F_U \left[  \log(w)  \right] \quad \text{for}  \ w >0 ,
\end{equation}
where $F_U (u) : =\Prob (  U \leq u ) $ and $U : = \log( W ) = \log(\tilde{W}) - X^\tau \beta$ is the logarithm of the homogenized payment.


It follows from Eq.\ (\ref{eq:Fpoly}) that $\pi(r)$ can be expressed as the following functional of $F_W$:\footnote{The last term on the right-hand side follows from the identities $d F_V^{\,n-1} = (n-1)F_V^{n-2} dF_V $ and $ n(n-1)F_V^{n-2}(1 - F_V)dF_V = dF_W $, with the latter arising from $F_W (w) = \lowmathcal{p} [F_V(w) ]$.}
\begin{equation}
\label{eq:piFW}
\pi (r  )  =  \lowmathcal{p}^{-1} [ F_W (r) ]^n   v_s +  n  \left\{ 1 - \lowmathcal{p}^{-1} [ F_W (r) ] \right\} \lowmathcal{p}^{-1} [ F_W (r) ] ^{n-1} r   +   \int_{r}^{\infty} w \ d F_W(w) .
\end{equation}
Then, Eqs.\ (\ref{eq:FW}) and (\ref{eq:piFW}) suggest the following three-step procedure:

\begin{enumerate}

\item Run an OLS regression of $\log(\tilde{W}_l)$ on $X_l$ and a constant.

Compute the estimated homogenized log-payments  $\hat{U}_l := \log(\tilde{W}_l) - X_l^\tau \hat\beta$, without subtracting the constant term, and construct the following empirical c.d.f.:
\begin{equation*}
\hat{F}_U (u )  \,   : =  \,  \frac{1}{L}\sum_{l=1}^{L}  \mathds{1} \left(    \hat{U}_l    \leq  u \right) .
\end{equation*}

\item Estimate  $F_W$ using Eq.\ (\ref{eq:FW}):
\begin{equation}
\hat{F}_W (w )   \,   : =  \,  \left\{ 
 \begin{array}{ll}
\hat{F}_U \left[    \log(w) \right]  & \text{if} \ w > 0  , \\
 0 & \text{otherwise}.
 \end{array}
 \right.
 \label{eq:estiG}
\end{equation}

\item Estimate $\pi (r)$ by the plug-in method:
\begin{equation}
\label{eq:defesti}
\hat{\pi} ( r)    \, : =  \,  \hat{F}_V ( r ) ^n   v_s +  n  \left[ 1 - \hat{F}_V ( r )  \right] \hat{F}_V ( r ) ^{n-1} r   +   \int_{r}^{\infty} w \ d \hat{F}_W (w)   \quad \text{for} \ r \geq 0 ,
\end{equation}
where $\hat{F}_V ( r )  : = \lowmathcal{p}^{-1} [ \hat{F}_W (r) ]$.

\end{enumerate}
Although the third step may appear computationally intensive due to the integral expression, it is not, as explained below.

Since $ \hat{F}_W$ takes values only in the discrete set $\{ 0, 1/L , 2/L , \dots , 1 \}$ with corresponding breakpoints at 
$$\hat{W}_{(1)} := \exp ( \hat{U}_{(1:L)} ) < \dots   <  \hat{W}_{(l)} : =  \exp ( \hat{U}_{(l:L)} )  < \dots <  \hat{W}_{(L)} : =  \exp ( \hat{U}_{(L:L)} ) ,$$
$\hat{\pi} $ is indeed a right-continuous piecewise linear function, having also the same break points.
To illustrate the computational convenience of this feature,  define $\xi_l : = \lowmathcal{p}^{-1} (  l / L)$ for $l = 0 , 1 , \dots , L $, as well as $ \hat{W}_{(0)} : = -  \infty$ and $\hat{W}_{(L+1)}: =  \infty$, noting that $\xi_l $ is indeed the unique root of the polynomial $\lowmathcal{p} ( \cdot )  - l/L$ in $[0,1]$.
Then, $\hat{F}_V$ can be simply expressed as
\begin{equation*}
\hat{F}_V ( v )   =  
\sum_{l=1}^{L} \, \xi_l \, \mathds{1}  \left(  \hat{W}_{(l)} \leq  v < \hat{W}_{(l+1)}    \right) \quad  \text{for} \ v \in \mathbb{R}  , \\
\end{equation*}
or equivalently as $ \hat{F}_V( v ) = \xi_{\iota(v)} $, where $\iota: \mathbb{R} \rightarrow \{0, 1, \dots , L \} $ is an index function defined by
\begin{equation}
\label{eq:defesti1}
\iota(v)  : =  
\left\{
\begin{array}{ll}
0  &  \text{if} \ v < \hat{W}_{(1)} , \\
\max \left\{ l  \in  \{ 1, \dots , L \} :  \hat{W}_{(l)} \leq v   \right\}  &   \text{otherwise} .
\end{array}
\right.
\end{equation}
Moreover, adopting the convention that $\int_{r}^{\infty}$ means integrating over $( r , \infty)$, we have the identity
\begin{equation}
\label{eq:defesti2}
\int_{r}^{\infty} w \ d \hat{F}_W (w)    =  \frac{1}{L} \sum_{l=\iota(r)+1}^L \hat{W}_{(l)} .
\end{equation}
Thus, combining together Eq.\ (\ref{eq:defesti}) with Eqs.\ (\ref{eq:defesti1})-(\ref{eq:defesti2}) yields
\begin{equation*}
\hat{\pi} ( r)  \   = \     \xi_{\iota(r)}^n \,  v_s \ + \ n \left( 1 -   \xi_{\iota(r)} \right)  \xi_{\iota(r)}^{n-1}r 
\ +  \  \frac{1}{L} \sum_{ l= \iota(r) +1}^{ L} \hat{W}_{(l)} 
\end{equation*}
or, equivalently,
\begin{equation}
\label{eq:piecelinear}
\hat{\pi} ( r)   \ = \  \sum_{l=0}^L \left( {A}_l   + b_l r  \right) \mathds{1} \left( \hat{W}_{(l)} \leq r < \hat{W}_{(l+1)}  \right)
\end{equation}
with the constant and slope coefficients given by 
\begin{equation*}
\left( {A}_l  , b_l \right)  :  =  \left\{
\begin{array}{ll}
\left( \xi_{l}^n \, v_s  + (1/L) \sum_{l^\prime=l+1}^L \hat{W}_{(l^\prime)} , \ n \left( 1 -   \xi_{l} \right)   \xi_{l}^{n-1}  \right)  & \text{for}  \  0 \leq l  < L  , \\
\left( v_s  , \ 0 \right)  & \text{for}  \  l  = L  .
\end{array}
\right.   
\end{equation*}
Algorithm 1 provides an efficient procedure for computing these coefficients.

Two features of the representation in Eq.\ (\ref{eq:piecelinear}) are worth highlighting. First, the integral originally appearing in (\ref{eq:piFW}) is eliminated, yielding a closed-form expression for the estimated expected revenue function.
Thus, the computation of $\hat\pi$ does not require numerical integration, thereby avoiding approximation errors and the introduction of additional tuning parameters, whose effects are often ignored in asymptotic analyses.
Second, the slopes $b_l$ are nonrandom and depend only on the roots $ \xi_{l}$: this deterministic property plays an important role in the bootstrap procedures developed in the following sections.

\begin{algorithm}\small
\caption{Computing breakpoints and coefficients for Eq.\ (\ref{eq:piecelinear})}
\algorithmicrequire   $(n ,v_s)$; $(  \tilde{W}_1 ,  X_1 )  , \dots ,   (  \tilde{W}_L ,  X_L ) $
\begin{algorithmic}[1]

\State $\hat{\beta} \ \gets$ slope coefficients from OLS regression of $\log(\tilde{W}_l)$ on $X_l$ and a constant

\For{$l = 1, \dots, L$}
    \State $\hat{W}_l \gets \tilde{W}_l \exp(-X_l^\prime \hat{\beta})$
\EndFor

\State $\hat{W}_{(1)} < \dots < \hat{W}_{(L)} \ \gets $ sort homogenized payments in ascending order

\For{$l = 0, \dots, L-1$}
    \State $\xi_l \gets \text{root of polynomial}\ n \, t^{n-1} - (n-1)t^n  - l/L \text{ in } [0,1]$
    \State $A_l  \gets \xi_l^n v_s + \text{sum}( \hat{W}_{(l+1)} , \dots ,  \hat{W}_{(L)} )/ L$
     \State $b_l \gets n (1 - \xi_l)\, \xi_l^{n-1}$
\EndFor
\State $(A_L , b_L)  \gets (v_s , 0)$
\end{algorithmic}
\algorithmicensure  breakpoints $\hat{W}_{(1)} < \dots < \hat{W}_{(L)}$; coefficients $(A_l , b_l)_{l=0}^L$
\end{algorithm}

\section{Asymptotics and inference}	\label{sec:asinf}

In this section, I develop the asymptotic theory and inference procedures for the proposed estimator. Specifically, Subsection \ref{ss:wc} establishes the weak convergence of $\hat{\pi}$, while Subsection \ref{ss:bootap} introduces a bootstrap method for approximating its limiting distribution. Finally, Subsection \ref{ss:inf} presents bootstrap-based inference procedures for the expected revenue, including confidence intervals and uniform-in-$r$ confidence bands.


The asymptotic analysis is conducted as $L \rightarrow \infty$, holding the data-generating process fixed, i.e., for fixed structural parameters $(  F_X  , F_V , \beta)$.
Moreover, $\pi$ and $\hat\pi$ are regarded as bounded functions on an interval $[\ushort r , \bar r]  \subset (v_s , \bar v)$, where $\ushort r < \bar r$ are arbitrarily chosen such that the set of optimal reserve prices is contained in $(\ushort r , \bar r)$, $\mathcal{R}  \subset (\ushort r , \bar r)  $. As noted in Subsection \ref{sec:modrev}, $\mathcal{R}$ is a compact subset of $ (v_s , \bar v)$; hence, this condition is satisfied by choosing $\ushort r$ and $\bar r$ sufficiently close to $v_s$ and $\bar v$, respectively. This leaves the case of a reserve price $r \leq v_s$ out of the analysis, but such a case is never optimal.

\subsection{Weak convergence}		\label{ss:wc}

In this subsection, I derive the asymptotic distribution of $\sqrt{L} ( \hat\pi - \pi )$ as a bounded function on $[\ushort r , \bar r]$, i.e., $\sqrt{L} ( \hat\pi - \pi ) \in \ell^\infty ( [\ushort r , \bar r]  )$. To this end, I first establish the weak convergence of $\sqrt{L}(\hat{F}_W  -  F_W)$ as a bounded function on $[ \ushort{r} , \infty ) $. Define the covariance function
\begin{equation*}
\Psi_\omega  (  w  , y )  :  =      F_W ( w  \wedge y )   -   F_W ( w   )   F_W ( y )    +  f_U \left[ \log(w) \right]   \mu_X^\tau \Sigma_X^{-1}   \mu_X f_U \left[ \log(y) \right] \sigma_U^2 
\end{equation*}
for $w , y > 0$, where $f_U$ denotes the p.d.f.\ of $U : = \log(W)  = \log ( \tilde{W}) - X ^\tau \beta$ and $\mu_X = \E ( X )$.

\begin{lemma}	\label{lem:weakconG}

Under Assumption \ref{assump:dgp}, $\sqrt{L}(\hat{F}_W  -  F_W ) \leadsto \mathbb{G}_{\omega}$ in $\ell^\infty (  [ \ushort{r} , \infty )  )$, where $\mathbb{G}_{\omega}$ is a tight centered Gaussian process with covariance function $\Psi_\omega $, i.e.,
$  \mathrm{cov} \left(  \mathbb{G}_{\omega} (w)  ,  \mathbb{G}_{\omega} (y)  \right)   =  \Psi_\omega  (  w  , y ) $.

\end{lemma}

Several remarks are noteworthy.
First, the covariance function $\Psi_\omega$ consists of two components: the term
$F_W(w\wedge y)-F_W(w)F_W(y)$ captures the sampling variation from estimating $F_W$ by the (pseudo-)empirical distribution function $\hat{F}_W$,
while $  f_U [ \log(w) ]   \mu_X^\tau \Sigma_X^{-1}  \mu_X \sloppy f_U [ \log(y) ] \sigma_U^2  \geq 0$ arises from the estimation of $\beta$ by OLS and vanishes as a function only when $\mu_X = 0$. As discussed above, here the estimation error generated by the homogenization step is non-negligible and must be accounted for in the asymptotic analysis. In contrast, the empirical auction literature often appropriately ignores this component when first-step estimators (e.g., OLS regression coefficients from the homogenization procedure) converge at a faster rate than second-step estimators (e.g., kernel estimators of the private-value p.d.f.\ in first-price auctions).

Second, this lemma might be of independent interest for conducting inference on $F_V$ when only the winning bid is observed.\footnote{E.g., see \cite{bugni26in} and \cite{ka26finauc} for recent developments in this regard.} Specifically, since  $F_V(v)=\lowmathcal{p}^{-1} [F_W(v)]$, the obtained weak convergence result can be used to construct confidence intervals and uniform confidence bands for $F_V$ based on the bootstrap method described in the next subsection.
Moreover, the result could be extended to weak convergence in $\ell^\infty(\mathbb{R}_{>0})$ under alternative regularity conditions. However, since this extension is not needed for the subsequent analysis and would complicate the exposition, I do not pursue it here.

The next theorem is the main technical result of the paper, as it establishes the weak convergence of $\sqrt{L} ( \hat\pi - \pi ) $ as a bounded function on $ [ \ushort{r} , \bar{r} ]$. To obtain this result, I apply the functional delta method, combining Lemma \ref{lem:weakconG} with the fact that $\hat\pi(r)$ can be expressed as a nonlinear functional of $\hat{F}_W$ through the representation in Eq.\ (\ref{eq:defesti}).

\begin{theorem}
\label{thm:main}

Suppose that Assumption \ref{assump:dgp} holds and let $\mathbb{G}_\pi$ be a tight centered Gaussian process in $\ell^\infty ( [ \ushort{r} , \bar{r} ] )$ with covariance function $\Psi_{\pi} $, i.e.\ $  \mathrm{cov} \left(  \mathbb{G}_{\pi} ( r )  ,  \mathbb{G}_{\pi} ( s )  \right)  =  \Psi_{\pi} (  r  , s )  $, where
\begin{eqnarray}
\Psi_{\pi} (  r  , s )    &  : =&  \varphi_{F_W}  (r )\varphi_{F_W} (s ) \Psi_{\omega} ( r , s ) \ - \   \varphi_{F_W} (r) \int_s^\infty \Psi_\omega (r, y) \, dy   \nonumber  \\
&  &  \quad  \quad - \  \varphi_{F_W} (s)  \int_r^\infty \Psi_\omega (t , s) \, dt  \ + \ \int_r^\infty \int_s^\infty \Psi_\omega (t , y) \, dt \,  dy  \quad \text{and}   \label{eq:defvarphi} \\
\varphi_{F_W}  (r)  & : =  &     n \,  \lowmathcal{p}^{-1} \left[{F_W} ( r ) \right]  ^{n-2} \, \frac{    (v_s - r) \, \lowmathcal{p}^{-1} \left[{F_W} ( r ) \right] +   (n-1) r \,   \left\{ 1 - \lowmathcal{p}^{-1} \left[{F_W} ( r ) \right]  \right\} }
{ \lowmathcal{p}^\prime \{ \lowmathcal{p}^{-1} \left[{F_W} ( r ) \right] \}  } 
\, - \,  r  .    \nonumber
\end{eqnarray}
Then, 

\begin{enumerate}

\item $\sqrt{L} ( \hat\pi - \pi ) \leadsto \mathbb{G}_\pi$ in $\ell^\infty ( [ \ushort{r} , \bar{r} ] )$. 

\item $\mathrm{V}_{\pi} (r)  : = \Psi_{\pi} (  r  , r )  > 0$ for all $r \in [ \ushort{r} , \bar{r} ]$.

\end{enumerate}

\end{theorem}

This theorem characterizes the asymptotic behavior of the proposed estimator and provides the foundation for the inference procedures developed below. Specifically, viewing $\hat{\pi}$ as an estimator of the entire function $\pi$ over $[\ushort{r},\bar{r}]$, this theorem implies uniform consistency at the parametric rate:
\begin{equation*}
\left\|\hat{\pi}-\pi \right\|_{[\ushort{r},\bar{r}],\infty} \, = \, O_p \left( L^{-1/2} \right) .
\end{equation*}
Moreover, for a fixed $r \in [\ushort{r},\bar{r}]$,
$\sqrt{L}[\hat{\pi}(r)-\pi(r)]$
converges (pointwise) in distribution to $ \mathbb{G}_\pi (r)$, a mean-zero normal random variable with variance $\mathrm{V}_{\pi}(r) >0$.
Therefore, for given $\alpha \in (0,1)$ and $r \in [\ushort{r},\bar{r}]$, an asymptotically valid pointwise confidence interval for $\pi(r)$ with level $1-\alpha$ is
\begin{equation*}
\left[
\hat{\pi}(r)  - z_{ 1 - \frac{\alpha}{2} } \, \sqrt{ \frac{\hat{\mathrm{V}}_{\pi} (r)}{L} }  \, , \, 
\hat{\pi}(r)  - z_{ \frac{\alpha}{2} } \, \sqrt{ \frac{\hat{\mathrm{V}}_{\pi} (r)}{L} }  
\right] ,
\end{equation*}
where $z_a$ denotes the $a$-quantile of a standard normal distribution, $a \in (0,1)$, and $\hat{\mathrm{V}}_{\pi}(r)$ is a consistent plug-in estimator of the asymptotic variance $\mathrm{V}_{\pi}(r)$.
However, I do not pursue this approach because the expression for the asymptotic variance is relatively complicated, and its estimation would require estimating nonparametric components that may adversely affect finite-sample performance. Instead, the inference procedures proposed below rely on the bootstrap method.

\subsection{Bootstrap approximation}	\label{ss:bootap}

In this subsection, I propose a smooth bootstrap method to approximate the distribution of $\mathbb{G}_\pi$ and establish its asymptotic validity. I begin by introducing two assumptions about the kernel function $\kappa$ and the bandwidth $\lowmathcal{h} > 0$.

\begin{assump}
\label{assump:ker}
Let $\kappa$ be a nonnegative kernel function supported on $[-1,1]$, symmetric around zero, and integrating to one. Assume further that $\kappa$ admits two continuous derivatives on $\mathbb{R}$.
\end{assump}

\begin{assump}
\label{assump:band}
Let $\lowmathcal{h} > 0$ be a bandwidth satisfying $\lowmathcal{h} \rightarrow 0$ and  $\lowmathcal{h}^4 L  \rightarrow \infty$.
\end{assump}

I now construct the bootstrap analogue of $\sqrt{L} ( \hat\pi - \pi )$ used to approximate the distribution of $\mathbb{G}_\pi$. First, consider
\begin{equation*}
\hat{f}_{U, \lowmathcal{h} } ( u )    : =    \frac{1}{ \lowmathcal{h}  L } \sum_{l=1}^L  \kappa  \left(  \frac{u -  \hat{U}_l  }{\lowmathcal{h}} \right) 
\end{equation*}
as a kernel-density estimator of $f_U $, and let $\{  {U}_1^{\ast} , \dots ,  {U}_L^\ast \}$ be conditionally independent draws from the distribution with c.d.f.
\begin{equation*}
\hat{F}_{U,  \lowmathcal{h} } ( u )  \ : = \  \int_{-\infty}^u   \hat{f}_{U, \lowmathcal{h}} ( t ) dt  ,
\end{equation*}
given the sample $\mathcal{S}_L$. Observe that, unlike the usual bootstrap procedure, which resamples directly from the empirical distribution of the observed data, here $\{  {U}_1^{\ast} , \dots ,  {U}_L^\ast \}$ is generated from a smoothed version of the empirical distribution of $\hat{U}_l $. Second, setting
\begin{equation*}
\tilde{U}_l^\ast  \, : = \,   X_l ^\tau \hat{\beta} +  {U}_l^\ast   \ \ \text{and}  \ \ 
\hat{U}^\ast_l   =   \tilde{U}_l^\ast  -  X_l ^\tau  \hat\beta^\ast  \ \ \text{for}  \ \   l=1, \dots , L,
\end{equation*}
where $\hat\beta^\ast$ is the vector of slope coefficients from the OLS regression of $\tilde{U}_l^\ast$ on $X_l$ and a constant,
define the empirical distribution function
\begin{equation*}
\hat{F}^\ast_U ( u )  \ := \ \frac{1}{L} \sum_{l=1}^L \mathds{1} \left( \hat{U}_l^\ast \leq u \right) .
\end{equation*}
Before proceeding, note that $X_l$ is held fixed and not bootstrapped, while the estimated homogenized bootstrap log-payments $\hat{U}^\ast_l$ satisfy the relationship
\begin{equation*}
\hat{U}^\ast_l   =   \tilde{U}_l^\ast  -  X_l ^\tau  \hat\beta^\ast   \,  = \,  {U}_l^\ast -  X_l ^\tau \left(   \hat\beta^\ast   - \hat\beta \right) .
\end{equation*}
Third and lastly, I use $\sqrt{L} \left[ \hat{\pi}^\ast ( r  )  -  \hat{\pi}_{\lowmathcal{h}} ( r  )   \right] $ as the bootstrap analog of $\sqrt{L} [\hat\pi ( r)   - \pi  (r ) ]$ for $ r \in  [ \ushort{r} ,  \bar{r} ] $,
where
\begin{eqnarray*}
\hat{\pi}^\ast (r  )  & : = &  \lowmathcal{p}^{-1} [ \hat{F}_W^\ast  ( r )   ]^n   v_s +  n  \left\{ 1 - \lowmathcal{p}^{-1} [  \hat{F}_W^\ast  ( r ) ] \right\}  \lowmathcal{p}^{-1} [  \hat{F}_W^\ast  ( r )  ] ^{n-1} r   +   \int_{r}^{\infty} w \ d  \hat{F}_W^\ast  (w)  ,   \\
\hat{\pi}_{\lowmathcal{h}} (r  ) &  : =  & \lowmathcal{p}^{-1} [  \hat{F}_{W ,  \lowmathcal{h}} (  r ) ] ^n   v_s +  n  \left\{ 1 - \lowmathcal{p}^{-1} [ \hat{F}_{W ,  \lowmathcal{h}} (  r )] \right\} \lowmathcal{p}^{-1} [ \hat{F}_{W , \lowmathcal{h}} (  r )] ^{n-1} r   +   \int_{r}^{\infty} w \ d \hat{F}_{W,   \lowmathcal{h} } (  w )  , 
\end{eqnarray*}
$\hat{F}_W^\ast  ( w ) : = \hat{F}^\ast_U [  \log (w) ]$, and $\hat{F}_{W,  \lowmathcal{h} } (  w ) : = \hat{F}_{U, \lowmathcal{h} }[ \log(w)]$ for $w >0$.
In words, $\hat{\pi}^\ast $ and $\hat{\pi}_{\lowmathcal{h}}$ are the expected revenue functions associated with the private-value distributions $\lowmathcal{p}^{-1} \circ \hat{F}_W^\ast $ and $\lowmathcal{p}^{-1} \circ  \hat{F}_{W ,  \lowmathcal{h}}$, respectively.

I remark that $\hat{\pi}^\ast$ can be computed by applying Algorithm 1 to the bootstrap observations $(\tilde{W}_l^\ast,X_l)$, where $\tilde{W}_l^\ast := \exp(\tilde{U}_l^\ast)$.
Moreover, Lines 7 and 9 need to be evaluated only once, since they do not depend on the sample. Hence, computing $\hat{\pi}^\ast$ only requires calculating the bootstrap analogues of the intercepts $(A_l )_{l=0}^L$.
In contrast, since $\hat{F}_{U, \lowmathcal{h}  }$ and therefore $\hat{F}_{W, \lowmathcal{h}  }$ are continuous, $\hat{\pi}_{\lowmathcal{h} }$ is not a piecewise linear function, so Algorithm 1 cannot be applied in this case. However, the computation of the integral expression $ \int_{r}^{\infty} w \ d \hat{F}_{W,   \lowmathcal{h} } (  w ) $ may not rely on numerical approximation, as several integrals of kernel functions --such as the triweight-- admit closed-form expressions. Moreover, this integral does not need to be recomputed across bootstrap replications.

The next lemma shows that $\sqrt{L}(  \hat{\pi}^\ast - \hat{\pi}_{\lowmathcal{h}} )$ consistently approximates the limiting process $\mathbb{G}_\pi$, and that the approximation carries over to the quantiles.
To state the result, fix $a\in(0,1)$ and $r\in[\ushort{r},\bar r]$, and let $q_a(r)$ denote the $a$-quantile of $\mathbb{G}_\pi(r)$, namely
\begin{equation*}
q_a (r ) \,  : = \,  \inf   \left\{ y \in \mathbb{R} : \,    \Prob (  \mathbb{G}_\pi (r) \leq y  ) \geq a \right\}  .
\end{equation*}
Let $\hat q_a(r)$ be its bootstrap analogue, defined as the conditional $a$-quantile of
$\sqrt{L} [ {\hat{\pi}^\ast(r)-\hat{\pi}_{\lowmathcal{h}}(r)} ]$ given the sample $\mathcal{S}_L$:
\begin{equation*}
 \hat{q}_a (r ) \,  : = \,  \inf \,  \left\{ y \in \mathbb{R} : \, 
 \Prob \left( \sqrt{L} \left[\hat{\pi}^\ast (r)  - \hat{\pi}_{\lowmathcal{h}} (r) \right] \leq y  \mid \mathcal{S}_L \right) \, \geq \, a \right\}  . 
\end{equation*}
Similarly, for a (possibly stochastic) nonempty subset $R\subseteq[\ushort{r},\bar r]$, let
\begin{equation*}
q_a^\infty ( R ) \,  : = \, 
\inf\left\{  y \in \mathbb{R}_{\geq 0} : \, \Prob \left( \left\|  \mathbb{G}_\pi  \right\|_{R,\infty } \leq y  \right)  \geq a   \right\}
\end{equation*}
denote the $a$-quantile of $  \|  \mathbb{G}_\pi  \|_{R,\infty }  $, and let
\begin{equation*}
 \hat{q}_{  a}^\infty ( R  )  \, : =  \,   \inf\left\{  y \in \mathbb{R}_{\geq 0} : \ \Prob \left(  \sqrt{L} \| \hat{\pi}^\ast   - \hat{\pi}_{\lowmathcal{h}}    \|_{ R , \infty  }   \leq y   \mid  \mathcal{S}_L \right)  \geq a   \right\} 
\end{equation*}
be the corresponding conditional bootstrap $a$-quantile.
Finally, let $\overset{\ast}{\leadsto}$ denote weak convergence in probability, conditional on the sample $\mathcal{S}_L$; a formal definition is given in Eq.\ (\ref{aeq:wcbequiv}) of the Appendix. 

\begin{lemma}
\label{lem:mainboot}
Suppose that Assumptions \ref{assump:dgp}-\ref{assump:band} are satisfied. Then,
\begin{enumerate}

\item $ \sqrt{L} \,  (\hat{F}_W^\ast -  \hat{F}_{W , \lowmathcal{h}} )    \,   \overset{\ast}{\leadsto}   \,   \mathbb{G}_{\omega}$ in $\ell^\infty ( [ \ushort{r} , \infty ) ) $.

\item $\sqrt{L} (\hat{\pi}^\ast   - \hat{\pi}_{ \lowmathcal{h} } ) \overset{\ast}{\leadsto} \mathbb{G}_\pi$ in $\ell^\infty (  [ \ushort{r} ,  \bar{r} ]  ) $.

\item $\hat{q}_{  a} ( r  )   \overset{p}{\rightarrow}   q_{  a} ( r  )$ and $\hat{q}_{  a}^\infty ( R  )  \overset{p}{\rightarrow}    q_{a}^\infty ( R  )$ for any $a \in (0,1)$, $r \in [ \ushort{r} ,  \bar{r} ]$, and any non-stochastic, nonempty, and compact subset $R \subseteq [ \ushort{r} ,  \bar{r} ]$.

\end{enumerate}
\end{lemma}

The first and second parts of this lemma establish the validity of the proposed smooth bootstrap procedure for approximating the limiting distributions $\mathbb{G}_{\omega}$ and $\mathbb{G}_\pi$, respectively. The third part shows that the bootstrap can also be used to consistently estimate the quantiles of $\mathbb{G}_\pi (r)$ and $ \|  \mathbb{G}_\pi  \|_{R,\infty }$. In particular, it is shown in the Appendix that, for any nonempty compact subset $R \subseteq [\ushort{r},\bar{r}]$, the c.d.f.\ of $\|\mathbb{G}_\pi \|_{R,\infty}$ is indeed continuous and strictly increasing on $\mathbb{R}_{\geq 0}$. Consequently, bootstrap-based approximations of these quantiles provide the foundation for the inference procedures developed in the next subsection.
I also remark that the third part can be extended to certain classes of random sets $R$, although this extension is not included in the lemma and is instead addressed in Section \ref{subsec:orpresti} below.

\subsection{Inference on the expected revenue}	\label{ss:inf}

Based on the preceding results, I construct pointwise confidence intervals and uniform confidence bands for $\pi$, and I establish their asymptotic validity. The corresponding critical values are obtained using the bootstrap method developed in the previous subsection.

For given $\alpha \in (0 , 1)$ and a reserve price $r \in [\ushort r, \bar r]$, I propose the following confidence interval of level $1-\alpha$ for $\pi (r)$ based on the bootstrap reverse-percentile method:
\begin{equation*}
\mathrm{CI}_{1-\alpha} (r) \,  : = \,  \left[   \hat{\pi} (r)   -  \frac{ \hat{q}_{1-{\alpha}/{2}} (r) }{\sqrt{L}} \,  ,  \,
 \ \hat{\pi} (r)   -   \frac{ \hat{q}_{{\alpha}/{2}} (r) }{\sqrt{L}} \right] ,
\end{equation*}
The validity of $\mathrm{CI}_{1-\alpha} (r) $ is established in the following proposition.
Before stating the result, I remark that Efron’s percentile method would not be valid in this setting because the bootstrap difference $\hat{\pi}^\ast-\hat{\pi}_{\lowmathcal{h}}$ is centered at $\hat{\pi}_{\lowmathcal{h}}$ rather than at $\hat{\pi}$.


\begin{propo}
\label{propo:pointboots}
Suppose Assumptions \ref{assump:dgp}-\ref{assump:band} hold. Then, the proposed confidence interval is asymptotically valid:
$\Prob \left\{ \pi (r)  \in   \mathrm{CI}_{1-\alpha} (r)  \right\}  \rightarrow 1   - \alpha$ for any $ r \in [\ushort r, \bar r]$.
\end{propo}


While the preceding pointwise confidence intervals quantify uncertainty in $\pi(r)$ at individual values of $r$, they do not account for the fact that $\pi$ might be examined over the entire interval $[\ushort{r},\bar{r}]$. Uniform confidence bands overcome this limitation by providing simultaneous coverage over $[\ushort{r},\bar{r}]$, allowing for valid inference on the function's overall shape and economically relevant features.

To address this issue, I propose the confidence band
\begin{equation*}
\mathrm{CB}_{1-\alpha} \,  : = \,  \left\{  \mathrm{CB}_{1-\alpha} (r) :  r \in [ \ushort r , \bar r]  \right\} ,
\end{equation*}
where
\begin{equation*}
\mathrm{CB}_{1-\alpha} (r)   \, : = \,    \left[    \hat{\pi} (r)   -  \frac{   \hat{q}_{1-{\alpha}}^\infty ( [ \ushort r , \bar r] )  }{\sqrt{L}} ,  \,
  \hat{\pi} (r)   +   \frac{   \hat{q}_{1-{\alpha}}^\infty ( [ \ushort r , \bar r] )   }{\sqrt{L}} \right] .
\end{equation*}
Its asymptotic validity is established in the following proposition.

\begin{propo}
\label{propo:cband}
If Assumptions \ref{assump:dgp}-\ref{assump:band} hold, then $\Prob \left\{  \forall r \in   \left[ \ushort r, \bar r \right]  , \, \pi (r)  \in  \mathrm{CB}_{1-\alpha} (r) \right\}      \rightarrow 1   - \alpha $.
\end{propo}

To conclude this section, I highlight that the proposed bootstrap-based procedures provide feasible inference on the expected revenue function while avoiding direct estimation of the cumbersome covariance function $\Psi_{\pi}$.

\section{Optimal reserve prices: set estimation and inference} 	\label{subsec:orpresti}

In this section, I propose estimators for the maximum expected revenue, $\pi_{\max}  : = \max_{r\geq0} \pi (r)$, and the set of optimal reserve prices, $\mathcal{R} = \arg\max_{r\geq0} \pi (r) \subset  [\ushort{r} , \bar{r}  ]$, without imposing additional conditions on the primitives of the auction model. I show that both estimators are consistent, and I also provide valid formulas to build a confidence interval for $\pi_{\max}$ and a confidence region for $\mathcal{R}$. 


Starting with the estimator of $\pi_{\max}$, note that $\pi_{\max}  =  \max_{r \in [\ushort r , \bar r]} \pi (r)$ and consider its empirical counterpart $\hat{\pi}_{\max} : = \sup_{r \in [\ushort r , \bar r]} \hat{\pi} (r)$. Since $\hat\pi$ is piecewise linear and nondecreasing within each segment, $\hat{\pi}_{\max}$ can be easily computed by the formula
\begin{equation*}
\hat{\pi}_{\max} \ = \ \max \left\{   A_l   + b_l \,\hat{W}_{(l+1)} :  \ l = 1, \dots , L-2 ,  \ \ushort{r} \leq  \hat{W}_{(l+1)} \leq \bar{r}  \right\} .
\end{equation*}
Thus, there is no need to rely on global maximization or grid search algorithms and, since Lemma \ref{lem:weakconG} implies that $\hat{W}_{(1)} < \ushort{r} < \bar{r}  < \hat{W}_{(L)}$ holds w.p.a.1, we can restrict the search to $l = 1, \dots , L-2 $.

To estimate the set $\mathcal{R}$, I follow \cite{cht07}'s approach and introduce the population criterion function $\mathcal{C} (r)  : = {\pi}_{\max}  - \pi( r) \geq 0$ together with its empirical counterpart $\hat{\mathcal{C}} (r  )  : = \hat{\pi}_{\max}  - \hat\pi( r) \geq 0$, noting that $\mathcal{R} = \{ r \in  [\ushort r , \bar r] : \,  \mathcal{C} (r)  = 0 \}$. In addition, I introduce a (possibly stochastic) sequence $c_L > 0$ that satisfies the next assumption, which guarantees that $c_L$ grows to infinity with $L$, but not too fast.

\begin{assump}		\label{ass:cL}
The sequence $c_L > 0$ satisfies $1/c_L = o_p( 1 )$ and $c_L = o_p(\sqrt{L} )$.
\end{assump}

For a given $c_L >0$, now I propose estimating $\mathcal{R}$ by a contour set of level $c_L$:
\begin{equation*}
\hat{\mathcal{R}} \left( c_L \right)  \ :  = \ \left\{ r \in [ \ushort{r}  , \bar{r}    ]  :  \sqrt{L} \hat{\mathcal{C}}(r)  \leq c_L \right\} \
 = \ \left\{  r \in [ \ushort{r}  , \bar{r}    ]  :  \, \hat\pi( r)     \geq     \hat{\pi}_{\max} -  \frac{c_L}{\sqrt{L}}    \right\}.
\end{equation*}
I emphasize that this set estimator can be expressed as a finite union of intervals of the form
\begin{equation*}
\hat{\mathcal{R}} \left( c_L \right)  \ = \ \bigcup_{l \in  \mathcal{J}}  \,  [ \ushort{r}_l , \bar{r}_l ) ,
\end{equation*}
where the endpoints $ \ushort{r}_l , \bar{r}_l $, as well as the subset $ \mathcal{J} \subset \{ 1 , \dots , L \}$, can be easily computed using Algorithm 2.  I remark that this algorithm avoids computational approximation error and does not rely on global search methods. Guidelines for selecting $c_L$ will be discussed in the next section.

\begin{algorithm}[hh]\small
\caption{Computing the estimated set of optimal reserve prices}
\algorithmicrequire $\ushort{r} , \bar{r}$;\, $c_L$;\, $\hat{W}_{(1)} < \dots < \hat{W}_{(L)}$ \& $(A_l , b_l)_{l=0}^L$ from Algorithm 1
\begin{algorithmic}[1]
\State $\hat{\pi}_{\max} \gets \max \{ \,  A_l   + b_l \,\hat{W}_{(l+1)} :  \ l = 1, \dots , L-2 ,  \ \ushort{r} \leq  \hat{W}_{(l+1)} \leq \bar{r} \,\} $
\State $\mathcal{J} \gets$ find all $l=1,\dots,L-1$ such that \ $\ushort{r} \leq  \hat{W}_{(l+1)} \leq \bar{r} $ \ \& \
$ A_l   + b_l \, \hat{W}_{(l+1)} \geq \hat{\pi}_{\max} - c_L /\sqrt{L}$
\For{$l \in \mathcal{J}$}
\State $\ushort{r}_l \gets \hat{W}_{(l)} \vee  [ ( \hat{\pi}_{\max} - c_L/\sqrt{L} - A_l   )/ b_l   ] $
\State $\bar{r}_l \gets \hat{W}_{(l+1)}$
\EndFor
\end{algorithmic}
\algorithmicensure  intervals $ [ \ushort{r}_l , \bar{r}_l )$, $l \in \mathcal{J}$
\end{algorithm}

The following lemma provides useful asymptotic properties of the proposed estimators for ${\pi}_{\max} $ and $\mathcal{R}$.
Let $\lowmathcal{d}_{H} (R_1 , R_2)$ denote the Hausdorff distance between the nonempty sets $R_1$ and $R_2$. Moreover, given a function $\phi \in \ell^\infty (  [ \ushort{r} , \bar{r} ] )$, denote its oscillation over a nonempty set $R \subseteq  [ \ushort{r} , \bar{r} ] $ by $\mathrm{osc}_{R} (\phi ) : = \sup_{r \in R}\phi (r)  - \inf_{r \in R}\phi (r) \, \geq \, 0$, noting that $\mathrm{osc}_{R} (\phi ) = 0$ when $R$ is a singleton.

\begin{lemma}
\label{lem:optiest}
Suppose Assumption \ref{assump:dgp} holds. Then, 
\begin{enumerate}

\item $\sqrt{L} (  \hat{\pi}_{\max} - \pi_{\max} )  \leadsto  \sup_{r \in \mathcal{R}} \mathbb{G}_\pi  (r)$ and therefore $\sqrt{L} (  \hat{\pi}_{\max} - \pi_{\max}  
) = O_p (1)  $.

\item  $\sqrt{L} \|  \hat{\mathcal{C}} \|_{ \mathcal{R} , \infty } \leadsto \mathrm{osc}_{\mathcal{R}} ( \mathbb{G}_\pi )$ and therefore, if Assumption \ref{ass:cL} also holds, we have that $\mathcal{R}  \subseteq \hat{\mathcal{R}} (c_L)$ w.p.a.1 and $\lowmathcal{d}_{H} (  \hat{\mathcal{R}} (c_L)  ,  \mathcal{R} ) = o_p (1)$.

\end{enumerate}

\end{lemma}

This lemma provides the asymptotic distributions of $\sqrt{L} (  \hat{\pi}_{\max} - \pi_{\max} ) $ and $\sqrt{L} \|  \hat{\mathcal{C}} \|_{ \mathcal{R} , \infty }$, from which the consistency of $\hat{\pi}_{\max}$ and $ \hat{\mathcal{R}} (c_L) $ follows.
The resulting limiting distributions serve as key building blocks for the inference procedures for $\pi_{\max}$ and $\mathcal{R}$ developed below.


Starting with $\pi_{\max}$, and given the results in Lemmas \ref{lem:mainboot}.2 and \ref{lem:optiest}.1, I propose a $(1\!-\!\alpha)$-confidence interval of the form
\begin{equation*}
\mathrm{CI}_{1-\alpha }^{\max} \  : = \  \left[  \hat{\pi}_{\max}  -  \frac{ \hat{q}_{ 1-\alpha/2 }^{\max}  }{\sqrt{L}} ,  \,
 \ \hat{\pi}_{\max}  -   \frac{ \hat{q}_{ \alpha/2 }^{\max}  }{\sqrt{L}} \right] ,
\end{equation*}
where $\alpha \in ( 0  ,  1)$ and 
\begin{equation*}
\hat{q}_{a}^{\max}  \   : = \    \inf\left\{  y \in \mathbb{R} : \ \Prob \left(   \sup_{r \in  \hat{\mathcal{R}} (c_L)} \sqrt{L} \left[ \hat{\pi}^\ast(r)   - \hat{\pi}_{ \lowmathcal{h} } (r) \right]  \leq y   \mid  \mathcal{S}_L \right)  \geq a   \right\} \quad \text{for} \ a \in ( 0 , 1) \, .
\end{equation*}
Turning to inference for $\mathcal{R}$, the goal is to construct a (conservative) confidence region $\mathrm{CR}_{1-\alpha}$ of level $1-\alpha$ satisfying
\begin{equation}
\label{eq:consercr}
\underset{L \rightarrow \infty}{\lim \! \inf} \, \Prob \left( \mathcal{R}  \subseteq \mathrm{CR}_{1-\alpha}  \right)  \, \geq \, 1 - \alpha  \,  .
\end{equation}
Since  $\{  \mathcal{R}  \subseteq \hat{\mathcal{R}}(c) \}  =   \{ \sqrt{L} \|  \hat{\mathcal{C}} \|_{ \mathcal{R} , \infty }  \leq c \}$ for any $c > 0$, it is natural to consider a confidence region of the form $\hat{\mathcal{R}}(c)$ with $c > 0$ chosen such that $\Prob  \{ \sqrt{L} \|  \hat{\mathcal{C}} \|_{ \mathcal{R} , \infty }  \leq c \} \approx 1 - \alpha$.
Lemma \ref{lem:optiest}.2 suggests that $c > 0$ can be taken as an estimate of the $(1\!-\!\alpha)$-quantile of $ \mathrm{osc}_{\mathcal{R}} ( \mathbb{G}_\pi )$, namely $c \approx q_{ 1 - \alpha }^{\mathrm{osc}} :=  \inf \{  y \in \mathbb{R}_{\geq0} : \, \Prob ( \mathrm{osc}_{\mathcal{R}} ( \mathbb{G}_\pi )\leq y   )  \geq 1\!-\!\alpha  \}$.
Then, Lemma \ref{lem:mainboot} motivates the bootstrap-based estimator of this quantile, leading to the confidence region
\begin{equation*}
\mathrm{CR}_{1-\alpha} \, :  =  \, \hat{\mathcal{R}} \left(  \hat{q}_{ 1 - \alpha }^{\mathrm{osc}}   \right)   \, , 
\end{equation*}
where
\begin{equation*}
\hat{q}_{ 1 - \alpha }^{\mathrm{osc}}  \, : =  \, \inf \left\{  y \in \mathbb{R}_{\geq0} : \, \Prob \left( \mathrm{osc}_{  \hat{\mathcal{R}} (c_L)  } \left( \sqrt{L} ( \hat{\pi}^\ast   - \hat{\pi}_{ \lowmathcal{h} }  ) \right)  \leq y  \mid \mathcal{S}_L \right)  \geq 1 - \alpha   \right\} 
\end{equation*}
is the bootstrap-based estimator of ${q}_{ 1 - \alpha }^{\mathrm{osc}}$, the $(1\!-\!\alpha)$-quantile of $ \mathrm{osc}_{\mathcal{R}} ( \mathbb{G}_\pi )$.

The next proposition establishes the asymptotic validity of $\mathrm{CI}_{1-\alpha }^{\max}$ and $\mathrm{CR}_{1-\alpha}$, with the latter result requiring that $\mathcal{R}$ contains more than one element.

\begin{propo}
\label{propo:infoptp}
Suppose that Assumptions \ref{assump:dgp}-\ref{ass:cL} hold. Then,
\begin{enumerate}

\item $\Prob \left( \pi_{\max}  \in   \mathrm{CI}_{1-\alpha }^{\max}  \right) \, \rightarrow \, 1   - \alpha$.

\item  $\Prob \left( \mathcal{R} \subseteq  \mathrm{CR}_{1-\alpha} \right)  \rightarrow 1  - \alpha$ if $\mathcal{R}$ contains more than one element.

\end{enumerate}

\end{propo}

Two remarks are noteworthy. First, this proposition involves a technical complication in comparison with Propositions \ref{propo:pointboots} and \ref{propo:cband}, whose conclusions follow from standard arguments once Lemma \ref{lem:mainboot} has been established. Specifically, here the critical values $\hat{q}_{ 1 - \alpha }^{\mathrm{max}}$ and $\hat{q}_{ 1 - \alpha }^{\mathrm{osc}}$ are computed using an estimator of the set $\mathcal{R}$, rather than the set itself. Hence, their validity requires controlling the effect of replacing $\mathcal{R}$ by its estimator. The proof of Proposition \ref{propo:infoptp} establishes that this replacement is asymptotically innocuous: the two bootstrap critical values consistently estimate the corresponding population quantiles.

Second, the coverage result in the second part is stated for the case in which $\mathcal{R}$ contains more than one element. This restriction is needed because, in that case, $\mathrm{osc}_{\mathcal{R}} ( \mathbb{G}_\pi )$ has a distribution that is continuous and strictly increasing on $\mathbb{R}_{\geq 0}$, as shown in the Appendix.
In contrast, when $\mathcal{R}$ is a singleton, we have $\Prob (\mathrm{osc}_{\mathcal{R}} ( \mathbb{G}_\pi ) = 0 ) = 1$ and therefore the limiting distribution of the oscillation of $ \mathbb{G}_\pi $ is degenerate at zero. As a consequence, the exact asymptotic coverage result of Proposition \ref{propo:infoptp}.2 cannot be obtained from the same argument.

This degeneracy essentially occurs because the value of $\pi$ on $\mathcal{R}$, namely $\pi_{\max}$, is unknown. By contrast, in certain moment-inequality models, the empirical criterion function is typically of the form $-\hat{\phi}(\cdot)$ instead of $\hat{\phi}^{\max}-\hat{\phi}(\cdot)$, where $\hat{\phi}$ is an estimator of some function $\phi \geq 0$ defined on the parameter space and that satisfies $\phi=0$ only on the identified set \cite[see, e.g.,][]{cht07,bug10}.
Thus, the value of $\phi$ on the identified set does not need to be estimated and, consequently, the limiting distribution of the criterion-function is not affected by an additional estimation step.
It is also worth noting that the possibility of the identified set being a singleton can pose complications for establishing the validity of certain inference (resampling) procedures in moment-inequality models. For instance, \cite{chor24} rule out this possibility by imposing the stronger assumption that the identified set has nonempty interior.

To address the difficulty that arises in our setting, one can instead construct conservative confidence regions that satisfy Eq.\ (\ref{eq:consercr}) in the general case.
The idea is to replace the critical value corresponding to the distribution of $\mathrm{osc}_{\mathcal{R}}(\mathbb{G}_\pi)$ with one based on a continuous and strictly increasing distribution that first-order stochastically dominates it.
Since $2\| \mathbb{G}_\pi  \|_{ \mathcal{R} , \infty } \geq \mathrm{osc}_{\mathcal{R}} ( \mathbb{G}_\pi ) $ holds regardless of the cardinality of $\mathcal{R}$,
I propose the following conservative confidence region:
\begin{equation*}
\mathrm{CR}_{1-\alpha}^{ \mathrm{con} } \, :  =  \, 
\hat{\mathcal{R}} \left( 2 \, \hat{q}_{ 1 - \alpha }^{\infty}  \left[   \hat{\mathcal{R}} (c_L)  \right]  \right)   \, . 
\end{equation*}

The next proposition establishes the validity of this construction as a conservative alternative. 

\begin{propo}
\label{propo:convers}
If Assumptions \ref{assump:dgp}-\ref{ass:cL} hold, then
\begin{equation*}
\underset{L \rightarrow \infty}{\lim \! \inf} \, \Prob \left( \mathcal{R}  \subseteq \mathrm{CR}_{1-\alpha}^{ \mathrm{con} }  \right)  \ \geq \ 1 - \alpha  \,  .
\end{equation*}

\end{propo}

To conclude this section, several remarks are noteworthy. First, inspection of the proof of Proposition \ref{propo:convers}, together with Lemma \ref{alem:supptG} in the Appendix, reveals that $\Prob(\mathcal{R}\subseteq\mathrm{CR}_{1-\alpha}^{\mathrm{con}})\rightarrow1$ when $\mathcal{R}$ is a singleton. This result follows from two observations. On the one hand, $\mathrm{osc}_{\mathcal{R}}(\mathbb{G}_\pi)$ degenerates at zero in this case. On the other hand, Lemma \ref{lem:optiest}.2 implies that $\hat{q}_{1-\alpha}^{\infty}[\hat{\mathcal{R}}(c_L)]\geq\hat{q}_{1-\alpha}^{\infty}(\mathcal{R})$ w.p.a.1, while the latter converges in probability to a strictly positive constant by Lemma \ref{lem:mainboot}. Consequently, the conservative critical value remains asymptotically bounded away from zero (and infinity), yielding coverage approaching one when $\mathcal{R}$ is a singleton.

Second, it may be possible to establish that the conservative confidence region is genuinely conservative, rather than merely satisfying a weak coverage bound. Specifically, the arguments developed in the Appendix might be extended to show that $q_{1-\alpha}^{\mathrm{osc}}<2q_{1-\alpha}^{\infty}(\mathcal{R})$ for any admissible specification of the structural parameters. This strict inequality would imply that the asymptotic coverage probability of $\mathrm{CR}_{1-\alpha}^{\mathrm{con}}$ exceeds $1-\alpha$ when $\mathcal{R}$ contains more than one element.

Finally, the asymptotic results established in Section \ref{sec:asinf} hold pointwise in the structural parameters $(\beta,F_V)$. By contrast, the literature on inference in partially identified models \cite[including][]{cht07,bug10,chor24} usually establishes uniform validity over classes of data-generating processes. Although uniform inference is less common in the empirical auction literature, with \cite{jz21} providing an exception, extending the present results in this direction would be valuable. Such an extension would require additional technical arguments and is left for future research.

\section{Monte Carlo experiments}	\label{sec:mc}

This section presents the results of Monte Carlo experiments evaluating the finite-sample performance of the proposed estimators and inference procedures. It also provides guidance on the choice of the tuning parameters $\lowmathcal{h}$ and $c_L$.


The data-generating process is specified as follows. I consider two distributions of private values:
\begin{itemize}

\item Unif.: uniform over $[0,10]$. 

\item Multiple Optimal Reserve Prices (MORP): private-value distribution given by the c.d.f.
\begin{eqnarray*}
F_V(v)
& = &
\mathds{1} ( 0\leq v\leq2 )
\left\{
1-\exp\left[-\frac{v^3}{8}+\frac{v^4}{32}\right]
\right\} \\
& & \ +  \,
\mathds{1} ( 2<v\leq3 )
\left\{ 1- \exp\left[-\frac{1}{3}-\frac{(3-v)^3}{6}\right]/(v-1) \right\} 
+
\mathds{1} ( 3<v\leq4 )
\left[
1-\frac{\exp(-1/3)}{v-1}
\right] \\
& & \  +  \,
\mathds{1} (4<v<10)
\left\{
1-
\frac{\exp(-1/3)}{v-1}
\exp\left[
-\left(\frac{v-4}{10-v}\right)^4
\right]
\right\}
+
\mathds{1} ( v\geq10 ) .
\end{eqnarray*}

\end{itemize}
Both distributions satisfy Assumption \ref{assump:dgp} and are supported on $[0,10]$.\footnote{The MORP distribution was constructed with the assistance of ChatGPT to provide an example in which the set of optimal reserve prices is not a singleton.} The seller's valuation is fixed at $v_s=1$, and I consider $n=2$ and $n=3$ bidders. Under both values of $n$, the set of optimal reserve prices is $\mathcal{R}=\{5.5\}$ for the uniform distribution and $\mathcal{R}=[3,4]$ for the MORP distribution. The number of covariates is $d=2$, where $X_1$ and $X_2$ are independently and identically distributed as $2\,\mathrm{Beta}(3/2,3/2)-1$, and the slope coefficients are $\beta=(1,1)$. The sample sizes considered are $L=250$ and $L=500$. 
Finally, the considered interval for reserve prices is taken to be $[\ushort{r},\bar{r}]=[1.5,6.5]$.

\begin{table}[tt]
\renewcommand{\arraystretch}{1.25}
	\caption{Estimated bias and std.\ deviation (sd) of $\hat{\pi}$ from Monte Carlo experiments.}
	\begin{center}
		\begin{scriptsize}

			\begin{tabular}{       l      l    l     l    ||     c  |  c   ||   l    l   ||  l     l }
				
			\thickhline				

				\multicolumn{1}{ c |}{\multirow{2}{*}{Dist.}}  &  \multicolumn{1}{c |}{\multirow{2}{*}{$n$}} & \multicolumn{1}{c |}{\multirow{2}{*}{$ \mathcal{R}$ }}  &  \multicolumn{1}{c ||}{\multirow{2}{*}{$\pi_{\max}$ }} &  \multicolumn{1}{c |}{\multirow{2}{*}{$r$ }}  & \multicolumn{1}{c ||}{\multirow{2}{*}{$ \pi  (r ) $ }}  & \multicolumn{2}{c||}{$L=250$} &  \multicolumn{2}{c }{$L=500$}  \\    \cline{7-10}	
				
		\multicolumn{1}{c |}{} 		 & \multicolumn{1}{ c |}{}   &   \multicolumn{1}{ c |}{}  &   &  \multicolumn{1}{ c |}{}  &  &  $\mathrm{bias} $    &  \multicolumn{1}{c||}{sd } & \multicolumn{1}{c}{bias} & \multicolumn{1}{c}{sd}   \\
				
\thickhline

Unif. &	2 & 	$\{ 5.5 \}$ &	4.443  &	2.000 & 3.667 & -0.006 & 0.133 & 0.003 & 0.092 \\	
&	&	&	&	3.000 & 3.963 & -0.008 & 0.127 & 0.003 & 0.088 \\	
&	&	&	&	4.000 & 4.240 & -0.010 & 0.138 & 0.002 & 0.095 \\	
&	&	&	&	5.000 & 4.417 & -0.020 & 0.165 & 0.001 & 0.112 \\	
&	&	&	&	6.000 & 4.413 & -0.033 & 0.209 & 0.001 & 0.143 \\	
&	&	&	&	7.000 & 4.150 & -0.032 & 0.271 & 0.007 & 0.195 \\	\cline{5-10}
&	&	&	&	 - & ${\pi}_{\max} $ &  0.078 & 0.179 & 0.063 & 0.124 \\	 \hline
					
Unif. &	3 & 	$\{ 5.5 \}$ &	 5.458  &	2.000 & 5.064 & 0.000 & 0.141 & -0.005 & 0.096 \\	
&	&	&	&	3.000 & 5.176 & -0.001 & 0.134 & -0.005 & 0.091 \\	
&	&	&	&	4.000 & 5.320 & -0.002 & 0.130 & -0.005 & 0.088 \\	
&	&	&	&	5.000 & 5.438 & -0.006 & 0.141 & -0.006 & 0.094 \\	
&	&	&	&	6.000 & 5.432 & -0.012 & 0.168 & -0.010 & 0.114 \\	
&	&	&	&	7.000 & 5.172 & -0.023 & 0.213 & -0.013 & 0.149 \\	\cline{5-10}
&	&	&	&	 - & ${\pi}_{\max} $ & 0.049 & 0.148 & 0.027 & 0.101 \\	 \hline
MORP &	2 & 	$ [ 3 , 4 ] $ &	 2.340  &	2.000 & 2.237 & -0.001 & 0.070 & -0.001 & 0.048 \\	
&	&	&	&	3.000 & 2.340 & -0.006 & 0.112 & -0.003 & 0.075 \\	
&	&	&	&	4.000 & 2.340 & -0.020 & 0.168 & -0.005 & 0.112 \\	
&	&	&	&	5.000 & 2.338 & -0.033 & 0.233 & -0.010 & 0.156 \\	
&	&	&	&	6.000 & 2.266 & -0.083 & 0.319 & -0.035 & 0.203 \\	
&	&	&	&	7.000 & 1.516 & -0.131 & 0.428 & -0.049 & 0.323 \\	\cline{5-10}
&	&	&	&	 - & ${\pi}_{\max} $ & 0.198 & 0.173 & 0.135 & 0.124 \\	 \hline
MORP &	3 & 	$ [ 3 , 4 ] $ &	2.883  &	2.000 & 2.810 & -0.006 & 0.087 & 0.000 & 0.057 \\	
&	&	&	&	3.000 & 2.883 & -0.009 & 0.106 & 0.000 & 0.070 \\	
&	&	&	&	4.000 & 2.883 & -0.016 & 0.145 & -0.006 & 0.097 \\	
&	&	&	&	5.000 & 2.880 & -0.027 & 0.190 & -0.015 & 0.128 \\	
&	&	&	&	6.000 & 2.788 & -0.072 & 0.255 & -0.034 & 0.168 \\	
&	&	&	&	7.000 & 1.757 & -0.049 & 0.410 & -0.016 & 0.266 \\	\cline{5-10}
&	&	&	&	 - & ${\pi}_{\max} $  & 0.129 & 0.156 & 0.088 & 0.102 \\	
 \thickhline					
				
			\end{tabular}
		\end{scriptsize}
	\end{center}
	
	{\raggedright \footnotesize{The last row of each scenario (marked with ${\pi}_{\max}$) provides the estimated bias and std.\ deviation of $\hat{\pi}_{\max} $.}}
	
	\label{table:bisd}
\end{table}

\begin{figure}[tt]

\caption{Performance of $\hat{\pi}$ when $L=250$.} 

\vspace{.1in}

{\centering

\includegraphics[scale=.76]{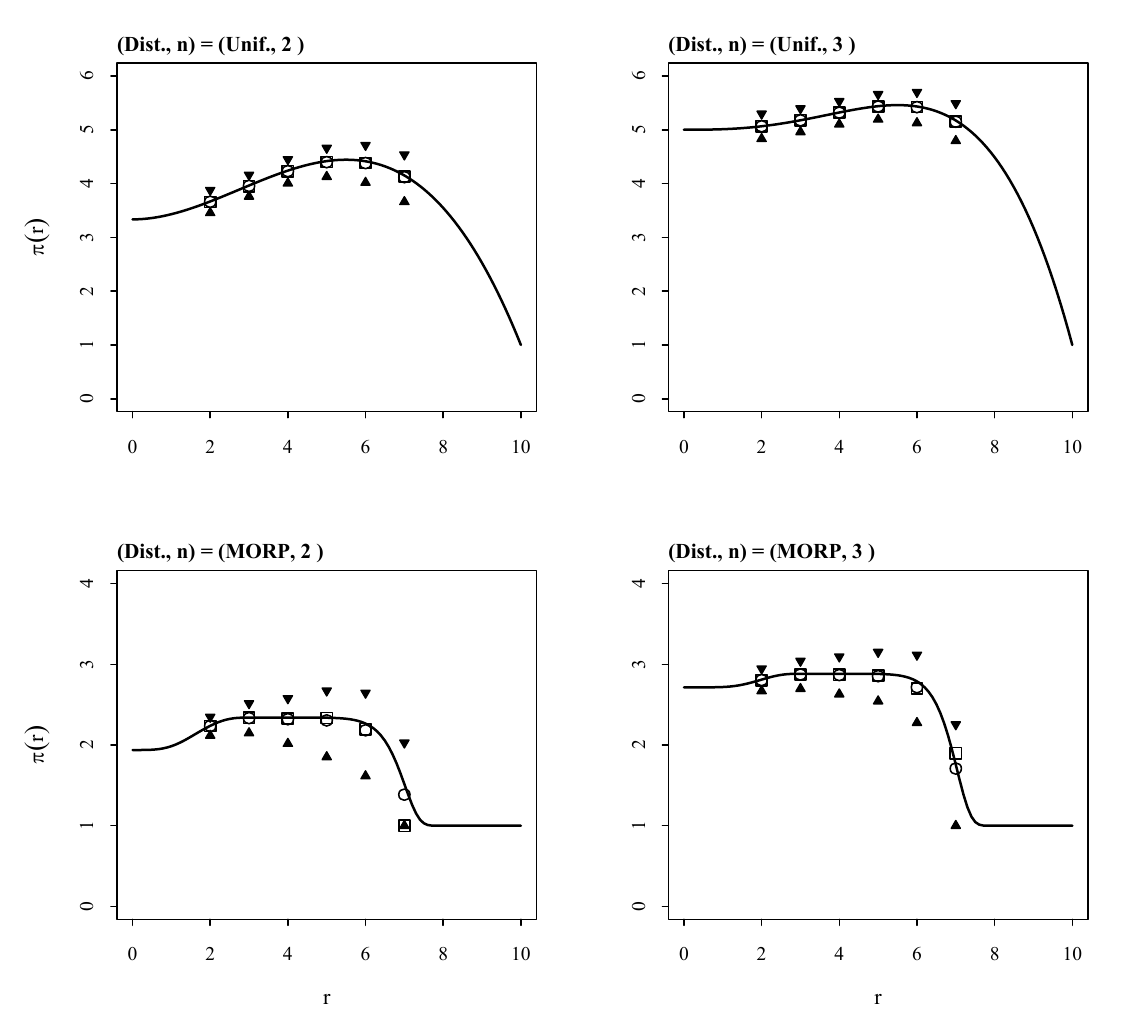}

}

{\raggedright \footnotesize{Solid curve depicts the true expected revenue function $\pi(r)$. Circles and squares are the estimated mean and median of $\hat\pi (r)$, respectively, from the simulations. Up- and down-pointing triangles are the 5th and 95th percentiles of $\hat\pi (r)$, respectively.}}

\label{fig1mc}

\end{figure}

\begin{figure}[tt]

\caption{Performance of $\hat{\pi}$ when $L=500$.} 

\vspace{.1in}

{\centering

\includegraphics[scale=.76]{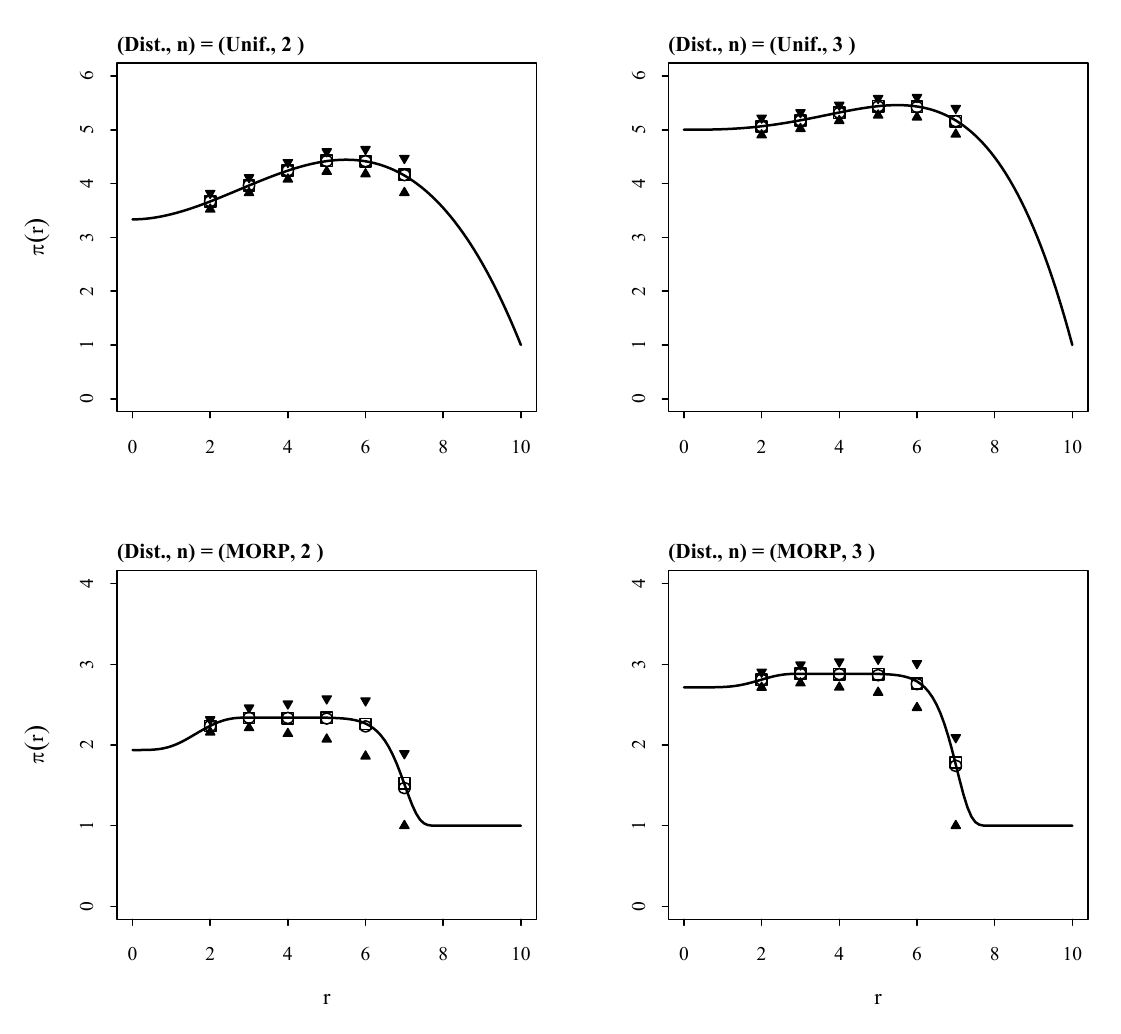}

}

{\raggedright \footnotesize{Solid curve depicts the true expected revenue function $\pi(r)$. Circles and squares are the estimated mean and median of $\hat\pi (r)$, respectively, from the simulations. Up- and down-pointing triangles are the 5th and 95th percentiles of $\hat\pi (r)$, respectively.}}

\label{fig2mc}

\end{figure}

\begin{table}[tt]
\renewcommand{\arraystretch}{1.25}
	\caption{ Average Hausdorff distance between $ \hat{\mathcal{R}} ( c_L ) $ and $\mathcal{R}$. }
	\begin{center}
		\begin{scriptsize}
			\begin{tabular}{ c |      c   c  |  c   c  |   c  c |   c   c  }

 \thickhline	
Dist. \ & \multicolumn{4}{c|}{ Unif. }  &  \multicolumn{4}{c}{ MORP }   \\  \hline
	
$n$   & \multicolumn{2}{c|}{ 2 }  &    \multicolumn{2}{c|}{ 3 }  &   \multicolumn{2}{c|}{ 2 }  &    \multicolumn{2}{c}{ 3 }  \\  

$L$  & 250 & 500  & 250 & 500  & 250 & 500  & 250 & 500 \\  \hline
\rule{0pt}{14pt} $\lowmathcal{d}_{H} (   \hat{\mathcal{R}} ( c_L ) ,   \mathcal{R} )$ &  
1.57 & 1.36 &
2.29 & 1.89 &
1.95 & 1.77 &
1.90 & 1.81 \\
 \thickhline	
			\end{tabular}
		\end{scriptsize}
	\end{center}
	
	\label{table:orp}
\end{table}

I conduct 1,000 Monte Carlo replications for each design, determined by the distribution of private values, the number of bidders $n$, and the sample size $L$. In each replication, I first generate $nL$ independent valuations from either the Unif.\ or the MORP distribution. I then compute the corresponding payments according to Assumption \ref{assump:dgp},
$ \tilde{W}_l=\exp(X_{l,1}+X_{l,2})V_{l,(n-1:n)}$,
where $V_{l,(n-1:n)}$ is the second-highest generated valuation in the $l$th auction.

Next, I apply Algorithm 1 to compute the expected revenue estimator $\hat{\pi}$. Following the procedure described in Section \ref{ss:bootap}, I then generate 560 bootstrap replications of
\[
\sqrt{L}\left[\hat{\pi}^\ast(r)-\hat{\pi}_{\lowmathcal{h}}(r)\right]
\]
over an equally spaced grid on $[1.5,6.5]$ with grid spacing $0.01$. The number of bootstrap replications (560) was chosen solely for computational efficiency, as it is a multiple of the available parallel computing cores. I employ the triweight kernel with bandwidth
$
\lowmathcal{h}=3.15 \, \hat{\sigma}_U L^{-1/3}
$,
where $\hat{\sigma}_U$ is the estimated standard deviation of $U$ based on the residuals $\hat{U}_l$ from each replication. This bandwidth corresponds to the asymptotically optimal rule for estimating a c.d.f.\ using the triweight kernel. These bootstrap replications are used to compute the critical values for the confidence intervals and bands, as well as confidence regions.


Next, I compute  $\hat{\pi}_{\max}$ and the set estimator $\hat{\mathcal{R}}(c_L)$ using $c_L=\hat{\pi}_{\max}\log[\log(L)]/2$ as the tuning parameter.
Finally, I construct confidence intervals for ${\pi}_{\max}$ and confidence regions for ${\mathcal{R}}$ as described in the previous section. In the absence of theoretical guidance, the $\log[\log(L)]$ rate is motivated by the simulation study of \citet{bug10}, among other candidate rates, while the scaling constant $\hat{\pi}_{\max}/2$ was selected based on its favorable finite-sample performance in the Monte Carlo experiments.

The simulations were conducted on the Holland Computing Center's Swan cluster at the University of Nebraska, using compute nodes equipped with two Intel Xeon Gold 6348 processors (56 cores per node) and 256 GB of memory, interconnected via HDR100 InfiniBand. Each Monte Carlo replication was parallelized across 28 CPU cores. The average computation time per replication was approximately 7 seconds for $L=250$ and 20 seconds for $L=500$. On a typical desktop computer (e.g., a Dell PC with an 8-core Intel Core i7 or AMD Ryzen 7 processor), the corresponding computation times are expected to be approximately 20-30 seconds and 1-2 minutes per replication, respectively, depending on the processor generation and available memory.

The results of the Monte Carlo experiments are reported in Tables \ref{table:bisd}-\ref{table:cov} and illustrated in Figures \ref{fig1mc}-\ref{fig2mc}. Table \ref{table:bisd} reports the estimated bias and standard deviation of $\hat{\pi}(r)$ at the evaluation points $r=2,3,\ldots,7$, together with the corresponding quantities for $\hat{\pi}_{\max}$.\footnote{The reserve price $r=7$, although outside $[\ushort{r},\bar{r}]=[1.5,6.5]$, is also considered to assess pointwise finite-sample performance where expected revenue declines sharply, particularly under the MORP design.} Across the designs, the bias of $\hat{\pi}(r)$ is generally small relative to the magnitude of $\pi(r)$ and decreases as the sample size increases. The bias is most pronounced under the MORP design at reserve prices above the optimal set, particularly at $r=6$ and $r=7$, where the expected revenue function declines sharply. The standard deviation of $\hat{\pi}(r)$ is also small relative to the corresponding value of $\pi(r)$ and decreases substantially as $L$ increases. It tends to increase with $r$, especially under the MORP design near the steeply declining portion of the expected revenue curve. The estimator $\hat{\pi}_{\max}$ exhibits a positive finite-sample bias, but both its bias and standard deviation decrease with the sample size.

Figures \ref{fig1mc} and \ref{fig2mc} illustrate the performance of the proposed estimator for $L=250$ and $L=500$, respectively. In each panel, the solid curve represents the true expected revenue function, while the circles and squares report the Monte Carlo mean and median of $\hat{\pi}(r)$, respectively, at the evaluation points. The mean and median are nearly indistinguishable in most designs, indicating little finite-sample skewness in the distribution of the estimator. The main exception occurs under the MORP design at $r=7$ when $L=250$. In this region, the sharp decline in the expected revenue function amplifies the effect of estimation error. In particular, when $n=2$, the Monte Carlo median is close to the seller's valuation, $v_s=1$, suggesting that in at least half of the replications the estimated model predicts that the object remains unsold at this reserve price. This behavior likely reflects finite-sample estimation of the upper tail and support of the valuation distribution. These discrepancies become substantially smaller when the sample size increases to $L=500$.

To evaluate the performance of $\hat{\mathcal{R}}(c_L)$, Table \ref{table:orp} reports the average Hausdorff distance between the estimated and true sets of optimal reserve prices. The average distance decreases with the sample size in every design, as predicted by the consistency result. Although the estimated distances are not negligible, the expected revenue curves in Figures \ref{fig1mc} and \ref{fig2mc} indicate that reserve prices in a neighborhood of $\mathcal{R}$ generate expected revenues close to the maximum. Consequently, the economic loss associated with the reported estimation error is considerably smaller than the Hausdorff distances alone might suggest. This distinction is relevant from a policy perspective, for which the revenue loss induced by the estimated reserve price may be more important than its distance from the set of exact maximizers.

\begin{table}[tt]
\renewcommand{\arraystretch}{1.25}
	\caption{Coverage probabilities of proposed confidence intervals, confidence bands, and confidence regions.}
	\begin{center}
		\begin{scriptsize}
			\begin{tabular}{    l  | l || c  c  c |   c  c  c |  c  c  c |  c  c  c   }
				
			\thickhline

\multirow{2}{*}{Dist.}	&  $ (n , L ) = $   &      \multicolumn{3}{c|}{ (2, 250) } &      \multicolumn{3}{c|}{ (2, 500) } &      \multicolumn{3}{c|}{ (3, 250) } &      \multicolumn{3}{c }{ (3, 500) }
  \\ \cline{2-14}
  &  Method $\backslash$ $1-\alpha$   & 0.90 & 0.95 & 0.99 & 0.90 & 0.95 & 0.99 & 0.90 & 0.95 & 0.99 & 0.90 & 0.95 & 0.99 \\

\thickhline				

Unif. &	$\mathrm{CI}_{1-\alpha} (2) $ &	0.93 & 0.96 & 0.99 & 0.92 & 0.96 & 0.99 & 0.90 & 0.94 & 0.99 & 0.90 & 0.95 & 0.99 \\
&	$\mathrm{CI}_{1-\alpha} (3) $ &	0.93 & 0.97 & 0.99 & 0.93 & 0.96 & 1.00 & 0.90 & 0.94 & 0.99 & 0.90 & 0.95 & 0.99 \\
&	$\mathrm{CI}_{1-\alpha} (4) $ &	0.92 & 0.96 & 0.99 & 0.91 & 0.96 & 1.00 & 0.90 & 0.95 & 0.99 & 0.92 & 0.95 & 1.00 \\
&	$\mathrm{CI}_{1-\alpha} (5) $ &	0.93 & 0.97 & 1.00 & 0.92 & 0.96 & 0.99 & 0.89 & 0.94 & 0.99 & 0.91 & 0.96 & 1.00 \\
&	$\mathrm{CI}_{1-\alpha} (6) $ &	0.92 & 0.97 & 0.99 & 0.92 & 0.96 & 0.99 & 0.89 & 0.94 & 0.99 & 0.90 & 0.95 & 0.99 \\
&	$\mathrm{CI}_{1-\alpha} (7) $ &	0.92 & 0.95 & 0.99 & 0.90 & 0.94 & 0.99 & 0.90 & 0.95 & 0.99 & 0.91 & 0.95 & 1.00 \\
&	$\mathrm{CI}_{1-\alpha}^{\max} $ &	0.92 & 0.95 & 0.99 & 0.92 & 0.95 & 0.99 & 0.91 & 0.95 & 0.99 & 0.90 & 0.95 & 0.99 \\
&	$\mathrm{CB}_{1-\alpha} $ &	0.90 & 0.95 & 0.99 & 0.90 & 0.95 & 0.99 & 0.90 & 0.95 & 1.00 & 0.91 & 0.96 & 1.00 \\
&	$\mathrm{CR}_{1-\alpha}  $ &	1.00 & 1.00 & 1.00 & 1.00 & 1.00 & 1.00 & 1.00 & 1.00 & 1.00 & 1.00 & 1.00 & 1.00 \\
	\hline	
MORP &	$\mathrm{CI}_{1-\alpha} (2) $ &	0.90 & 0.94 & 0.99 & 0.90 & 0.95 & 0.98 & 0.88 & 0.93 & 0.98 & 0.90 & 0.95 & 0.99 \\
&	$\mathrm{CI}_{1-\alpha} (3) $ &	0.88 & 0.93 & 0.98 & 0.89 & 0.94 & 0.98 & 0.89 & 0.93 & 0.98 & 0.91 & 0.95 & 0.99 \\
&	$\mathrm{CI}_{1-\alpha} (4) $ &	0.89 & 0.95 & 0.99 & 0.90 & 0.94 & 0.99 & 0.90 & 0.94 & 0.99 & 0.91 & 0.96 & 0.99 \\
&	$\mathrm{CI}_{1-\alpha} (5) $ &	0.89 & 0.95 & 0.99 & 0.90 & 0.95 & 0.99 & 0.89 & 0.94 & 0.99 & 0.92 & 0.95 & 0.99 \\
&	$\mathrm{CI}_{1-\alpha} (6) $ &	0.92 & 0.94 & 0.97 & 0.90 & 0.95 & 0.98 & 0.88 & 0.93 & 0.99 & 0.90 & 0.94 & 0.98 \\
&	$\mathrm{CI}_{1-\alpha} (7) $ &	0.75 & 0.81 & 0.84 & 0.83 & 0.89 & 0.93 & 0.85 & 0.88 & 0.92 & 0.85 & 0.91 & 0.96 \\
&	$\mathrm{CI}_{1-\alpha}^{\max} $ &	0.86 & 0.92 & 0.97 & 0.87 & 0.92 & 0.97 & 0.88 & 0.93 & 0.98 & 0.91 & 0.95 & 0.98 \\
&	$\mathrm{CB}_{1-\alpha} $ &	0.83 & 0.87 & 0.92 & 0.90 & 0.94 & 0.97 & 0.88 & 0.94 & 0.99 & 0.89 & 0.94 & 0.99 \\
&	$\mathrm{CR}_{1-\alpha}  $ &	0.90 & 0.95 & 0.99 & 0.87 & 0.91 & 0.97 & 0.95 & 0.98 & 1.00 & 0.95 & 0.98 & 0.99 \\
	\hline	

\thickhline

				
			\end{tabular}
		\end{scriptsize}
	\end{center}
	
	\label{table:cov}
\end{table}

Finally, Table \ref{table:cov} reports the estimated coverage probabilities of the proposed inference procedures. Under the uniform design, the pointwise confidence intervals $\mathrm{CI}_{1-\alpha} (r)$ and uniform confidence bands $\mathrm{CB}_{1-\alpha}$ have coverage probabilities close to their nominal levels, even when $L=250$. Under the MORP design, the pointwise confidence intervals also perform well throughout most of the reserve-price space, although they exhibit noticeable undercoverage at $r=7$, especially in the smaller sample. This deterioration is consistent with the larger bias and variability reported in Table \ref{table:bisd} for this portion of the expected revenue curve. The confidence interval $\mathrm{CI}_{1-\alpha}^{\max} $ for $\pi_{\max}$ displays moderate undercoverage in some of the smaller-sample MORP designs, but its performance improves with $L$. The confidence bands also improve with the sample size. The confidence regions $\mathrm{CR}_{1-\alpha}  $ for $\mathcal{R}$ are conservative under the uniform design, where $\mathcal{R}$ is a singleton, and attain coverage probabilities closer to their nominal levels under the MORP design. In this regard, I recall that the asymptotic results established in the previous section do not guarantee correct coverage of the confidence regions when $\mathcal{R}$ is a singleton.

\section{Empirical application}	\label{sec:ei}

In this section, I apply the proposed estimators and inference procedures to a sample of timber auctions conducted by the U.S.\ Forest Service. Specifically, I use the publicly available dataset analyzed in \cite{lp08}, which also provides a detailed description of the data and institutional setting.

The sample consists of ascending auctions with nonbinding reserve prices, in which the object for sale is a timber tract. I separately analyze auctions with two and three bidders. For each auction, the observed outcome is the transaction payment for the timber tract, expressed in thousands of 1979 U.S.\ dollars. In an ascending auction, this payment corresponds to the second-highest valuation. The auction-specific covariates are the volume of the timber tract, measured in thousand board feet (MBF), and its appraisal value, measured in 1979 U.S.\ dollars per MBF. The sample medians of these two covariates are 967 MBF and 68 dollars per MBF, respectively.

Using these data, I estimate the expected revenue function and the set of optimal reserve prices, both expressed in thousands of 1979 U.S.\ dollars, for a timber tract with a volume of 967 MBF and an appraisal value of 68, measured in 1979 U.S. dollars per MBF. I set the seller's valuation equal to 66 thousand 1979 U.S.\ dollars, obtained by multiplying the tract volume by its appraisal value: $66 \approx 967 \times 68 /1000$. The considered interval for the reserve prices is set to $[ \ushort{r} , \bar{r} ] = [70 ,300]$. The tuning parameters are selected following the suggestions from the previous section: $\lowmathcal{h}=3.15 \, \hat{\sigma}_U L^{-1/3}$ and $c_L=\hat{\pi}_{\max}\log[\log(L)]/2$, while the number of bootstrap replications is set to 560.

\begin{figure}[hhh]

\caption{Estimated expected revenue as a function of the reserve price $r$ (expressed in thousands of 1979 U.S. dollars) and set of optimal reserve prices for a timber tract with a volume of 967 MBF and an appraisal value of 68, measured in 1979 U.S. dollars per MBF. Seller's valuation is set at $v_s = 66$.}

\vspace{.1in}

{\centering

\includegraphics[scale=.82]{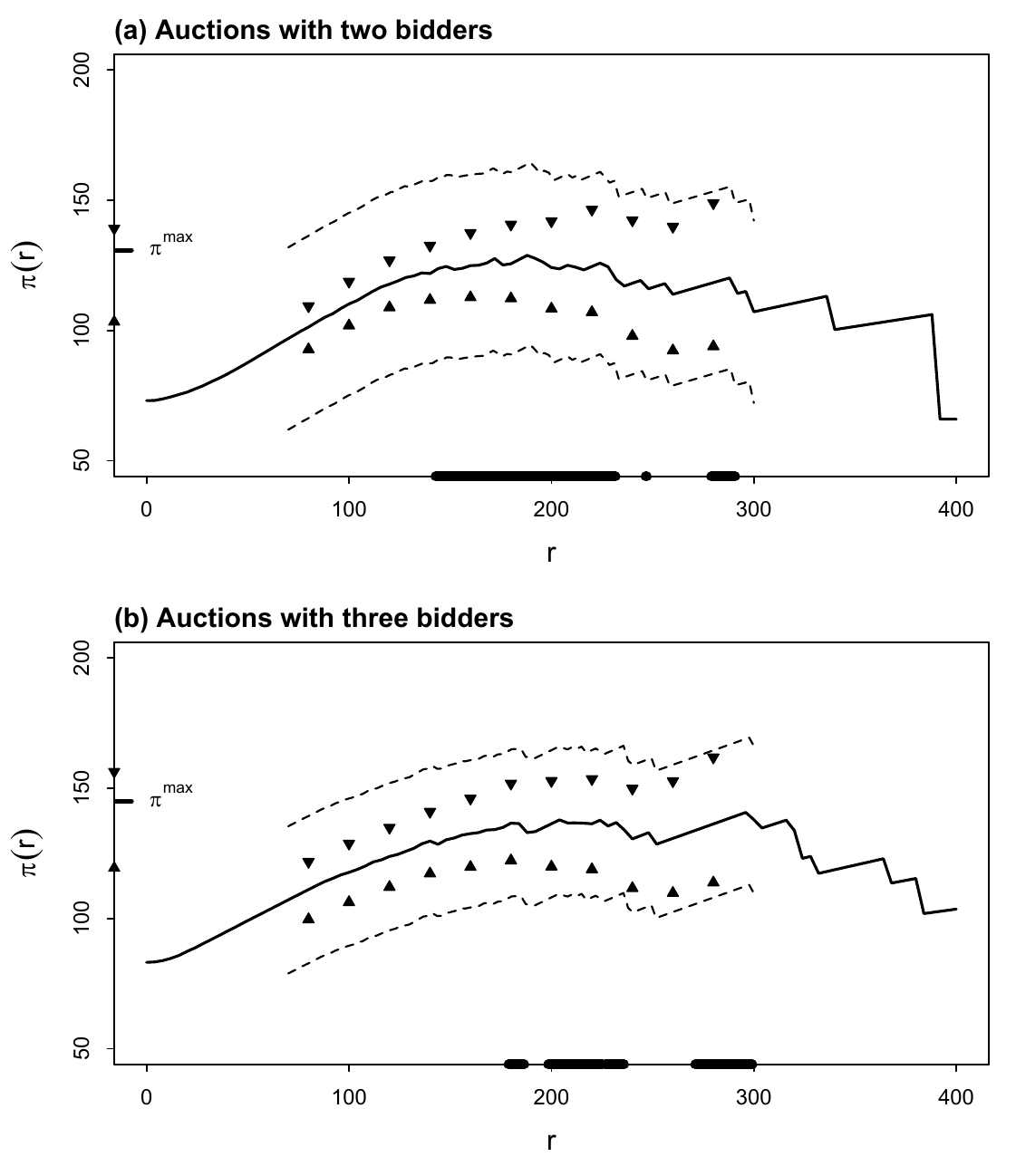}

}

{\raggedright \footnotesize{The solid curve represents the estimated expected revenue function, while the mark on the vertical axis is the estimated maximum expected revenue. Upward- and downward-pointing triangles indicate the lower and upper endpoints of 95\% pointwise confidence intervals. Dashed curves denote the 95\% confidence band over $[70 ,300]$. Thick segments on the horizontal axis represent the estimated set of optimal reserve prices.}}

\label{fig1ea}

\end{figure}

Figure \ref{fig1ea} presents the estimation results. In both cases, the estimated expected revenue (reported by the solid curve) initially increases with the reserve price and then becomes relatively flat over a broad range of intermediate reserve prices. At sufficiently high reserve prices, expected revenue begins to decline. The estimated expected revenue function is uniformly higher for auctions with three bidders than for auctions with two bidders: this difference is consistent with the stronger competition generated by an additional bidder.
The estimated maximum expected revenue is approximately 130 thousand 1979 U.S.\ dollars in the two-bidder sample and 145 thousand 1979 U.S.\ dollars in the three-bidder sample. Moreover, the estimated revenue function for three-bidder auctions remains close to its maximum over a wider range of reserve prices, suggesting that expected revenue is less sensitive to the precise choice of the reserve price in this case.


In both panels, the estimated set of optimal reserve prices (marked by thick segments on the horizontal axis) consists of several intervals concentrated in the relatively flat portion of the estimated expected revenue function, where different reserve prices yield very similar revenue levels.
This finding illustrates the usefulness of allowing the set of optimal reserve prices to contain multiple values rather than imposing uniqueness. From a practical perspective, the seller therefore appears to have substantial flexibility in choosing a reserve price without incurring a meaningful loss in expected revenue. The results also highlight the value of examining the entire expected revenue function when making such a decision. An increase in the reserve price may have only a modest effect on expected revenue while reducing bidder participation. Although this participation effect is not explicitly incorporated into the model, it may be relevant from a policy perspective, since unusually low participation could raise concerns about the competitiveness or perceived integrity of the auction process. 

The pointwise confidence intervals (marked by upward- and downward-pointing triangles) are noticeably narrower than the uniform confidence bands (marked by the dashed curves), as the latter account for simultaneous coverage of the expected revenue function over the entire interval $[70,300]$.
The substantial overlap of the pointwise confidence intervals across neighboring reserve prices suggests limited precision in distinguishing expected revenues near the estimated maximum. The confidence interval for the maximum expected revenue is asymmetric around its point estimate. This asymmetry arises from the reverse-percentile bootstrap procedure used to construct the interval, which is based directly on quantiles of the bootstrap distribution and therefore need not be symmetric around the estimated maximum expected revenue.

\section{Concluding remarks}		\label{sec:con}

In the context of an ascending auction with independent private values, symmetric bidders, and observable auction-specific covariates, this paper proposes estimators of the seller’s expected revenue function and the set of optimal reserve prices. I establish  weak convergence of the expected revenue estimator and develop a bootstrap method that consistently approximates its limiting distribution. From these results, I construct pointwise confidence intervals and uniform-in-$r$ confidence bands for the expected revenue function, as well as confidence regions for the set of optimal reserve prices.

I conclude by suggesting several directions for future research.
First, it would be interesting to consider a setting in which the econometrician observes only the transaction price $W$, but not the number of bidders $n$. In this case, the proposed framework could be complemented by a sensitivity analysis assessing how the results vary across alternative assumptions about the number of bidders.
Second, the framework could be extended to accommodate seller risk aversion, as in \cite{gim25id}. Although the asymptotic results and inference procedures developed in this paper could be extended to this setting, the simple computational procedures for the proposed estimators in Algorithms 1 and 2 would need to be modified.

Finally, more broadly, the approach developed in this paper could potentially be extended beyond the symmetric IPV paradigm with observed auction heterogeneity to settings in which the expected revenue function, or bounds on it, admit a suitable linear approximation. One possible direction is the correlated private-values framework of \citet[Assumption 3]{agq13}, which encompasses symmetric affiliated private values, the symmetric conditional IPV paradigm, and IPV with unobserved heterogeneity. In this framework, the authors derive bounds on the seller's expected revenue and expected surplus. Another potential application is the setting considered by \cite{powe26}, which allows for unobserved auction-level heterogeneity and a potentially unknown number of bidders.

These settings also highlight a broader motivation for the approach developed here: inference on expected revenue and optimal reserve prices may be feasible without fully recovering the underlying distribution of private values. Exploring whether the  proposed bootstrap procedure --generating residuals from a smoothed distribution while holding the covariates fixed-- can be adapted to such settings might be a direction for future research.

\clearpage

\appendix

\pdfbookmark[1]{Appendix}{Appendix}

\section*{Appendix: Proofs}

\counterwithin{equation}{section}

\renewcommand{\theequation}{A.\arabic{equation}} \setcounter{equation}{0}

\renewcommand{\thealemma}{A.\arabic{alemma}} \setcounter{lemma}{0}

\renewcommand{\thesection}{\Alph{section}.\arabic{equation}} \setcounter{equation}{0}

\renewcommand{\thesubsection}{A.\arabic{subsection}} \setcounter{subsection}{0}



This section provides the proofs of lemmas, propositions, and theorems stated in the body of the text. These proofs are provided in Section \ref{asubsec:proof} and rely on the auxiliary lemmas stated in Section \ref{subsec:alemmas} below. The auxiliary lemmas do not rely on results from the body of the text and their proofs are relegated to the Supplementary Material. 

Before stating the auxiliary lemmas, I introduce additional definitions and notation used throughout the Appendix and Supplementary Material.

Let $(\Omega,\mathscr{A},\Prob)$ denote the underlying
probability space, and let $\Prob_{\mathrm{i}}$ and
$\Prob^{\mathrm{o}}$ denote inner and outer probability,
respectively. To accommodate potentially nonmeasurable
mappings, I extend the notion of convergence in probability
used in the main text as follows. Consider a sequence of (not necessarily measurable) mappings
$\mathbb{Y}_L:\Omega\rightarrow\mathbb{R}$,
$L\in\mathbb{N}$. We say that $\mathbb{Y}_L$ converges
in probability to $c\in\mathbb{R}$ if every subsequence
$(\mathbb{Y}_{L_m})_{m\in\mathbb{N}}$ admits a further
subsequence $(\mathbb{Y}_{L_{m_k}})_{k\in\mathbb{N}}$
such that
$
\Prob_{\mathrm{i}}\{
\lim_{k\rightarrow\infty}\mathbb{Y}_{L_{m_k}}=c
\}=1.
$
In this case, we also write $\mathbb{Y}_L\overset{p}{\rightarrow}c$
or $\mathbb{Y}_L-c=o_p(1)$.


\paragraph{Weak convergence in probability.}

For each $L\in\mathbb{N}$, let $\mathcal{Z}_L\in\mathbb{R}^L$ be a random vector, independent of the sample $\mathcal{S}_L$, serving as the bootstrap randomization device.
Consider a normed space $( \mathcal{V} , \| \cdot \|_{ \mathcal{V} } )$ together with a sequence of mappings $\mathbb{Y}_L^\ast := \tilde{\phi}_L(\mathcal{S}_L,\mathcal{Z}_L) : \Omega \rightarrow \mathcal{V} $, where $L \in \mathbb{N}$ and each $\tilde{\phi}_L$ is a mapping from the support of $(\mathcal{S}_L,\mathcal{Z}_L)$ into $\mathcal{V}$. Let $\mathbb{Y}: \Omega \rightarrow \mathcal{V}$ be Borel measurable and separable.

We say that $\mathbb{Y}_L^\ast$ converges weakly to $\mathbb{Y}$ in probability, conditional on the sample $\mathcal{S}_L$, if every subsequence $(\mathbb{Y}_{L_m}^\ast)_{m\in\mathbb{N}}$ admits a further subsequence $(\mathbb{Y}_{L_{m_k}}^\ast)_{k\in\mathbb{N}}$ such that
\begin{equation}
\label{aeq:wcbequiv}
\Prob_{\mathrm{i}}
\Bigg\{
\forall \phi\in\mathscr{C}_{\mathrm{b}}(\mathcal{V}),
\ 
\lim_{k\rightarrow\infty}
\E^{\mathrm{o}}\left[
\phi(\mathbb{Y}_{L_{m_k}}^\ast)
\mid\mathcal{S}_{L_{m_k}}
\right]
\, = \,
\E[\phi(\mathbb{Y})]
\Bigg\} \  = \ 1 \, ,
\end{equation}
where  $\E^{\mathrm{o}}( \cdot \mid \mathcal{S}_L)$ denotes the outer expectation treating $\mathcal{S}_L$ as fixed.
In such a case, we write $\mathbb{Y}_L^\ast \overset{\ast}{\leadsto} \mathbb{Y}$ in $\mathcal{V}$ and, with a slight abuse of notation, we also write
$\mathbb{Y}_L^\ast\overset{\ast}{\leadsto}c$ when $\Prob(\mathbb{Y}=c)=1$ for some $c\in\mathcal{V}$.
Since the limiting process $\mathbb{Y}$ is separable, I remark that 
\begin{equation*}
\mathbb{Y}_L^\ast \overset{\ast}{\leadsto} \mathbb{Y}
\ \ \text{if and only if}  \ \ 
\hat{d}_{L} \left( \mathbb{Y}_L^\ast , \mathbb{Y}  \right) \, =\,  o_p(1) \, , 
\end{equation*}
where $\hat{d}_{L} $ denotes the bounded-Lipschitz metric treating $\mathcal{S}_L$ as fixed, i.e.,
\begin{equation*}
\hat{d}_{L} \left( \mathbb{Y}_L^\ast , \mathbb{Y}  \right) \, : = \, \sup_{\phi \in \mathrm{BL} (  \mathcal{V}   )} \, 
\left|  \E^{\mathrm{o}}\left[
\phi(\mathbb{Y}_{L}^\ast)
\mid\mathcal{S}_{L}
\right]    -  \E[\phi(\mathbb{Y})] \right| \, 
\end{equation*}
with $ \mathrm{BL} (  \mathcal{V}   )$ being the collection of mappings $\phi : \mathcal{V} \rightarrow \mathbb{R}$ that satisfy $| \phi (w) - \phi (v)  | \leq \| w - v  \|_{\mathcal{V}  }$ and $| \phi (v)   | \leq 1$ for all $v ,w \in\mathcal{V} $; see Section 1.12 in \cite{vw23}.

The definition in Eq.\ (\ref{aeq:wcbequiv}) differs slightly from the conventional formulation of bootstrap weak convergence that involves convergence in \emph{outer} probability of the bounded-Lipschitz metric.\footnote{Such a formulation can be obtained by establishing additional asymptotic measurability conditions; see \citet[Sections 1.9 and 1.12]{vw23}.} Nevertheless, it is sufficient for the inference procedures developed in this paper because it implies convergence of the conditional distributions of the relevant functionals, as explained below.

Let $\phi:\mathcal{V}\rightarrow\mathbb{R}$ be a continuous mapping such that $\phi(\mathbb{Y}_L^\ast) = \phi \circ \tilde{\phi}_L(\mathcal{S}_L,\mathcal{Z}_L)$ is Borel measurable on $(\Omega, \mathscr{A})$ for every $L\in\mathbb{N}$.
Then,
\begin{equation}
\label{eqa:impli}
\mathbb{Y}_L^\ast\overset{\ast}{\leadsto}
\mathbb{Y} \ \text{in} \  \mathcal{V}  \ \ \text{implies} \ \ 
\Prob\left(
\phi(\mathbb{Y}_L^\ast)\leq y
\mid\mathcal{S}_L
\right)
\ \overset{p}{\rightarrow} \
\Prob\left(
\phi(\mathbb{Y})\leq y
\right)
\end{equation}
at every continuity point $y$ of the c.d.f.\
of $\phi(\mathbb{Y})$. This implication follows by combining the continuous mapping theorem with the Portmanteau lemma, and it provides the basis for establishing consistency of the bootstrap critical values used throughout the paper.
Moreover, we have that 
\begin{equation}
\label{eqa:characc}
\phi(\mathbb{Y}_L^\ast) \overset{\ast}{\leadsto} 0  \  \text{in} \ \mathbb{R} \ \ \text{if and only if} \ \  
\forall \epsilon > 0 \, ,
\ \Prob \left( | \phi(\mathbb{Y}_L^\ast) | \geq \epsilon \mid\mathcal{S}_L \right)   \overset{p}{\rightarrow}  0 \, ; 
\end{equation}
this equivalence corresponds to the conventional definition of convergence to zero in bootstrap probability and, in such a case, we write $ \phi(\mathbb{Y}_L^\ast) = o_{p^\ast} (1)$.

%

\paragraph{Additional notation and definitions.}

Denote $\ushort{u} := \log ( \ushort{v} )$ and $\bar{u} : = \log ( \bar{v} )$, with the convention that $\log(0 ) = -\infty$. Moreover, let $\mathbb{D}_\Pi \subset \ell^\infty ( \mathbb{R} )$ be the collection of univariate c.d.f.'s supported on a compact subset,
and define the map $\Pi : \mathbb{D}_\Pi  \rightarrow \ell^\infty ( [ \ushort{r} , \bar{r} ]  )$ by $\Pi ( H  )  : = \Pi ( H , \cdot)$ with
\begin{equation}
\Pi ( H  , r )  \, : =  \, \lowmathcal{p}^{-1} [ H ( r) ]^n   v_s  \, + \,  n  \left\{ 1 - \lowmathcal{p}^{-1} [ H ( r ) ] \right\} \lowmathcal{p}^{-1} [ H( r ) ] ^{n-1} r  \, +  \,  \int_{r }^{\infty} w \, d H (w) .
\label{eqa:clfunc}
\end{equation}
Note that we can write $\Pi ( F_W  ) = \pi $, $\Pi ( \hat{F}_W  ) = \hat\pi $, $\Pi ( \hat{F}_W^\ast  ) = \hat\pi^\ast $, and $\Pi ( \hat{F}_{W, \lowmathcal{h} }  ) = \hat{\pi}_{\lowmathcal{h}} $. For a given $H \in  \mathbb{D}_\Pi$ and a real function $h$ satisfying $H + h \in  \mathbb{D}_\Pi $,  define also the (bounded) function $\Pi^\prime_{H} ( h ) : =  \Pi^\prime_{H} ( h  , \cdot )  \in \ell^\infty ( [ \ushort{r} , \bar{r} ]  ) $ by
\begin{equation*}
\Pi^\prime_{H} ( h  , r )  \, := \,  \varphi_{ H } (r) \, h( r)  \, - \,   \int_{r }^{\infty} h( w)  \, d w  \, , 
\end{equation*}
where $\varphi_H$ is defined by generalizing the definition in Eq.\ (\ref{eq:defvarphi}) to allow an arbitrary c.d.f.\ $H$ satisfying $H( \ushort{r} ) >0$, instead of $F_W$. Finally,  let $\mathbb{D}_0$ denote the collection of continuous functions $\phi : [ \ushort{r} , \infty )  \rightarrow \mathbb{R}$ that satisfy $\phi (w) = 0$ for all $w > \bar{v}$, noting that $\Pi^\prime_{F_W}$ is linear and continuous on $\mathbb{D}_0$.

\subsection{Auxiliary lemmas}	\label{subsec:alemmas}

The first auxiliary lemma establishes key properties of the preliminary estimators needed to derive the limiting distribution of the proposed expected-revenue estimator and its bootstrap analog.

\begin{alemma}
\label{alem:approxG}
The following statements hold under Assumptions \ref{assump:dgp}-\ref{assump:band}. 

\begin{enumerate}

\item
$f_U$ vanishes on $( \bar{u} , \infty)$, and it is bounded and Lipschitz continuous on $(  \ushort{u} , \infty)$.
Hence, the (covariance) function
\begin{equation*}
\Psi_\nu  (  u  , u^\prime)  \,   : =   \,     F_U ( u  \wedge u^\prime )   -   F_U ( u   )    \,F_U ( u^\prime )   
 +   f_U (u) \left( \mu_X^\tau  \,  \Sigma_X^{-1}  \mu_X \right)   f_U (u^\prime )  \, \sigma_U^2  
\end{equation*}
is Lipschitz continuous on $[ \log ( \ushort{r} ) , \infty ) \times [ \log ( \ushort{r} ) , \infty ) $.

\item 

\begin{enumerate}

\item 
$
\sup_{u \geq \log ( \ushort{r} ) }\,  \left|  \sqrt{L} \left[ \hat{F}_U ( u  )  -  F_U ( u ) \right]    
-   L^{-1/2} \sum_{l=1}^L \psi_l ( u )  \right| \, = \, o_p (1) 
$, where

$ \psi_l ( u )   : =     \mathds{1} \left( U_l  \leq u      \right)  -  F_U (u)  +    f_U (u) \mu_X^\tau \Sigma_X^{-1} ( X_l  - \mu_X )( U_l  - \mu_U )$.

\item $ L^{-1/2} \sum_{l} \psi_l    \,   \leadsto    \,  \mathbb{G}_{\nu}$ in $\ell^\infty ( [ \log ( \ushort{r} ) , \infty ) ) $,
where $\mathbb{G}_{\nu}$ is a tight  centered Gaussian process with covariance function given by $ \Psi_\nu $.

\item $  \sqrt{L}\, ( \hat{F}_U-  F_U  )  \,  \leadsto  \,   \mathbb{G}_{\nu} $ in $\ell^\infty ( [ \log ( \ushort{r} ) , \infty ) ) $.

\item  We have that $\Prob ( \mathbb{G}_{\nu} \in  \tilde{\mathbb{D}}_0) = 1$, where $\tilde{\mathbb{D}}_0$ denotes the subset of continuous functions $\phi :  [ \log ( \ushort{r} ) , \infty ) \rightarrow \mathbb{R}$ that satisfy $\phi ( u ) = 0$ for all $u >  \bar{u}$. Hence, $\mathbb{G}_{\nu}$ is separable.

\end{enumerate}

\item 

\begin{enumerate}

\item There exists a constant $\bar{c}_U >0$ such that the inequality
\begin{equation*}
\sup_{\tilde{u} , u   } \,
\frac{\left|  \hat{f}_{U , \lowmathcal{h} } ( \tilde{u}  ) -  \hat{f}_{U , \lowmathcal{h} } ( u )   \right|}{| \tilde{u} - u  |}  
 \,  \leq \,  \bar{c}_U \quad \text{holds w.p.a.1,}
\end{equation*}
where the supremum is taken over all distinct $\tilde{u} , u  \geq \ushort{u}^\prime : = \log[ ( \ushort{v} + \ushort{r} )  /2] $.

\item $\| \hat{f}_{U , \lowmathcal{h} }  - f_U  \|_{ ( \ushort{u} + 2 \lowmathcal{h}   , \infty) , \infty} 
\, = \, O_p (  \sqrt{ \log(  \lowmathcal{h}^{-1}  )/ ( \lowmathcal{h} L) } + \lowmathcal{h} )$.

\item $ \| \hat{F}_{U , \lowmathcal{h} }  - F_U  \|_{ \mathbb{R} , \infty} \, = \, O_p (    \sqrt{\log(  \lowmathcal{h}^{-1}   ) / (  \lowmathcal{h}  L  )  } +  \lowmathcal{h}  )$.

\item $ \hat{\mu}_{U ,  \lowmathcal{h}  } - \mu_U= o_p (1)$ and $\sup_{ c  \in \mathbb{R} } 
 \, |   \int_{-\infty}^c  u^2 \hat{f}_{U,\lowmathcal{h}} (u) \, du  - \E [ U^2\mathds{1} (   U\leq c ) ] |   
 =  o_p ( 1) $,
 
 where $\hat{\mu}_{U ,  \lowmathcal{h}  } :=  \int_{-\infty}^\infty   u \, d \hat{F}_{U , \lowmathcal{h} } (u) $.
 

\end{enumerate}

 \item  $\E (  \| \hat\beta^\ast - \hat\beta  \|_2^2 \mid \mathcal{S}_L ) = O_p ( L^{-1} )$.
 

\item Denote $ \psi_l^\ast (  u  )   : =   \mathds{1} \left( U_l^\ast  \leq u      \right)  -  \hat{F}_{U , \lowmathcal{h} } (u)   +     f_U (u) \, \bar{X}^\tau \hat{\Sigma}_X^{-1} \, ( X_l  - \bar{X} )( U_l^\ast  - \hat{\mu}_{U ,  \lowmathcal{h}  } ) $.
\begin{enumerate}

\item 
$
\sup_{ u \geq \log ( \ushort{r} ) }  \, 
\left|  \sqrt{L} \left[ \hat{F}_U^\ast(  u  )  -  \hat{F}_{U , \lowmathcal{h}} ( u ) \right]    
-    L^{-1/2}  \sum_{l=1}^{L} \psi_l^\ast ( u )  \right|
 \,  =  \, o_{p^\ast} (1)   
$.

\item $L^{-1/2} \sum_{l} \psi_l^\ast \,   \overset{\ast}{\leadsto}   \,   \mathbb{G}_{\nu}$ in $\ell^\infty ( [ \log ( \ushort{r} ) , \infty ) ) $.

\item As a result, $ \sqrt{L} \,  (\hat{F}_U^\ast -  \hat{F}_{U , \lowmathcal{h}} )    \,   \overset{\ast}{\leadsto}   \,   \mathbb{G}_{\nu}$ in $\ell^\infty ( [ \log ( \ushort{r} ) , \infty ) ) $.

\end{enumerate}

\end{enumerate}

\end{alemma}

The next auxiliary lemma outlines key differentiability properties of $\Pi$.

\begin{alemma}	\label{alem:approxGboot}

Suppose that Assumption \ref{assump:dgp} is satisfied.
Then, there exists a constant $ \bar{c}_{\pi} > 0$ such that the inequality
\begin{equation*}
\left\|  \Pi ( \tilde{H}  ) -  \Pi ( H  )  - \Pi^\prime_{H} (\tilde{H}  - H  )   \right\|_{ [ \ushort{r} , \bar{r} ] , \infty  } 
\   \leq  \ \bar{c}_{\pi}  \, \left\|     \tilde{H}  - H   \right\|_{ [ \ushort{r} ,  \bar{r} ]  , \infty  }^2  
\end{equation*}
 holds for any $(H , \tilde{H} ) \in  \mathbb{D}_\Pi^2$ satisfying
\begin{equation*} 
\frac{ F_W (  \ushort{r} ) }{ 2 }  \ \leq \  H ( \ushort{r} ) \wedge   \tilde{H} ( \ushort{r} )
  \ \leq \  H ( \bar{r} ) \vee   \tilde{H} ( \bar{r} ) \ \leq \  \frac{1+ F_W (  \bar{r} ) }{ 2 }  \, .
\end{equation*}
Hence, considering $\Pi$ as a function on $\mathbb{D}_\Pi \subset \ell^\infty ( \mathbb{R} )$, $\Pi^\prime_{F_W} :  \mathbb{D}_0 \subset \ell^{\infty} (\mathbb{R} ) \rightarrow \ell^\infty ( [ \ushort{r} , \bar{r} ]  )  $ is the Hadamard derivative of $\Pi $ at $F_W$ tangentially to $\mathbb{D}_0$.

\end{alemma}

%

The next two auxiliary lemmas provide useful properties of Gaussian processes satisfying certain regularity conditions.

\begin{alemma}		\label{alem:contG}

Consider a nonempty and compact subset $R \subset \mathbb{R}^m$, where $m \in \mathbb{N}$. Let $\mathbb{G}$ be a centered Gaussian process in $\mathscr{C}( R )$ with Lipschitz-continuous covariance function $\Psi$ satisfying $\inf_{r \in R} \Psi (r,r) > 0$. Then,  the c.d.f.\ of $\| \mathbb{G}   \|_{R, \infty}$ is continuous on $\mathbb{R}$ and strictly increasing on $\mathbb{R}_{\geq0}$.

\end{alemma}

\begin{alemma}		\label{alem:supptG}
Let $\mathbb{B}$ be a Brownian Bridge and $\mathrm{Z}$ be a standard normal random variable. Assume that $\mathbb{B}$ and $\mathrm{Z}$ are independent, and consider 
\begin{equation*}
\tilde{\mathbb{G}} ( r) \ := \ \mathbb{B} \circ F_W ( r) \, + \, 
 f_U \left( \log( r ) \right)  \left( \mu_X^\tau  \,  \Sigma_X^{-1}  \mu_X \right)^{1/2} \sigma_U \,  \mathrm{Z}
\end{equation*}
as a Gaussian process indexed by $r \in [ \ushort{r} , \infty)$.
The following statements hold under Assumption \ref{assump:dgp}.
\begin{enumerate}

\item For every $r \in [ \ushort{r} , \bar{r} ] $, $\Prob \left(   \Pi^\prime_{F_W} ( \tilde{\mathbb{G}}  ,  r )   = 0 \right) = 0$.

\item For every $c \in \mathbb{R}$ and any distinct $r_1 , r_2 \in [ \ushort{r} , \bar{r} ] $, $\Prob \left(   \Pi^\prime_{F_W} ( \tilde{\mathbb{G}}  ,  r_2 )   - \Pi^\prime_{F_W} ( \tilde{\mathbb{G}}  ,  r_1 )  = c \right) = 0$.

\item For any $(r_1 , r_2) \in [ \ushort{r} , \bar{r} ]^2 $, $\mathrm{cov} \left(  \Pi^\prime_{F_W} ( \tilde{\mathbb{G}}  ,  r_1 )   , \Pi^\prime_{F_W} ( \tilde{\mathbb{G}}  ,  r_2 )   \right)  =  \Psi_\pi ( r_1 , r_2)$.

\item  For any nonempty and closed subset $R \subset [ \ushort{r} , \bar{r} ]$,
the c.d.f.\ of $\sup_{r \in R} \,  \Pi^\prime_{F_W} ( \tilde{\mathbb{G}}  ,  r )  $ is continuous and strictly increasing on $\mathbb{R}$.

\item  For any closed subset $R \subseteq [ \ushort{r} , \bar{r} ] $ that contains more than one element, the c.d.f.\ of the oscillation $ \mathrm{osc}_{R} \left( \Pi^\prime_{F_W} ( \tilde{\mathbb{G}}  ,  \cdot ) \right)$ is continuous on $\mathbb{R}$ and strictly increasing on $\mathbb{R}_{\geq0}$.

\end{enumerate}

\end{alemma}

%

\subsection{Proofs of the results in the main text}	\label{asubsec:proof}


\subsubsection*{Proof of Lemma \ref{lem:weakconG}}

Since $\hat{F}_W (w) = \hat{F}_U [ \log(w)]$ and $ F_W (w) = F_U [ \log(w)] $ for all $w \in \mathbb{R}_{>0}$, the desired result follows by combining the Continuous mapping theorem with Lemma \ref{alem:approxG}.2, which yields
\begin{equation*}
\sqrt{L} \left(\hat{F}_W   -  F_W \right)  = \sqrt{L}\left(\hat{F}_U \circ \log  -  F_U \circ \log \right)   \  \leadsto  \   \mathbb{G}_{\nu} \circ \log \quad \text{in}  \ \ell^\infty \left(  [ \ushort{r},  \infty ) \right) , 
\end{equation*}
and then by noting that the covariance function of $\mathbb{G}_{\omega} : =  \mathbb{G}_{\nu} \circ \log $ is $\Psi_\omega  (  w  , y )   = \Psi_\nu  (  \log( w )  ,\log( y )) $. Finally, tightness of $\mathbb{G}_{\omega}$ follows directly from the tightness of $\mathbb{G}_{\nu}$ (Lemma \ref{alem:approxG}.2.(b)) and continuity of $\log$ on $\mathbb{R}_{ >0 }$.
\qed

\subsubsection*{Proof of Theorem \ref{thm:main}}

\paragraph{1.}
The proof of this part is divided into three steps. In the first, I show that $\mathbb{G}_{\omega} = \mathbb{G}_{\nu} \circ \log$ takes its values in $\mathbb{D}_0$ w.p.1; hence, it is tight by separability and completeness of $\mathbb{D}_0$ \cite[Lemma 1.3.2]{vw23}. To establish that $\Prob ( \mathbb{G}_{\omega} \in \mathbb{D}_0 ) = 1$, note that  $\{  \tilde\phi \circ \log : [ \ushort{r} , \infty) \rightarrow \mathbb{R}  :  \tilde\phi \in \tilde{\mathbb{D}}_0 \} \subseteq \mathbb{D}_0$ and  that $\mathbb{G}_{\nu}$ takes its values in $\tilde{\mathbb{D}}_0$ w.p.1 by Lemma \ref{alem:approxG}.2.(d).


In the second step, I prove that $\sqrt{L} ( \hat\pi - \pi ) $ converges weakly to a tight centered Gaussian process in $\ell^\infty ( [ \ushort{r} , \bar{r} ] )$. To do so, recall that $\mathbb{G}_{\omega} $ is separable (first step) and note that $\Pi^\prime_{F_W}$ is the Hadamard differential of $\Pi$ at $F_W$ tangentially to $\mathbb{D}_0$ (Lemma \ref{alem:approxGboot}).
Thus, after applying the Delta method \cite[Theorem 3.10.4]{vw23}, we obtain
\begin{equation*}
\sqrt{L} \left( \hat\pi - \pi \right)  \, =  \,     \sqrt{L}  \, \left[ \Pi ( \hat{F}_W  )  - \Pi ( {F}_W  )   \right] \   \leadsto  \    
\Pi^\prime_{F_W} ( \mathbb{G}_{\omega}  ) 
\, = : \, \mathbb{G}_{\pi}  \quad \text{in}  \ \ell^\infty \left(  \left[ \ushort{r},  \bar{r} \right] \right) ,
\end{equation*}
where $\mathbb{G}_{\pi}$ is a tight Gaussian process by Proposition 2.7 in \cite{du14clt}; recall that $\mathbb{G}_{\omega}$ is a tight Gaussian process in $\ell^\infty \left(  [ \ushort{r},  \infty ) \right)$ by Lemma \ref{lem:weakconG}, while $\Pi^\prime_{F_W} $ is linear and continuous on $\mathbb{D}_0$.

In the third step, I derive the covariance function of $\mathbb{G}_{\pi}$. To this end, I simply invoke Lemma \ref{alem:supptG}.3 and note that $ \mathbb{G}_{\omega}$ and $\tilde{\mathbb{G}}$ are equal in law.
\paragraph{2.} 
By contradiction, suppose that $\mathrm{V}_\pi (r)  = 0$ for some $r \in [  \ushort{r}  , \bar{r} ]$, so that $\Prob \left(   \Pi^\prime_{F_W} ( \mathbb{G}_{\omega}   ,  r )   = 0 \right)  = 1$.
But, by Lemma \ref{alem:supptG}, this is impossible because $\mathbb{G}_{\omega}$ is equal in law to $\tilde{\mathbb{G}}$.
\qed

\paragraph{Proof of Lemma \ref{lem:mainboot}}

\paragraph{1.}
The desired result follows by choosing an arbitrary increasing sequence $L_m \in \mathbb{N}$ and then applying the arguments used in the proof of Lemma \ref{lem:weakconG} along a properly chosen subsequence $(L_{m_k})$.

\paragraph{2.}
By Lemma \ref{alem:approxGboot}, for any $\epsilon >0$, note that we can bound
\begin{multline*}
\Prob \left\{ \left\|
\sqrt{L} ( \hat\pi^\ast - \hat\pi_{ \lowmathcal{h} }  ) 
\, - \,
\Pi_{ \hat{F}_{W , \lowmathcal{h} }}^\prime  \left( \sqrt{L}   \left(  \hat{F}_W^\ast - \hat{F}_{W , \lowmathcal{h} }  \right) \right)
\right\|_{ [ \ushort{r}  , \bar{r} ]  ,  \infty} > \epsilon  \mid \mathcal{S}_L
\right\}  \\
\leq \, 
\Prob \left\{ \bar{c}_\pi \left\|  \sqrt{L} \left(  \hat{F}_W^\ast - \hat{F}_{W , \lowmathcal{h} }  \right)
\right\|_{ [ \ushort{r}  , \bar{r} ]  ,  \infty}^2  >  \epsilon \sqrt{L}   \mid \mathcal{S}_L  \right\} + o_p (1) \, = \, o_p (1) \,,
\end{multline*}
with the equality following by the first part of this lemma, the implication in Eq.\ (\ref{eqa:impli}), and tightness of $ \mathbb{G}_{\omega}$.
To establish the desired result, write $\hat{\bar{v}}  : = \exp ( \max\{ \hat{U}_1 , \dots , \hat{U}_L  \} + 2\lowmathcal{h} ) \vee \bar{r}$ and 
\begin{equation*}
\hat{\Pi}^\prime  ( h  , r )  \, := \,  \varphi_{ \hat{F}_{W , \lowmathcal{h} } } (r) \, h( r)  \, - \,   \int_{r }^{2 \hat{\bar{v}}   } h( w)  \, d w   \, .
\end{equation*}
Then, since $\hat{F}_W^\ast $ is supported on a subset of $( - \infty , \hat{\bar{v}} \exp( \max_{l} \| X_l \|_2 \,  \| \hat\beta^\ast - \hat\beta  \|_2 )  ]$, observe that Lemma \ref{alem:approxG}.4 implies
\begin{multline*}
\Prob \left\{ \left\|
\Pi_{ \hat{F}_{W , \lowmathcal{h} }}^\prime  
\left( \sqrt{L}   \left(  \hat{F}_W^\ast - \hat{F}_{W , \lowmathcal{h} }  \right) \right)
\, - \,
\hat{\Pi}^\prime    \left( \sqrt{L}   \left(  \hat{F}_W^\ast - \hat{F}_{W , \lowmathcal{h} }  \right) \right)
\right\|_{ [ \ushort{r}  , \bar{r} ]  ,  \infty} > \epsilon  \mid \mathcal{S}_L
\right\} \\
\leq \,
\Prob \left\{ 
\int_{2 \hat{\bar{v}} }^\infty \left|  \sqrt{L}   \left(  \hat{F}_W^\ast - \hat{F}_{W , \lowmathcal{h} }  \right) \right|
  > \epsilon  \mid \mathcal{S}_L
 \right\} 
\, \leq \,     
\Prob \left\{ 
\sup_{x \in \mathscr{S}_X } \| x \|_2 \,  \left\| \hat\beta^\ast - \hat\beta  \right\|_2   > \log (2)  \mid \mathcal{S}_L
\right\}  \,  = \, o_p (1) \, ,
\end{multline*}
while $\hat{\Pi}^\prime  ( \sqrt{L}   (  \hat{F}_W^\ast - \hat{F}_{W , \lowmathcal{h} }  ) )  \overset{\ast}{\leadsto}  \mathbb{G}_\pi $ follows by the first part of this lemma and the extended version of the continuous mapping theorem \cite[Theorem 1.11.1.(i)]{vw23}.
\qed

\paragraph{3.}
Since the c.d.f.\ of $ \|  \mathbb{G}_\pi  \|_{R,\infty }$ is continuous and strictly increasing on its support by Theorem \ref{thm:main}.2 and Lemma \ref{alem:contG}, the desired result follows as a corollary of the second part of this lemma; see, e.g., \citet[Lemma 11.2.1.(ii)]{lehmammromano05}.
\qed

\subsubsection*{Proof of Proposition \ref{propo:pointboots}}
By standard arguments, note that we can write
\begin{multline*}
\Prob \left\{  \pi(r) \in \mathrm{CI}_{1-\alpha} (r)  \right\}  \ = \ \Prob \left\{ \sqrt{L} \left[ \hat{\pi}(r) - \pi(r) \right]  \,+ \, {q}_{1-{\alpha}/{2}} (r)  - \hat{q}_{1-{\alpha}/{2}} (r)  \, \leq \, {q}_{1-{\alpha}/{2}} (r)  \right\}  \\
- \  \Prob \left\{ \sqrt{L} \left[ \hat{\pi}(r) - \pi(r) \right]  \,+ \, {q}_{{\alpha}/{2}} (r)  - \hat{q}_{{\alpha}/{2}} (r)  \, < \, {q}_{{\alpha}/{2}} (r)  \right\} \, ,
\end{multline*}
so the desired result follows by Theorem \ref{thm:main}, Lemma \ref{lem:mainboot}.3, and Slutsky’s theorem.
\qed

\subsubsection*{Proof of Proposition \ref{propo:cband}}
Since $\Prob  \{  \forall r \in   [ \ushort r, \bar r ] , \, \pi (r)  \in  \mathrm{CB}_{1-\alpha} (r)    \}  = \Prob 
 \{  \| \sqrt{L} (  \hat\pi  - \pi )  \|_{ [ \ushort{r} , \bar{r} ]  ,\infty }  \leq  \hat{q}_{1-{\alpha}}^\infty ( [ \ushort r , \bar r] )   \} $, the desired result follows by arguments similar to those used in the preceding proof.
\qed

\subsubsection*{Proof of Lemma \ref{lem:optiest}}

\paragraph{1.} 
Consider the functional $\lowmathcal{m}: \ell^\infty  ( [ \ushort{r} , \bar{r}])\rightarrow \mathbb{R}$ defined by $\lowmathcal{m} ( \phi ) :=  \sup_{r \in  [ \ushort{r} , \bar{r}] } \phi(r)$. Note that it is Hadamard directionally differentiable tangentially to $\mathscr{C}( [ \ushort{r} , \bar{r}] )$ at any $\phi \in \mathscr{C}( [ \ushort{r} , \bar{r}] )$, and the corresponding derivative is given by
\begin{equation*}
\lowmathcal{m}_\phi^\prime ( h ) \ := \  \sup_{r\in \mathcal{M} (\phi)}  h(r)   ,
\end{equation*}
where $ h \in   \mathscr{C}( [ \ushort{r} , \bar{r}] )  $ and $\mathcal{M} (\phi) : = {\arg\!\max}_{r \in [ \ushort{r} , \bar{r}]}  \ \phi(r) $;
see Lemma S.4.9 in \citet[Supplementary Appendix]{fang19inf}.
Hence, since $ \mathbb{G}_{\omega}   \in  \mathbb{D}_0 $ w.p.1 and $ \Pi^\prime_{F_W} ( \mathbb{D}_0 )  \subset \mathscr{C}( [ \ushort{r} , \bar{r}] )$ (see the proof of Theorem \ref{thm:main} above), we have that the limiting Gaussian process satisfies $\mathbb{G}_{\pi}  = \Pi^\prime_{F_W} ( \mathbb{G}_{\omega}  ) \in \mathscr{C}( [ \ushort{r} , \bar{r}] )$ w.p.1 and therefore the desired result follows from the Delta method for Hadamard directionally differentiable functionals:
\begin{equation*}
\sqrt{L} \left( \hat{\pi}_{\max} -  {\pi}_{\max}  \right)
 \,  = \,  \sqrt{L} \, \left[ \lowmathcal{m} ( \hat{\pi} )  -  \lowmathcal{m}  ({\pi} )  \right]  
   \  \leadsto \  \lowmathcal{m}_\pi^\prime ( \mathbb{G}_{\pi}  ) \, = \,  \sup_{r \in \mathcal{R}} \, \mathbb{G}_{\pi} (r)  \, ;
\end{equation*}
recall that $\sqrt{L} ( \hat{\pi} - {\pi} ) \leadsto \mathbb{G}_{\pi} $ (Theorem \ref{thm:main}.1) and see \citet[Theorem 2.1]{fang19inf}.

\paragraph{2.} 
To establish weak convergence, let $ \mathcal{T} (\phi) : = \sup_{r \in [  \ushort{r}  , \bar{r}  ]}  {\phi} (r)    -   \inf_{r \in \mathcal{R}}  {\phi} (r) \geq 0$, $\phi \in \ell^\infty (  [  \ushort{r}  , \bar{r}  ] ) $, and observe that
\begin{equation*}
\|  \hat{\mathcal{C}} \|_{ \mathcal{R} , \infty }  \  =  \ \sup_{r \in \mathcal{R}}  \, \hat{\mathcal{C}} (r) 
 =  \  \left[ \sup_{r \in [  \ushort{r}  , \bar{r}  ]}  \hat{\pi} (r)    -   \inf_{r \in \mathcal{R}}  \hat{\pi} (r)  \right] 
\,  - \,  \underbrace{\left[ \sup_{r \in [  \ushort{r}  , \bar{r}  ]}  {\pi} (r)    -   \inf_{r \in \mathcal{R}}  {\pi} (r)  \right]}_{= \, {\pi}_{\max} - {\pi}_{\max} \, = \, 0}  \ = \ \mathcal{T} (\hat\pi)  - \mathcal{T} (\pi) \, .
\end{equation*}
Then, by arguments similar to those used in the preceding proof, the desired result follows by noting that the Hadamard directional derivative of  $ \mathcal{T}$ at $\pi \in \mathscr{C}( [ \ushort{r} , \bar{r}] )$, tangentially to $\mathscr{C}( [ \ushort{r} , \bar{r}] )$, is given by 
\begin{equation*}
 \mathcal{T}_{  \pi  }^\prime ( h ) \ : = \  \sup_{r \in \mathcal{R}} \, h (r)  \, - \, \inf_{r \in \mathcal{R}}  \, h (r)  \ = \ \mathrm{osc}_{\mathcal{R}} (h )
\end{equation*}
because $\pi (r) = \pi_{\max} \ \forall \, r \in \mathcal{R}$ and therefore $\arg\!\min_{r \in \mathcal{R}} \pi (r) = \mathcal{R}$.
To prove that $\mathcal{R}  \subseteq \hat{\mathcal{R}} (c_L)$ w.p.a.1 and $\lowmathcal{d}_{H} (  \hat{\mathcal{R}} (c_L)  ,  \mathcal{R} ) = o_p (1)$, simply invoke Theorem 3.1.(1) in  \cite{cht07}, noting that their Condition C.1 is clearly satisfied because $ \sqrt{L} \|  \hat{\mathcal{C}} \|_{ \mathcal{R} , \infty } = O_p (1)$.
\qed


\subsubsection*{Proof of Proposition \ref{propo:infoptp}}

\paragraph{1.} 
Note that we can write
\begin{multline*}
\Prob \left(  \pi_{\max} \in \mathrm{CI}_{1-\alpha}^{\max}  \right)  \ = \ \Prob \left\{ \sqrt{L} \left( \hat{\pi}_{\max}  - \pi_{\max}  \right)  \,+ \, {q}_{1-{\alpha}/{2}}^{\max}   - \hat{q}_{1-{\alpha}/{2}}^{\max}   \, \leq \, {q}_{1-{\alpha}/{2}}^{\max}  \right\}  \\
- \  \Prob \left\{ \sqrt{L}\left( \hat{\pi}_{\max}  - \pi_{\max}  \right)  \,+ \, {q}_{{\alpha}/{2}}^{\max}   - \hat{q}_{{\alpha}/{2}}^{\max}   \, < \, {q}_{{\alpha}/{2}}^{\max}  \right\} \, ,
\end{multline*}
where
\begin{equation*}
{q}_{a}^{\max}  \   : = \    \inf\left\{  y \in \mathbb{R} : \ \Prob \left(   \sup_{r \in \mathcal{R}} \, \mathbb{G}_\pi (r) \leq y \right)  \geq a   \right\}  .
\end{equation*}
Then, by arguments analogous to those in the proof of Proposition \ref{propo:pointboots} and the fact that $ \sup_{r \in \mathcal{R}} \,  \mathbb{G}_\pi (r) $ has a continuous and strictly increasing distribution (Theorem \ref{thm:main} and Lemma \ref{alem:supptG}), the desired result follows by establishing
\begin{equation}
\label{eqa:boot00}
\sup_{r \in  \hat{\mathcal{R}} (c_L)} \sqrt{L} \left[ \hat{\pi}^\ast(r)   - \hat{\pi}_{ \lowmathcal{h} }(r) \right] 
 \, \overset{\ast}{\leadsto} \, 
 \sup_{r \in \mathcal{R}} \,  \mathbb{G}_\pi (r)  .
\end{equation}
To prove this result, first, choose any increasing sequence $( L_m )_{m \in \mathbb{N}}$ of positive integers and pick a subsequence $( L_{m_k} )_{k \in \mathbb{N}}  $ such that
$\Prob_{\mathrm{i}} (  \Omega_{ \mathcal{R} }  \cap \Omega_{ wc } ) = 1$,
where 
\begin{equation*}
\Omega_{ \mathcal{R} }  \, : = \, \left\{  \lim_{k \rightarrow \infty  } \lowmathcal{d}_{H} \left(  \hat{\mathcal{R}} (c_{  L_{m_k} })  ,  \mathcal{R}  \right)  = 0  \right\}
\  \    \text{and}  \  \ 
\Omega_{ wc }  \,  :  =  \,     \left\{
\lim_{k \rightarrow \infty  }  \hat{d}_{L_{m_k} } 
\left( \sqrt{L_{m_k}} \left( \hat{\pi}^\ast   - \hat{\pi}_{ \lowmathcal{h} }  \right)  , \mathbb{G}_\pi    \right)
\, = \, 0
 \right\}   ;
\end{equation*}
such a sub-subsequence exists due to Lemmas \ref{lem:mainboot}.2 and \ref{lem:optiest}.2.
Second, observe that $\Omega_{ \mathcal{R} } \subseteq \Omega_{ cont }  $, where
\begin{multline*}
\Omega_{ cont }    : =   \bigg\{
\forall \  \text{sub-subsequence} \ h_{ L_{m_k} } \in \ell^\infty \left( [ \ushort{r} , \bar{r}  ]    \right)
\  \text{satisfying} \  h_{ L_{m_k} }  \rightarrow   \tilde{h}  \ \text{for some} \  \tilde{h} \in  \mathcal{C} ( [ \ushort{r} , \bar{r}  ]  ), \\
\sup_{r \in  \hat{\mathcal{R}} (c_{  L_{m_k} })} h_{ L_{m_k} } (r)  \, \rightarrow \, \sup_{r \in  \mathcal{R}} \tilde{h}(r) 
\bigg\} .
\end{multline*}
Third, the result in Eq.\ (\ref{eqa:boot00}) follows by noting that
\begin{equation*}
\Omega_{ \mathcal{R} }  \cap \Omega_{ wc }  \subseteq   \Omega_{ cont }  \cap \Omega_{ wc } \subseteq  
\left\{   
\lim_{k \rightarrow \infty  }  \, \hat{d}_{L_{m_k} } 
 \left( \sup_{r \in  \hat{\mathcal{R}} (c_{L_{m_k} })} \sqrt{L_{m_k}} \left[ \hat{\pi}^\ast(r)   - \hat{\pi}_{ \lowmathcal{h} }(r) \right]   , 
 \sup_{r \in \mathcal{R}} \,  \mathbb{G}_\pi (r)   \right)
\, = \, 0
\right\} ,
\end{equation*}
where the second inclusion is a consequence of the extended version of the continuous mapping theorem \cite[Theorem 1.11.1.(i)]{vw23}.

\paragraph{2.} 
It follows from 
$
\mathrm{osc}_{\hat{\mathcal{R}} (c_L)} \left( \sqrt{L} ( \hat{\pi}^\ast   - \hat{\pi}_{ \lowmathcal{h} }  ) \right)
 \overset{\ast}{\leadsto}  
 \mathrm{osc}_{\mathcal{R}} \left( \mathbb{G}_\pi \right)  
$,
which can be established using arguments analogous to those used in the preceding proof, and Lemma \ref{alem:supptG}.5.
\qed

\subsubsection*{Proof of Proposition \ref{propo:convers}}
Observe that 
\begin{eqnarray*}
\Prob \left( \mathcal{R}  \subseteq \mathrm{CR}_{1-\alpha}^{ \mathrm{con} }  \right)  
& = &  
\Prob \left\{ \sqrt{L} \|  \hat{\mathcal{C}} \|_{ \mathcal{R} , \infty }  \leq  2 \, \hat{q}_{ 1 - \alpha }^{\infty}  \left[   \hat{\mathcal{R}} (c_L)  \right]    \right\}  \\
 & \underset{ \text{Lemma \ref{lem:optiest}} }{\geq}  &   \Prob \left\{ \sqrt{L} \|  \hat{\mathcal{C}} \|_{ \mathcal{R} , \infty }  \leq  2 \, \hat{q}_{ 1 - \alpha }^{\infty}  \left( \mathcal{R} \right)  \, , \, \mathcal{R}  \subseteq \hat{\mathcal{R}} (c_L)     \right\}   \, + \, o (1) \\
& = &  \Prob \left\{ \sqrt{L} \|  \hat{\mathcal{C}} \|_{ \mathcal{R} , \infty }  \leq  2 \, \hat{q}_{ 1 - \alpha }^{\infty}  \left( \mathcal{R} \right)     \right\} \, + \, o(1) \\
& = & \Prob \left\{ \sqrt{L} \|  \hat{\mathcal{C}} \|_{ \mathcal{R} , \infty }  +
2 \left[{q}_{ 1 - \alpha }^{\infty}  \left( \mathcal{R} \right) -  \hat{q}_{ 1 - \alpha }^{\infty}  \left( \mathcal{R} \right) \right]
\, \leq \, 2 \, {q}_{ 1 - \alpha }^{\infty}  \left( \mathcal{R} \right)     \right\}  \, + \, o(1) .
\end{eqnarray*}
Then, the desired result follows by arguments similar to those used in the proof of Proposition \ref{propo:pointboots}, as well as from $2{q}_{ 1 - \alpha }^{\infty}  \left( \mathcal{R} \right) \geq {q}_{ 1 - \alpha }^{ \mathrm{osc} }$ and ${q}_{ 1 - \alpha }^{\infty}  \left( \mathcal{R} \right)   > 0$ (Lemma \ref{alem:contG}).
\qed

\vspace{.25in}

\section*{Acknowledgements}

I thank Fede Bugni, Gaurab Aryal, and Ruli Xiao for valuable discussions and insightful comments, as well as the participants at the SEA 95th Annual Meeting and the 2026 MEG Annual Meeting for their helpful feedback and suggestions. 
This work was completed utilizing the Holland Computing Center of the University of Nebraska, which receives support from the Nebraska Research Initiative. This research did not receive any specific grant from funding agencies in the public, commercial, or nonprofit sectors.

\clearpage

\section*{Supplementary Material: Proofs of Auxiliary Lemmas}

\renewcommand{\thepage}{S.\arabic{page}}\setcounter{page}{1}

\renewcommand{\theequation}{S.\arabic{equation}}
\renewcommand{\theHequation}{S.\arabic{equation}}
\setcounter{equation}{0}


\subsubsection*{Proof of Lemma \ref{alem:approxG}}

\paragraph{1.}   By the transformation formula, observe that $f_U$ takes the form of
\begin{equation}
f_U (u) \, = \,
\left\{
\begin{array}{ll}
n(n-1)\,
e^u\,
[F_V(e^u)]^{\,n-2}
\bigl[1-F_V(e^u)\bigr]
f_V(e^u)
& \text{for} \  u \in \left( \ushort{u}  \, , \, \bar{u}  \right) , \\
0& \text{otherwise} .
\end{array}
\right.
\label{eqa:fu}
\end{equation}
So, Lipschitz continuity of $f_U$ on
$( \ushort{u}  , \infty ) $ arises from the fact that $f_V$ is Lipschitz continuous (Assumption \ref{assump:dgp}.3) and because $f_U (u)$ vanishes smoothly as $u \rightarrow \bar{u}$. Then, the consequence follows since any finite sum of products of bounded Lipschitz continuous functions is itself Lipschitz continuous.

\paragraph{2.}

\paragraph{(a)} Define the function
\begin{equation*}
\check{G} ( u  , b ) \,  : =  \, \int \, F_U  \left( u + x^\tau ( b - \beta) \right) \,  d F_X (x)  \,  = \,
\E \left[ \mathds{1} ( U    \leq u  +  X^\tau ( b - \beta) )  \right]
\end{equation*}
for $(  u ,  b ) \in \mathbb{R} \times \mathbb{R} ^d$, where the second equality follows by the Law of Iterated Expectations and by independence between $U$ and $X$ (Assumption \ref{assump:dgp}). Denote also
\begin{equation*}
\mathcal{R}_l (  u ) \,  : = \,  \mathds{1} \left( \hat{U}_l  \leq u      \right)   - \check{G} ( u  , \hat\beta  )
 -\left[ \mathds{1} \left( U_l  \leq u      \right)   -  \check{G} ( u  , \beta  ) \right] 
\end{equation*}
and observe that $ \check{G} ( u  , \beta  ) = F_U (u) $, so we can write
\begin{eqnarray*}
 \hat{F}_U ( u  )  -  F_U ( u )     & = &  \frac{1}{L} \sum_{l=1}^L \left[   \mathcal{R}_l ( u)  +   \check{G} ( u  , \hat\beta  )  +   \mathds{1} \left( U_l  \leq u      \right)    -   2 \check{G} ( u  , \beta  )  \right] \\
&  =  &    \frac{1}{L} \sum_{l=1}^L    \mathds{1} \left( U_l  \leq u      \right)  - F_U ( u )      +    \check{G} ( u  , \hat\beta  )   -   \check{G} ( u  , \beta  )   +  \frac{1}{L} \sum_{l=1}^L  \mathcal{R}_l ( u )   .
\end{eqnarray*}
To obtain the desired result, it suffices to show that both
\begin{eqnarray}
 \check{G} ( u  , \hat\beta  )   -   \check{G} ( u  , \beta  )&   =  &  f_U (u)  \mu_X^\tau \, \Sigma_X^{-1} \, \frac{1}{L} \sum_{l=1}^L (X_l - \mu_X) (U_l  - \mu_U) \,   +  \,   o_p \left( L^{-1/2} \right) 
 \  \   \text{and}      \label{eqa:olsap}  \\
 L^{-1/2} \sum_{l=1}^L  \mathcal{R}_l (u ) &   =  &  o_p ( 1)   
 \label{eqa:stoap} 
\end{eqnarray}
hold uniformly in $u \geq \log ( \ushort{r} )$.

Starting with Eq.\ (\ref{eqa:olsap}), note that $\partial \check{G} ( u  , b ) / \partial  b= \int f_U (  u + x^\tau ( b - \beta)  ) x \, d F_X  (x)$ for $u > \log( \ushort{r}  )$ and recall that $f_U$ is Lipschitz continuous.
Thus, w.p.a.1, we can bound
\begin{multline*}
\sup_{u \in [ \log ( \ushort{r} ) , \infty )} \left|  \check{G} ( u  , \hat\beta )  - \check{G} ( u  , \beta )  - f_U ( u) \mu_X^\tau \left(    \hat\beta - \beta \right)  \right| 
\  \leq \   \int \left( c_U \int_0^{| x^\tau ( \hat\beta - \beta) |}  t  \, dt  \right) d F_X (x)  \\
 =  \ O_p \left(  L^{-1} \right)  ,
\end{multline*}
where $ c_U$ denotes the Lipschitz constant of $f_U$ from part 1. As a result, Eq.\ (\ref{eqa:olsap}) follows by standard arguments:
\begin{eqnarray*}
\hat\beta   & = &  \left[\sum_{l=1}^L ( X_l - \bar{X} ) (  X_l - \bar{X} )^\tau \right]^{-1} \sum_{l=1}^L (  X_l - \bar{X} ) \log(\tilde{W}_l)  \\
 & =  &  \beta \, + \,  \left[ \frac{1}{L} \sum_{l=1}^L ( X_l - \bar{X} ) (  X_l - \bar{X} )^\tau \right]^{-1} \frac{1}{L} \sum_{l=1}^L   (  X_l - \bar{X} )  (  U_l  - \mu_U) \\
& = &   \beta   \, + \,     \Sigma_X^{-1}  \frac{1}{L} \sum_{l=1}^L (  X_l -   \mu_X )  (  U_l  - \mu_U)  \, + \,    o_p \left( L^{-1/2} \right) .
\end{eqnarray*}

To prove the result in Eq.\ (\ref{eqa:stoap}), let $\mathscr{F}_R$ be the collection of functions $\lowmathcal{r}_{ (u, \delta ) } \in \ell^\infty (  \mathbb{R} \times \mathscr{S}_X )$ defined by
$\lowmathcal{r}_{ (u,\delta) } ( \tilde{u} , \tilde{x} )  : =    \mathds{1} (  \tilde{u}    \leq u  +  \tilde{x}^\tau\delta )$ and indexed by $(u , \delta )\in [ \log( \ushort{r} ) ,  \bar{u} + 2 c_X  ] \times \mathcal{B}(0, \tilde{\varepsilon} )$,  where $c_X : = 1 \vee \sup_{x \in \mathscr{S}_X } \| x \|_2$, $\mathcal{B}(0,   \tilde{\varepsilon}  )\subset \mathbb{R}^d$ denotes the open ball of radius $\tilde{\varepsilon}$ centered at the origin, and $0< \tilde{\varepsilon} <1$ can be taken to be any fixed constant such that $ \log ( \ushort{r} ) - c_X  \tilde{\varepsilon} > \ushort{u}$.
For $\delta >0$,  denote also
\begin{equation*}
\mathscr{F}_{R, \delta} \ : = \  \left\{  (\phi_1 , \phi_2 ) \in \mathscr{F}_R^2 :  \,  \rho (\phi_2 -\phi_1  )  <  \delta \right\},
\end{equation*}
where $\rho (\phi) := \sqrt{ \E[  \phi (  {U} , {X} ) ^2 ]}$ denotes the $\mathcal{L}_2 ( P_{UX} )$-norm.
Consider the  $\mathscr{F}_R$-indexed empirical process 
\begin{equation*}
\mathbb{G}_L ( \phi ) \  : = \ \frac{1}{\sqrt{L}} \sum_{l=1}^L  \left\{ \phi (  U_l , X_l )   -  \E [ \phi (  U , X )   ]  \right\} \, .
\end{equation*}
Observe that $\lowmathcal{r}_{(u,\hat\beta -\beta )} (U_l ,X_l ) =  \mathds{1} ( \hat{U}_l  \leq u      ) $ and $ \lowmathcal{r}_{(u,0)} (U_l,X_l ) = \mathds{1} ( {U}_l  \leq u      ) $, so whenever $\| \hat\beta - \beta \|_2 < \tilde{\varepsilon} $ we can write
\begin{equation*}
L^{-1/2} \sum_{l=1}^L  \mathcal{R}_l (u) \ = \ \mathbb{G}_L \left( \lowmathcal{r}_{(u,\hat\beta - \beta)}   - \lowmathcal{r}_{(u,0)}  \right)  \ \ \text{for}  \  u \in[  \log( \ushort{r} ) ,  \bar{u} + 2 c_X  ] ,
\end{equation*}
whereas $ \mathcal{R}_l (u)  = 0$ for $u >\bar{u} + 2 c_X $.
Now choose any $\epsilon , \delta >0$ and, since $F_U$ is Lipschitz continuous on $[  \log ( \ushort{r} ) - c_X  \tilde{\varepsilon} , \infty)$, pick $0 < \delta^\prime <  \tilde{\varepsilon} $ sufficiently small such that
\begin{equation*}
\left\{ (  \lowmathcal{r}_{ (u, t ) }  ,  \lowmathcal{r}_{ (u, 0 ) }  )  \in  \mathscr{F}_{ R }^2  : \,  u \in  [ \log( \ushort{r} ) ,  \bar{u} + 2 c_X  ]  ,  \,  \|   t \|_2 < \delta^\prime \right\} \  \subset \  \mathscr{F}_{ R , \delta }\,  ,
\end{equation*}
 so we can bound
\begin{eqnarray*}
& &   \Prob
\left( \sup_{u \in [ \log ( \ushort{r} ) , \infty ) } \left| \frac{1}{\sqrt{L}} \sum_{l=1}^L  \mathcal{R}_l (u) \right|    >   \epsilon  \right) 
  \\
   &    &  \quad \leq \ \ 
   \Prob  \left(   \left\| \mathbb{G}_L ( \lowmathcal{r}_{( \cdot ,\hat\beta - \beta )}   - \lowmathcal{r}_{( \cdot , 0 )}  ) \right\|_{  [ \log ( \ushort{r} ) , \infty ) , \infty  }     >   \epsilon , \,   \|  \hat\beta  -  \beta \|_2   < \delta^\prime \right)   \ +  \    \Prob \left( \|  \hat\beta  -  \beta \|_2   \geq \delta^\prime \right)  \\
   &    &  \quad \leq \ \   \Prob \left(    \sup_{ (\phi_1 , \phi_2) \in \mathscr{F}_{ R , \delta}  } \, \left| \mathbb{G}_L \left( \phi_2  -  \phi_1 \right)    \right|   \, > \,   \epsilon \right)  \   +  \   \Prob \left( \|  \hat\beta  -  \beta \|_2   \geq \delta^\prime \right)  .
\end{eqnarray*}
To establish the result in Eq.\ (\ref{eqa:stoap}), just note that the equality
\begin{equation*}
\lim_{\delta \downarrow 0} \,  \underset{L \rightarrow \infty} {\lim \! \sup} \  \Prob   \left( \sup_{ (\phi_1 , \phi_2) \in \mathscr{F}_{R , \delta}    } \left| \mathbb{G}_L ( \phi_2  -  \phi_1 )    \right| \geq \epsilon\right) = 0 
\end{equation*}
holds because $\mathscr{F}_R $ constitutes a $P_{UX}$-Donsker class: see \citet[Sections 2.1.2 and 3.13.4.1, Problem 2.1.2, as well as Theorem 2.5.6]{vw23} and the arguments in the proof of Lemma 1 in \cite{akr01}.

\paragraph{(b)} 

Let $\mathscr{F}_S$ be the collection of functions $\phi_u $ on $ \mathbb{R} \times \mathbb{R}^d $ of the form
\begin{equation*}
\phi_u \left( \tilde{u} , \tilde{x} \right) =      \mathds{1} \left(\tilde{u} \leq u      \right)  -  F_U (u)  +    f_U (u) \mu_X^\tau \Sigma_X^{-1} ( \tilde{x} - \mu_X )( \tilde{u}  - \mu_U ) 
\end{equation*}
with $( \tilde{u} , \tilde{x} ) \in  \mathbb{R} \times \mathbb{R}^d $ and indexed by $u \in   [ \log(\ushort{r} )  , \infty )$.
Note that there exists a constant $\bar{c}_S > 0$ such that the function $\bar{\phi} \left( \tilde{u} , \tilde{x} \right)  : =     1  + \bar{c}_S  \,  \| \tilde{x} - \mu_X \|_{1} \,  | \tilde{u}  - \mu_U | $ can be taken as the envelope of $\mathscr{F}_S$.
Then, the proof can be completed in two steps.

First, for any finite dimensional vector $(u_1, \dots , u_m)$, observe that
$ L^{-1/2} \sum_{l} ( \psi_l (u_1) , \dots ,  \psi_l (u_m) )$ converges in distribution to a multivariate normal with variance-covariance matrix $(\Psi_\nu  (  u_j  , u_k) )_{j,k}$.
This result follows by the multivariate Central Limit Theorem \cite[see, e.g.,][Theorem 29.5]{bil95}, for which I remark that $\psi_l (u) = \phi_u ( U_l , X_l) $ and $\E [ \bar{\phi} ( U , X )^2  ] <\infty$ by Assumption \ref{assump:dgp}.

Second, note that $ L^{-1/2} \sum_{l} \psi_l  $ converges weakly to a tight random element in $\ell^{\infty} ( [ \log(\ushort{r} )  , \infty ) )$.
To see this, by Theorem 2.5.2 in \cite{vw23}, it suffices to verify that $\mathscr{F}_S$ satisfies the uniform entropy condition in Eq.\ (2.5.1) of that reference. This can be done by inspecting Example 2.5.4, as well as Lemmas 2.6.16 and 2.6.20.(v)-(vi), in that reference, for which I remark that $\phi_u ( \cdot_1 , \cdot_2  )  \in \mathscr{F}_S$ can be expressed as the sum of $\mathds{1} \left(\cdot_1 \leq u      \right) -  F_U (u)  $ and the product of two linear transformations of $( \cdot_2 - \mu_X )$ and $( \cdot_1  - \mu_U ) $; see also Section 3.13.4.1 in \cite{vw23}.

\paragraph{(c)}  This part follows directly by combining together parts (a) and (b) with Lemma 1.10.2.(i) in \cite{vw23}.

\paragraph{(d)} This part can be deduced from the following two observations. 
First, since $\V (  \mathbb{G}_{\nu}(u)  ) = 0$ for all $u > \bar{u} $ and $\mathbb{G}_{\nu}$ is tight, observe that
\begin{equation*}
\Prob \left( \| \mathbb{G}_{\nu} \|_{ ( \bar{u}   , \infty) }  = 0  \right) = 1
\end{equation*}
because the covariance function of $ \mathbb{G}_{\nu}$ coincides with that of a degenerate Gaussian process that equals zero on $ ( \bar{u}  , \infty)$ w.p.1; see Lemma 1.5.3 in \cite{vw23}.
Second, note that $\mathbb{G}_{\nu} $ admits a continuous version on $[\log (\ushort{r}),  \bar{u}  + 1]$ due to part 1 of this lemma and  \citet[Theorem 2.3.7]{gn16}.

\paragraph{3.} 
Hereafter consider $L$ sufficiently large so that $\lowmathcal{h} < 1/2$, and set $\bar{u}^\prime : = \bar{u} + 1$.

\paragraph{(a)} Note that we can bound
\begin{equation*}
\left| \hat{f}_{U , \lowmathcal{h} }  (  \tilde{u} )   -   \hat{f}_{U , \lowmathcal{h} }  (  u  ) \right| 
\  \leq  \   \left\| \hat{f}_{U , \lowmathcal{h} }^\prime   \right\|_{ [  \ushort{u}^\prime ,\bar{u}^\prime ]  , \infty}
\,  \left| \tilde{u}    -  u \right|  \quad   \quad \forall \   \tilde{u}  , u  \geq \ushort{u}^\prime .
\end{equation*}
To obtain the desired result, then it suffices to prove that there exists a constant $\bar{c}_U >0$ such that $\Prob (  \| \hat{f}_{U , \lowmathcal{h} }^\prime   \|_{ [  \ushort{u}^\prime ,\bar{u}^\prime ]  , \infty} \leq \bar{c}_U ) \rightarrow 1$. For this purpose, first, let  $ \mathscr{K}_L^\prime$ be the class of functions $\phi_{( u , \delta )} : \mathbb{R} \times \mathscr{S}_X \rightarrow  \mathbb{R}$
of the form
\begin{equation*}
\phi_{( u , \delta )}  ( \tilde{u} , \tilde{x}  ) \  = \
\frac{1}{   \sqrt{ \lowmathcal{h} } } \, \kappa^\prime \left( \frac{u - \tilde{u} -  \tilde{x}^\tau \, \delta  }{  \lowmathcal{h} } \right)   
\end{equation*}
and indexed by $( u ,  \delta )  \in [  \ushort{u}^\prime ,  \bar{u}^\prime ] \times \mathcal{B}  (0 ,  L^{-1/3} )$. Second, since  $ \hat{f}_{U , \lowmathcal{h} }^\prime ( u ) =  ( \lowmathcal{h}^{3/2} L )^{-1} \sum_l  \phi_{( u , \hat\beta - \beta )}  ( U_l , X_l  )  $, observe that we can bound
\begin{eqnarray}
\left\| \hat{f}_{U , \lowmathcal{h} }^\prime   \right\|_{  [  \ushort{u}^\prime ,\bar{u}^\prime ]  , \infty }
& \underset{\text{w.p.a.1}}{\leq} &
\sup_{ \phi \in   \mathscr{K}_{L}^\prime }   \, 
\frac{1}{ \lowmathcal{h}^{3/2} L }\left|  \sum_{l=1}^L \phi  ( U_l , X_l  )   \right|  \nonumber \\
& \leq  &
\frac{1}{ \lowmathcal{h}^{3/2} L }  \,  \sup_{ \phi \in   \mathscr{K}_{L}^\prime }   \, 
\left|  \sum_{l=1}^L \phi  ( U_l , X_l  )  - \E [ \phi  ( U_l , X_l  ) ] \right|  \label{eqa:boundfhatp1} \\
& & \quad   +  \
\sup_{ \phi \in   \mathscr{K}_{L }^\prime }   \, 
\left| \frac{1}{ \lowmathcal{h}^{3/2} L } \,  \sum_{l=1}^L \E [ \phi  ( U_l , X_l  ) ] \  \right|.  \label{eqa:boundfhatp2}
\end{eqnarray}
So, the desired result follows by showing that the terms in Eqs.\ (\ref{eqa:boundfhatp1}) and (\ref{eqa:boundfhatp2}) are $o_p (1)$ and $O(1)$, respectively.

Starting with the term in Eq.\ (\ref{eqa:boundfhatp1}), by Proposition 3.5.15 in \cite{gn16} and since $\sup_L \sup_{\phi \in  \mathscr{K}^\prime_{L }} \E [ \phi ( U, X )^2 ] < \infty$, the desired result follows if we prove that there is a constant $C_{B} >0$ such that the inequality
\begin{equation}
N_{[]}  \left(    \mathscr{K}_{L }^\prime ,  \mathcal{L}^2(  P_{UX})  , \epsilon \right) \ \leq \ 
\left( \, 1 \, + \, \left\lceil   \frac{ C_{B}}{\epsilon \, \lowmathcal{h} }  \right\rceil \,   \right)^{d+1}   
\label{eqa:bracket}
\end{equation}
holds when $L$ is sufficiently large and $\epsilon >0$ sufficiently small, where $N_{[]} (    \mathscr{K}_{L }^\prime ,  \mathcal{L}^2(  P_{UX })  , \epsilon )$ denotes the $\mathcal{L}^2(  P_{UX} )$-bracketing number of $ \mathscr{K}_{L }^\prime$; see \citet[p.\ 195]{gn16} for a detailed definition.
To prove the inequality in Eq.\ (\ref{eqa:bracket}), for a given $0 < \epsilon < c_\kappa := 4 \| \kappa^{\prime\prime} \|_{\mathbb{R} , \infty }  \| f_U \|_{\mathbb{R} , \infty }^{1/2} < \infty$ and $ \lowmathcal{h}  < 1/2 $, consider the following bracketing functions:
\begin{eqnarray*}
\ushort{\kappa}_{jk} (  \tilde{u} ,  \tilde{x} ) & = &    
\frac{1}{ \sqrt{ \lowmathcal{h} } }   \kappa^\prime \left( \frac{u_j - \tilde{u} -  \tilde{x}^\tau \, \delta_k  }{  \lowmathcal{h} } \right)     
\ - \    \frac{ \epsilon \| \kappa^{\prime\prime} \|_{\mathbb{R} , \infty }   }{ c_\kappa \sqrt{ \lowmathcal{h} }  }  
\, \mathds{1} \left\{  \left|    u_j - \tilde{u}  - \tilde{x}^\tau \delta_k  \right|  \leq  \lowmathcal{h} \left(    1   +  \frac{\epsilon}{c_\kappa}  \right)  \right\}   \ \text{and}   \\
\bar{\kappa}_{jk} (  \tilde{u} ,  \tilde{x} )  & = & \frac{1}{ \sqrt{ \lowmathcal{h} } }   \kappa^\prime \left( \frac{u_j - \tilde{u} -  \tilde{x}^\tau \, \delta_k  }{  \lowmathcal{h} } \right)     
\ + \    \frac{ \epsilon \| \kappa^{\prime\prime} \|_{\mathbb{R} , \infty }   }{ c_\kappa \sqrt{ \lowmathcal{h} }  }  
\, \mathds{1} \left\{  \left|    u_j - \tilde{u}  - \tilde{x}^\tau \delta_k  \right|  \leq  \lowmathcal{h} \left(    1   +  \frac{\epsilon}{c_\kappa}  \right)  \right\} ,
\end{eqnarray*}
where $\{ ( u_j , \delta_k ) \}$ are equally spaced grid points on $[  \ushort{u}^\prime  , \bar{u}^\prime ] \times   \mathcal{B} ( 0 ,1 )$ such that, for any $( u , \delta ) \in [  \ushort{u}^\prime  , \bar{u}^\prime ] \times \mathcal{B} ( 0 , 1 )$, there is $( u_j , \delta_k )$ so that $| u - u_j | <  \lowmathcal{h} \epsilon /(2 c_\kappa ) $ and $\| \delta  - \delta_k  \|_2 <  \lowmathcal{h} \epsilon /(2 c_X c_\kappa ) $. Then, observe that the following conditions are satisfied, from which the inequality in Eq.\ (\ref{eqa:bracket}) follows with $C_B = [ 2 c_\kappa (\bar{u}^\prime - \ushort{u}^\prime) ] \vee (  4 c_\kappa c_X ) $:
\begin{itemize}

\item[C1)] By construction of the bracketing functions and the mean value theorem, for any $\phi \in  \mathscr{K}_{L}^\prime$, there exists a pair $(j,k)$ such that  $\ushort{\kappa}_{jk} ( \tilde{u} , \tilde{x}  )  \leq  \phi   ( \tilde{u} , \tilde{x}  ) \leq  \bar{\kappa}_{jk}  ( \tilde{u} , \tilde{x}  )$ for all $ ( \tilde{u} , \tilde{x}  ) \in \mathbb{R} \times \mathscr{S}_X $.

\item[C2)]  The  $\mathcal{L}^2(  P_{UX} )$-distance between $\bar{\kappa}_{jk}$ and $\ushort{\kappa}_{jk}$ can be bounded from above by $\epsilon$ because
\begin{multline*}
\E \left[ \left( \bar{\kappa}_{jk} (  U , X)  - \ushort{\kappa}_{jk} ( U  , X) \right)^2  \right] \ \leq \
4   \frac{ \epsilon^2 \| \kappa^{\prime\prime} \|_{\mathbb{R} , \infty }^2   }{ c_\kappa^2  \lowmathcal{h}   }  \,
\Prob \left\{  \left|    u_j - U  - X^\tau \delta_k  \right|  \leq  \lowmathcal{h} \left(    1   +  \frac{\epsilon}{c_\kappa}  \right)  \right\} \\
\leq \   4   \frac{ \epsilon^2 \| \kappa^{\prime\prime} \|_{\mathbb{R} , \infty }^2   }{ c_\kappa^2  \lowmathcal{h}   }  \, 2  \lowmathcal{h} \left(    1   +  \frac{\epsilon}{c_\kappa}  \right) \left\| f_U \right\|_{\mathbb{R} , \infty } \ \leq \ \epsilon^2 .
\end{multline*}

\item[C3)] The set $[  \ushort{u}^\prime  , \bar{u}^\prime ] \times   \mathcal{B} ( 0 ,1 )$ can be covered by 
\begin{equation*}
\left( 1 + \left\lceil \frac{2 c_\kappa ( \bar{u}^\prime - \ushort{u}^\prime )}{\varepsilon \, \lowmathcal{h}  } \right\rceil \right) \,
\left( 1 + \left\lceil  \frac{4 c_\kappa c_X}{\varepsilon \,  \lowmathcal{h} }  \right\rceil    \right)^d 
\end{equation*}
balls of radius $\epsilon$ centered at the grid points $\{ ( u_j , \delta_k ) \}$.

\end{itemize}

In view of the term in Eq.\ (\ref{eqa:boundfhatp2}), consider $L$ sufficiently large so that $ (\ushort{u} + c_X L^{-1/3} -  \ushort{u}^\prime ) /\lowmathcal{h} < - 1$.
First, for a given $\phi  \in \mathscr{K}_L^\prime$ and $x \in \mathscr{S}_X$, note that we can bound
\begin{eqnarray*}
\left| \E   \left[ \frac{  \phi  ( U , X  )   }{ \lowmathcal{h}^{3/2} } \mid X = x \right]    \right|
&   = & 
 \left| \E \left[   \frac{1}{ \lowmathcal{h}^2  }  \kappa ^\prime \left(  \frac{u -  {U} - x^\tau \delta  }{\lowmathcal{h}} \right) \right]   \right|  \\
& \leq &
\frac{1}{ \lowmathcal{h}  } \int_{-1}^{(\bar{u} + x^\tau \delta -u)/\lowmathcal{h}} 
   \left|  \kappa ^\prime ( t)  \right|    \left|  f_U ( u +  \lowmathcal{h} t  - x^\tau \delta  )  -    f_U  ( u) \right|  \,  dt   \\
&  &    \quad + \ \ 
   \frac{ f_U  ( u) }{ \lowmathcal{h} }
   \left|   \int_{-1}^{(\bar{u} + x^\tau \delta -u)/\lowmathcal{h}} 
   \kappa ^\prime ( t)      \,  dt  \right| .
\end{eqnarray*}
Second, letting $c_U >0$ denote the Lipschitz constant of $f_U$ associated with part 1, observe that the first term on the right-hand side of the inequality can be bounded from above by 
\begin{equation*}
\frac{ 2  c_U   }{ \lowmathcal{h}  } 
   \left\| \kappa ^\prime \right\|_{ \mathbb{R} , \infty }   \left(  \lowmathcal{h}  +   c_X\, L^{-1/3} \right)  ,
\end{equation*}
while the second is $O(1)$ uniformly in $x \in \mathscr{S}_X$. To see this, note that $f_U  (u) \leq c_U \, |\bar{u}  - u| \leq c_U  ( \lowmathcal{h} + c_X \, L^{-1/3} )$ when $u > \bar{u} + x^\tau \delta  -  \lowmathcal{h}$ by Lipschitz continuity of $f_U$, whereas
$\int_{ -1 }^{(\bar{u} + x^\tau \delta -u)/\lowmathcal{h}}   \kappa ^\prime ( t) \, dt    = 0 $ when $u \leq \bar{u} + x^\tau \delta  -  \lowmathcal{h}$ by symmetry of the kernel. As a result, the term in Eq.\ (\ref{eqa:boundfhatp2}) is $O (1)$.

\paragraph{(b)}
Let  $ \mathscr{K}_L$ be the class of functions $\phi_{( u , \delta )} : \mathbb{R} \times \mathscr{S}_X \rightarrow  \mathbb{R}$
of the form
\begin{equation*}
\phi_{( u , \delta )}  ( \tilde{u} , \tilde{x}  ) \  = \ \kappa \left( \frac{u - \tilde{u} -  \tilde{x}^\tau \, \delta  }{  \lowmathcal{h} } \right)   
\end{equation*}
and indexed by $( u ,  \delta )  \in (   \ushort{u} + 2 \lowmathcal{h}   , \infty ) \times \mathcal{B}  (0 ,  L^{-1/3} )$. 
Before proceeding, note that $ \mathscr{K}_L$ is of VC type with respect to some constant envelope function; see Definition 3.6.10 in \cite{gn16}.
To see this, recall that $\kappa$ is continuously differentiable on $\mathbb R$ and supported on $[-1,1]$, so it admits a Jordan decomposition into two bounded nondecreasing functions: $\kappa = :  \kappa_{(1)} - \kappa_{(2)}$. Then, write
$\mathscr{F}_{\text{aff}}  : = \{  ( \tilde{u}  ,  \tilde{x} ) \mapsto b_1 - b_2\tilde{u} + \tilde{x}^\tau b_3   : \, 
( b_1 , b_2 , b_3) \in \mathbb{R}_{>0}  \times \mathbb{R}_{>0} \times  \mathcal{B}  (0 , 1 )\}$ and note that
\begin{equation*}
\mathscr{K}_L  \ \subset  \mathscr{K}_{ (1) }  -\mathscr{K}_{ (2) } \, , 
\end{equation*}
where both $\mathscr{K}_{ (m) } : = \{  \kappa_{(m)}\circ \varphi  : \, \varphi \in \mathscr{F}_{\text{aff}} \}  $, $m=1,2$,
are VC-subgraph by Lemmas 2.6.16 and 2.6.20.(viii) in \citet{vw23}.

Now write 
\begin{eqnarray*}
\left\| \hat{f}_{U , \lowmathcal{h} } - f_U   \right\|_{   [ \ushort{u} + 2 \lowmathcal{h}   ,  \infty)   , \infty }
& \underset{\text{w.p.a.1}}{\leq} &
\frac{1}{ \lowmathcal{h}  L}  \sup_{ \phi \in   \mathscr{K}_{L} }   \, 
\left|  \,  \sum_{l=1}^L \phi  ( U_l , X_l  )  \, - \, \E [ \phi  ( U_l , X_l  ) ] \, \right|   \\
& & \quad   +  \
\sup_{ ( u ,  \delta )   }   \,  \left| \, \E \left[ \frac{1}{    \lowmathcal{h}  } \, \kappa \left( \frac{u - U- X^\tau \, \delta  }{  \lowmathcal{h} } \right) \, - \, f_U (u)   \right] \, \right|  ,
\end{eqnarray*}
with the second supremum taken over  $(   \ushort{u} + 2 \lowmathcal{h}   , \infty ) \times \mathcal{B}  (0 ,  L^{-1/3} )$, and note that the first term on the right-hand side is $O_p ( \sqrt{ \log(  \lowmathcal{h}^{-1}  )/ ( \lowmathcal{h} L) } )$ by Corollary 3.5.8 in \cite{gn16}. Considering now the second term, and taking $L$ sufficiently large so that $\lowmathcal{h}  > c_X \, L^{-1/3}$,  we can bound
\begin{eqnarray*}
\left| 
 \E \left[ \frac{1}{    \lowmathcal{h}  }  \kappa \left( \frac{u - U- X^\tau \delta  }{  \lowmathcal{h} } \right)   -  f_U (u) \mid X= x \right]
 \right|
& \leq &  
 \int_{-1}^{(\bar{u} + x^\tau \delta -u)/\lowmathcal{h}} 
    \kappa  ( t)    \left|  f_U ( u +  \lowmathcal{h} t  - x^\tau \delta  )  -    f_U  ( u) \right|  dt   \\
&  &    \ \ \ + \ \
f_U (u) \left|  \, 
 \int_{-1}^{(\bar{u} + x^\tau \delta -u)/\lowmathcal{h}}     \kappa  ( t)   \,  dt \,  -  \, 1 
  \,  \right| .
\end{eqnarray*}
Then, observe that both terms on the right-hand side are bounded from above by $c_U (  \lowmathcal{h}  +   c_X\, L^{-1/3})  $. In regard to the second term, note that  $f_U  (u) \leq c_U  ( \lowmathcal{h} + c_X \, L^{-1/3})$ when $u > \bar{u} + x^\tau \delta  -  \lowmathcal{h}$ by Lipschitz continuity of $f_U$, whereas
$\int_{-1}^{(\bar{u} + x^\tau \delta -u)/\lowmathcal{h}}     \kappa  ( t)    dt  = 1 $ when $u \leq \bar{u} + x^\tau \delta  -  \lowmathcal{h}$.
As a result, we have that $\| \hat{f}_{U , \lowmathcal{h} }  - f_U  \|_{ [ \ushort{u} + 2 \lowmathcal{h}   ,  \bar{u} + 2 \lowmathcal{h}   ] , \infty}  =  O_p ( \lowmathcal{h} )$ and therefore $\| \hat{f}_{U , \lowmathcal{h} }  - f_U  \|_{ [ \ushort{u} + 2 \lowmathcal{h}   ,  \infty  ) , \infty}  =  O_p ( \lowmathcal{h} )$.

\paragraph{(c)} Observe that we can bound
\begin{eqnarray*}
\left\| \hat{F}_{U , \lowmathcal{h} } - F_U   \right\|_{  \mathbb{R}  , \infty }
& \underset{\text{w.p.a.1}}{\leq} &
 \frac{1}{  L}   \sup_{  ( u ,  \delta )   }   \, 
\left|  \,  \sum_{l=1}^L \mathcal{K}   \left( \frac{u - U_l- X_l^\tau \, \delta  }{  \lowmathcal{h} } \right) \, - \, \E \left[   \mathcal{K}  \left( \frac{u - U_l- X_l^\tau \, \delta  }{  \lowmathcal{h} } \right) \right]  \, \right|   \\
& & \quad   +  \
\sup_{ ( u ,  \delta )   }   \,  \left| \, \E \left[  \mathcal{K}  \left( \frac{u - U- X^\tau \, \delta  }{  \lowmathcal{h} } \right) \right] \, - \, F_U (u)    \, \right|  ,
\end{eqnarray*}
where the suprema are taken over $\mathbb{R} \times \mathcal{B}  (0 ,  L^{-1/3} )$ and $\mathcal{K} (y ) : = \int_{-\infty}^y \kappa (t) dt$.  Then, by arguments similar to those used in the proof of part (b), it can be shown that the first term on the right-hand side is $O_p  ( \sqrt{\log(  \lowmathcal{h}^{-1}   ) / (  \lowmathcal{h}  L  ) }  )$. For the second, by integration by parts, write
\begin{equation*}
\E \left[  \mathcal{K}  \left( \frac{u - U- X^\tau  \delta  }{  \lowmathcal{h} } \right) \mid X = x \right]   = \int_{-1}^1 F_U ( u  - x^\tau  \delta -   \lowmathcal{h} t  )  \kappa (t) \, dt 
\end{equation*}
and note that the desired result follows by Lipschitz continuity of $F_U$ on $\mathbb{R}$.

\paragraph{(d)}
To prove the first result, just note that 
\begin{equation*}
\int_{-\infty}^\infty \, u \hat{f}_{U,\lowmathcal{h}} (u) \, du 
 =  \frac{1}{L} \sum_{l=1}^L \hat{U}_l 
\, = \,  \frac{1}{L} \sum_{l=1}^L {U}_l  - \bar{X}^\tau \left( \hat\beta - \beta  \right) 
\, = \,  \mu_U + O_p \left( L^{-1/2} \right)  \, .
\end{equation*}
For the second, by change of variables, observe that we can write 
\begin{equation*}
\int_{-\infty}^c  u^2 \hat{f}_{U, \lowmathcal{h} } (u) \, du
\, = \,
\frac{1}{L}\sum_{l=1}^L \hat{U}_l^2 \mathcal{K} \left(\frac{c-\hat{U}_l }{\lowmathcal{h}  }\right) 
+  O_p \left(  \lowmathcal{h}  \right) 
\, = \,
\frac{1}{L}\sum_{l=1}^L {U}_l^2 \mathcal{K} \left(\frac{c-\hat{U}_l }{\lowmathcal{h}  }\right) 
+ O_p \left( L^{-1/2} +  \lowmathcal{h}  \right) ,
\end{equation*}
with the $O_p$-terms being uniform in $ c  \in \mathbb{R}  $, and we can also bound 
\begin{equation*}
\frac{1}{L}\sum_{l=1}^L 
{U}_l^2 \left| \mathcal{K}  \left(\frac{c-\hat{U}_l }{\lowmathcal{h}  }\right)  
-  \mathcal{K}  \left(\frac{c-{U}_l }{\lowmathcal{h}  } \right)  
\right|
\, \leq \, 
 \frac{1}{L}\sum_{l=1}^L 
{U}_l^2  \|  \kappa \|_{ \mathbb{R} , \infty } \frac{ \| X_l \|_2 \| \hat\beta - \beta \|_2 }{ \lowmathcal{h}  } 
\, =  \, o_p ( 1 ). 
\end{equation*}
To prove the desired result, then it suffices to show that 
\begin{equation}
\label{eqa:kkk}
\sup_{c \in \mathbb{R}} \,  \left| \frac{1}{L}\sum_{l=1}^L   {U}_l^2 \mathcal{K}  \left(\frac{c-{U}_l }{\lowmathcal{h}  } \right)  
\, - \,  \E \left[ U^2\mathds{1} \left(   U\leq c \right) \right] \right| 
 \ = \ o_p ( 1) \, .
\end{equation}
So, for any $c\in \mathbb{R}$, note that we can bound
\begin{equation*}
\left|
\mathcal{K}\left(\frac{c-U_l}{\lowmathcal{h}}\right)
-\mathds{1} ( U_l \leq c )
\right|
\, \leq \,
\mathds{1} \left( \left| U_l-c \right| \leq\lowmathcal{h} \right) 
\end{equation*}
and therefore
\begin{multline*}
\left|
\frac{1}{L}\sum_{l=1}^L U_l^2
\mathcal{K}\left(\frac{c-U_l}{\lowmathcal{h}}\right)
-
\frac{1}{L}\sum_{l=1}^L
U_l^2 \mathds{1} \left( U_l \leq c \right)
\right|
\, \leq \,
\frac{1}{L}\sum_{l=1}^L
U_l^2 \mathds{1} \left( |U_l-c|\leq\lowmathcal{h} \right)
\\
\leq \  
\underbrace{ \frac{1}{L}\sum_{l=1}^L
U_l^2 \mathds{1} \left( |U_l-c|\leq\lowmathcal{h} \right)
- \E \left[   U^2 \mathds{1} \left(|U-c|\leq\lowmathcal{h} \right) \right] }_{= o_p(1) \ \text{unif.\ in} \ c\in \mathbb{R} }
\ +  \  \underbrace{\E \left[   U^2 \mathds{1} \left( |U-c|\leq\lowmathcal{h} \right) \right]}_{
= o(1) \ \text{unif.\ in} \ c\in \mathbb{R} } \, ,
\end{multline*}
where the result under the first term on the second line follows by Theorem 2.4.1 in \cite{vw23}, while the result under the second is a consequence of the fact that
\begin{equation*}
\ \sup_{c \in  \mathbb{R}} \, \E \left[   U^2 \mathds{1} \left( |U-c|\leq\lowmathcal{h} \right) \right] 
\ \leq \ C^2 \, \sup_{ c \in  \mathbb{R} }  \, \Prob \left( |U-c|\leq\lowmathcal{h} \right)  
\ + \ 
\E \left[ U^2\mathds{1} \left(   |U| > C \right) \right] 	\quad  \forall \, C > 0 \,.
\end{equation*}
Finally, the result in Eq.\ (\ref{eqa:kkk}) arises from
\begin{equation*}
\sup_{c \in  \mathbb{R}} \,
  \left| \frac{1}{L}\sum_{l=1}^L  U_l^2\mathds{1} \left(  U_l\leq c \right)
- 
\E \left[ U^2\mathds{1} \left(   U\leq c \right) \right] \right|
\ = \ o_p(1)  \, ,
\end{equation*}
which also follows by Theorem 2.4.1 in \cite{vw23}.

\paragraph{4.}
Write $\tilde{\beta} := ( \,  \E (U) , \, \beta  \,)$ and let $\check{\beta} := ( \, \hat{\mu}_{U, \lowmathcal{h} } , \, \hat\beta  \,) \in \mathbb{R}^{d+1}$ and $\check{\beta}^\ast \in \mathbb{R}^{d+1}$ be the OLS estimators from the regressions in Sections \ref{sec:estier} and \ref{ss:bootap}, respectively, both including the intercept as first element. To obtain the desired result, since  part 3.(d) of this lemma implies $\mathrm{var}(U_l^\ast  | \mathcal{S}_L ) =  \int u^2 d \hat{F}_{ U , \lowmathcal{h} } (u)  - \hat{\mu}_{  U , \lowmathcal{h}  }^2 = O_p (1)$, just note that 
\begin{equation*}
 \E \left(  \left\|  \check\beta^\ast - \check\beta  \right\|_2^2  \mid  \mathcal{S}_L  \right)   
\,  = \,  \mathrm{trace} \left\{  \E \left[ (\check\beta^\ast - \check\beta ) (\check\beta^\ast - \check\beta )^\tau  \mid   \mathcal{S}_L \right]  \right\}
\, \leq  \,  O_p (L^{-1})   \, \mathrm{trace} \left\{ L ( \mathbf{X}^\tau \mathbf{X}  )^{-1}  \right\}  ,
\end{equation*}
where $\mathbf{X}$ is a $L \times (d+1)$ matrix that has $(  1  ,  \, X_l )^\tau$ in its $l$th row.

\paragraph{5.}

\paragraph{(a)}
This part follows by arguments similar to those used in the proof of 2. Letting $\hat{F}_X$ denote the empirical c.d.f.\ of $X$, define the functions
\begin{equation*}
\check{G}_{ \lowmathcal{h} } ( u  , b  ) 
\, : = \,  \int  \hat{F}_{ U ,   \lowmathcal{h}  } \left( u +x^\tau \left( b - \hat\beta \right) \right)  \,   d \hat{F}_X (x) 
\, = \, 
\frac{1}{L} \sum_{l=1}^L  \hat{F}_{ U ,   \lowmathcal{h}  } \left( u + X_l^\tau \left( b - \hat\beta \right) \right) 
\end{equation*}
and $ \mathcal{R}_{ l  }^\ast (  u )  : =    \mathds{1} \left( \hat{U}_l^\ast  \leq u      \right)   - \check{G}_{ \lowmathcal{h} } ( u  , \hat\beta^\ast  )
 -\left[ \mathds{1} \left( U_l^\ast  \leq u      \right)   -  \check{G}_{ \lowmathcal{h} } ( u  , \hat\beta  ) \right]  $. To obtain the desired result, since
\begin{equation*}
 \hat{F}_U^\ast ( u  )  -  \hat{F}_{ U ,   \lowmathcal{h}  } ( u )    
  \, = \,    \frac{1}{L} \sum_{l=1}^L 
 \mathds{1} \left( U_l^\ast  \leq u      \right)  - \hat{F}_{ U ,   \lowmathcal{h}  } ( u )   
 +    \check{G}_{ \lowmathcal{h} } ( u  , \hat\beta^\ast  )   
 -   \check{G}_{ \lowmathcal{h} } ( u  , \hat\beta  )   
 +  \frac{1}{L} \sum_{l=1}^L  \mathcal{R}_l^\ast ( u )  \,  ,
\end{equation*}
it suffices to show that
\begin{eqnarray}
& & \sup_{u } \left|  \check{G}_{ \lowmathcal{h} } ( u  , \hat{\beta}^\ast  )   -   \check{G}_{ \lowmathcal{h} } ( u  , \hat{\beta}  ) -  {f}_{ U } (u)  \bar{X}^\tau \, \hat{\Sigma}_X^{-1} \, \frac{1}{L} \sum_{l=1}^L \left( X_l - \bar{X} \right) \left( U_l^\ast  - \hat{\mu}_{U ,  \lowmathcal{h} } \right) \right| 
\,   =  \,   o_{p^\ast} ( L^{-1/2} )  , 
     \label{eqa:olsapb}  \\
& & \sup_{u} \left|  L^{-1/2} \sum_{l=1}^L  \mathcal{R}_l^\ast (u )    \right| 
\,   =  \,   o_{p^\ast} \left( 1 \right)  
 \label{eqa:stoapb} 
\end{eqnarray} 
where the suprema are taken over $u \geq \log ( \ushort{r} ) $.

Starting with Eq.\ (\ref{eqa:olsapb}), note that we can bound
\begin{multline*}
\sup_{u \geq  \log ( \ushort{r} ) } \, \left|  \check{G}_{ \lowmathcal{h} } ( u  , \hat{\beta}^\ast )  
- \check{G}_{ \lowmathcal{h} } ( u  , \hat{\beta} ) 
- \hat{f}_{U , \lowmathcal{h} } ( u) \, \bar{X}^\tau \left(  \hat{\beta}^\ast - \hat{\beta}  \right)  \right| 
\  \leq \   \int \left\{\bar{c}_U \int_0^{| x^\tau ( \hat{\beta}^\ast - \hat{\beta} ) |}  t  \, dt  \right\} d \hat{F}_X (x)  \\
 \ \leq \   \bar{c}_U \,   c_X^2 \left\|  \hat{\beta}^\ast - \hat{\beta}   \right\|_2^2  \ =  \ o_{p^\ast} \left(  L^{-1/2} \right)  \, , 
\end{multline*}
where $\bar{c}_U$ is the Lipschitz constant from part 3.(a) of this auxiliary lemma, while the equality follows by part 4 of this lemma.
Since parts 3-4 of this auxiliary lemma and Jensen's inequality imply
\begin{multline*}
\E  \left\{ \left|  \hat{f}_{U , \lowmathcal{h} }(u) - {f}_{U } (u)  \right|  \,   \left|  \bar{X}^\tau \left(  \hat{\beta}^\ast - \hat{\beta}  \right)   \right| \mid \mathcal{S}_L  \right\}  \\
\leq \
 \left\| \hat{f}_{U , \lowmathcal{h} } - {f}_{U } \right\|_{ [ \log ( \ushort{r} ) , \infty )  ,  \infty }  \,
\left\|  \bar{X}   \right\|_2  \E  \left(  \left\|   \hat{\beta}^\ast - \hat{\beta}    \right\|_2   \mid \mathcal{S}_L  \right) \,
 =  \,  o_{p} \left(  L^{-1/2} \right)  \, ,
 \end{multline*}
Eq.\ (\ref{eqa:olsapb}) follows immediately because
\begin{equation*}
 \hat{\beta}^\ast - \hat{\beta}   \ = \   
   \hat{\Sigma}_X^{-1}  \frac{1}{L} \sum_{l=1}^L (  X_l -   \bar{X} )  (  U_l^\ast  - \hat{\mu}_{U ,  \lowmathcal{h} })  \, .
\end{equation*}
To obtain the result in Eq.\ (\ref{eqa:stoapb}), by Theorem 9.2.1 in \cite{du02real}, it suffices to show that for any $\epsilon  > 0$ and any increasing  sequence $(L_m)_{m \in \mathbb{N}}$ of positive integers there exists a subsequence $(L_{m_k} )_{k \in \mathbb{N}}$ such that 
\begin{equation}
\lim_{k \rightarrow \infty} \, 
\Prob \left(  \sup_{u \geq \log ( \ushort{r} ) } \, \left| \frac{1}{\sqrt{ L_{m_k} }} \sum_{l=1}^{L_{m_k}}  \mathcal{R}_l^\ast (u) \right|    
 > \epsilon \, \middle| \, \mathcal{S}_{L_{m_k}}  \right) = 0 \quad \mathrm{a.s.}
\end{equation}
This result is established in four steps as follows.

\emph{Step 1.} Denote 
\begin{equation*}
\mathcal{T}_{ L}^{  \mathrm{r}1 }  \  : =  \   \sup_{ (\phi_1 , \phi_2) \in {\mathscr{F}}_{ R }^2 , \,  x \in \mathscr{S}_X} \,
\left| \,
\int_{ \log( \ushort{r} ) - c_X }^{\bar{u} + 3 c_X }  \left[  \phi_2 ( \cdot , x )  -  \phi_1 (\cdot ,x )   \right]^2 \, 
d \left( \hat{F}_{U,\lowmathcal{h}}   - F_U \right) 
\, \right| 
\end{equation*}
and, since the total variation of $ \left[  \phi_2 ( \cdot , x )  -  \phi_1 (\cdot ,x )   \right]^2$ over a compact interval cannot exceed $8$, observe that we can bound 
\begin{equation*}
\mathcal{T}_{ L}^{  \mathrm{r}1 }  \ \leq \  10 \,  \| \hat{F}_{U , \lowmathcal{h} }  - F_U  \|_{ \mathbb{R} , \infty} \,  
\end{equation*}
due to Theorems 2.21 (integration by parts) and 2.24 in \cite{wz77}.
In addition, write 
\begin{multline*}
\mathcal{T}_{ L}^{  \mathrm{r}2 }  \  : =  \  
 \sup_{ (\phi_1 , \phi_2) \in \mathscr{F}_{ R}^2 } \,
\bigg| \,
\frac{1}{  L }  \sum_{l=1}^{ L }   \int_{ \log( \ushort{r} ) - c_X }^{\bar{u} + 3 c_X }  \left[  \phi_2 \left( u , X_l \right)  -  \phi_1 \left(u , X_l \right)   \right]^2 \, d F_{U} ( u ) \\
 - \  \int_{\mathscr{S}_X}   \int_{ \log( \ushort{r} ) - c_X }^{\bar{u} + 3 c_X }  \left[  \phi_2 ( u , x )  -  \phi_1 (u , x )   \right]^2 \, d F_{U} ( u )  d F_X(x)
\, \bigg| 
\end{multline*}
and note that $\mathcal{T}^{  \mathrm{r}2 }_{L}  = o_p (1)$. To see this, first, let $\mathscr{F}_{\mathrm{co}}$ denote the collection of functions on $\mathscr{S}_X$ of the form $x \mapsto | F_U ( u_2 + x^\tau \delta_2) -  F_U ( u_1 + x^\tau \delta_1) |$ and indexed by $(u_1 , u_2 , \delta_1 , \delta_2 ) \in [ \log( \ushort{r} ) ,  \bar{u} + 2 c_X  ]^2 \times \mathcal{B}(0, \tilde{\varepsilon} )^2$.
Second, noting that $ \phi_2 ( u , \cdot )  = \phi_1 (u , \cdot )$ when $u \notin [  \log( \ushort{r} ) - c_X  , \bar{u} + 3 c_X ]$ for any $( \phi_1  , \phi_2 ) \in \mathscr{F}_R^2$ by construction of $\mathscr{F}_R$, observe that we can bound
\begin{multline*}
\mathcal{T}_{ L}^{  \mathrm{r}2 }  \   =  \  
 \sup_{ (\phi_1 , \phi_2) \in \mathscr{F}_{ R}^2 } \,
\left|
\left\{ \int_{ -\infty }^{ \infty } 
 \left[  \phi_2 \left( u , \cdot \right)  -  \phi_1 \left(u , \cdot \right)   \right]^2 \, d F_{U} ( u ) 
 \right\} d \left( \hat{F}_{X}   - F_X \right) 
\right|  \\
   \leq   \  \sup_{  \phi \in  \mathscr{F}_{\mathrm{co}} }
\left|
\int \, \phi  \,  d \left( \hat{F}_{X}   - F_X \right) 
\right| \, .
\end{multline*}
Finally, observe that the term on the utmost right-hand side is $o_p(1)$ because $\mathscr{F}_{\mathrm{co}}$ is VC-type with constant envelope function equal to 2, as it consists of the absolute differences between any two functions of a VC-type class.

\emph{Step 2.} Consider the $\mathscr{F}_R$-indexed bootstrap empirical process 
\begin{equation*}
\mathbb{G}_L^\ast (  \phi  ) \  : = \ 
\frac{1}{\sqrt{L}} \sum_{l=1}^L  \left\{ \phi (  U_l^\ast , X_l )  
 -  
 \E [ \phi (  U_l^\ast , X_l ) \mid \mathcal{S}_L  ]  \right\} \, , 
\end{equation*}
and observe that we can write $\lowmathcal{r}_{(u,\hat\beta^\ast -\hat\beta )} (U_l^\ast ,X_l ) =  \mathds{1} ( \hat{U}_l^\ast  \leq u      ) $, as well as $ \lowmathcal{r}_{(u,0)} (U_l^\ast ,X_l ) = \mathds{1} ( {U}_l^\ast  \leq u      ) $.
Consequently, when $\| \hat\beta^\ast - \hat\beta \|_2 <\tilde{\varepsilon} $, we have that
\begin{equation*}
L^{-1/2} \sum_{l=1}^L  \mathcal{R}_l^\ast (u) \ = \ \mathbb{G}_L^\ast \left( \lowmathcal{r}_{(u,\hat\beta^\ast - \hat\beta)}   - \lowmathcal{r}_{(u,0)}  \right) \ \ \text{for}  \  u \in[  \log( \ushort{r} ) ,  \bar{u} + 2 c_X  ] \, ,
\end{equation*}
whereas $ \mathcal{R}_l^\ast (u)  = 0$ for $u >\bar{u} + 2 c_X $.

\emph{Step 3.}
Now pick an arbitrary increasing subsequence $L_m \in \mathbb{N}$ and let $(L_{m_k} )_{k \in \mathbb{N}}$ be a sub-subsequence such that
\begin{equation*}
\lim_{ k \rightarrow \infty } \,  \max \left\{  \,
L^{2/5}\|  \hat\beta - \beta  \|_2 
\,  ,  \,     \E  \left(  \|  \hat\beta^\ast - \hat\beta  \|_2   \mid   \mathcal{S}_{L_{m_k}}  \right) 
\,  ,  \,  \mathcal{T}_{ L_{m_k} }^{  \mathrm{r}1 } 
\,  ,  \,  \mathcal{T}_{ L_{m_k} }^{  \mathrm{r}2 } 
 \, \right\} \ = \ 0 \ \ \mathrm{a.s.}
\end{equation*}
Observe that such a sub-subsequence exists due to part 4 of this auxiliary lemma and Step 1 above. Then, for given $\epsilon > 0$ and $\delta \in ( 0 , 1/2)$, note that we can bound
\begin{eqnarray*}
  & & \Prob \left( \sup_{u \in [ \log ( \ushort{r} ) , \infty ) } \left| \frac{1}{\sqrt{ L_{m_k} }} \sum_{l=1}^{L_{m_k}}   \mathcal{R}_l^\ast (u) \right|    >   \epsilon \,  \middle| \, \mathcal{S}_{L_{m_k}}  \right) \\
     &   &    \ \  \leq \  
   \Prob  \left(   \sup_{u \in [ \log ( \ushort{r} ) , \infty ) } \left| \mathbb{G}_{L_{m_k}}^\ast  ( \lowmathcal{r}_{(u, \hat\beta^\ast - \hat\beta )}   - \lowmathcal{r}_{(u,0)}  ) \right|    >   \epsilon , \,   \|  \hat\beta^\ast - \hat\beta \|_2   < \delta^\prime   \,  \middle| \, \mathcal{S}_{L_{m_k}}  \right)    
  +      \Prob \left( \|  \hat\beta^\ast - \hat\beta  \|_2   \geq \delta^\prime   \,  \middle| \, \mathcal{S}_{L_{m_k}} \right)  \\
  &   &  \ \ \leq \    \Prob \left(    \sup_{ (\phi_1 , \phi_2) \in {\mathscr{F}}_{ R , \delta} } \, \left| \mathbb{G}_{L_{m_k}}^\ast  \left( \phi_2  -  \phi_1 \right)    \right|   \, > \,   \epsilon  \,  \middle| \, \mathcal{S}_{L_{m_k}}  \right)  \,   
  +  \, \frac{ \E  \left(  \|  \hat\beta^\ast - \hat\beta  \|_2   \mid   \mathcal{S}_{L_{m_k}}  \right) }{ \delta^\prime } \, .
\end{eqnarray*}
where $0 < \delta^\prime < 1$ has been obtained in the proof of part (b) above, and depends on $\delta $. To obtain the desired result, then it suffices to show that
\begin{equation*}
\lim_{\delta \downarrow 0} \,  \underset{k \rightarrow \infty} {\lim \! \sup} \,  \Prob \, \left(    \sup_{ (\phi_1 , \phi_2) \in \mathscr{F}_{ R , \delta}  } \, \left| \mathbb{G}_{L_{m_k}}^\ast  \left( \phi_2  -  \phi_1 \right)    \right|   \, > \,   \epsilon  \,  \middle| \, \mathcal{S}_{L_{m_k}}  \right)  \ = \ 0  \quad \text{a.s.}
\end{equation*}
Since  $(  \mathscr{F}_R , \rho )$ is a totally bounded metric space, by  Theorem 2.11.1 and Lemma 2.11.6 in \cite{vw23},
this stochastic equicontinuity result can be obtained by verifying the following conditions: 
\begin{eqnarray}
& & \Prob \left\{ \forall \delta_k \downarrow 0 , \,  \sup_{ (\phi_1 , \phi_2) \in \mathscr{F}_{ R , \delta_k }  }
\frac{1}{  L_{m_k} }  \sum_{l=1}^{ L_{m_k} } \E  \left\{  \left[  \phi_2 \left( U^\ast_l , X_l \right)  -  \phi_1 \left(U^\ast_l , X_l\right)   \right]^2  
 \, \middle| \,  \mathcal{S}_{L_{m_k}}   \right\}  \rightarrow   0  \right\} = 1 \, , 
 \label{eqa:bse1} \\
  & &  \int_0^\infty  \, \sup_{P }   \, \sqrt{  \log  N  \left(  \varepsilon   , \mathscr{F}_{ R }   , \mathcal{L}_2 (P)  \right)  } \, d \varepsilon \, < \,  \infty ,
   \label{eqa:bse2}
\end{eqnarray}
where $N$ denotes the covering number \cite[Definition 2.1.5]{vw23} and the supremum in Condition (\ref{eqa:bse2}) is taken over the probability measures $P$ on $\mathbb{R}^{1+d}$.


\emph{Step 4.}
To verify Condition (\ref{eqa:bse1}), recalling that $ \phi_2 ( u , \cdot )  = \phi_1 (u , \cdot )$ when $u \notin [  \log( \ushort{r} ) - c_X  , \bar{u} + 3 c_X ]$ for any $( \phi_1  , \phi_2 ) \in \mathscr{F}_R^2$, write
\begin{equation*}
\E  \left\{ 
\left[  \phi_2 ( U^\ast_l , X_l )  -  \phi_1 (U^\ast_l , X_l)   \right]^2   \middle| \,  \mathcal{S}_{L} 
  \right\} 
   \, = \, \int_{ \log( \ushort{r} ) - c_X }^{\bar{u} + 3 c_X }  \left[  \phi_2 ( u , X_l )  -  \phi_1 (u , X_l)   \right]^2 \, d \hat{F}_{U,\lowmathcal{h}} ( u ) .
\end{equation*}
Then, note that we can bound 
\begin{equation*}
\frac{1}{  L_{m_k} }  \sum_{l=1}^{ L_{m_k} }
\E  \left\{ 
\left[  \phi_2 ( U^\ast_l , X_l )  -  \phi_1 (U^\ast_l , X_l)   \right]^2   \middle| \,  \mathcal{S}_{L_{m_k}} 
  \right\}
 \ \leq \ 
\mathcal{T}_{ L_{m_k} }^{  \mathrm{r}1 }  \, + \,  \mathcal{T}_{ L_{m_k} }^{  \mathrm{r}2 } \, + \,  \mathcal{T}^{  \mathrm{r}3   }_{k} \, 
\end{equation*}
for any $( \phi_1  , \phi_2 )  \in  \mathscr{F}_{ R , \delta_k }$,
where
\begin{equation*}
\mathcal{T}^{ \mathrm{r}3 }_{k} 
\  : =  \   \sup_{ (\phi_1 , \phi_2) \in {\mathscr{F}}_{ R , \delta_k} } \,
\int_{\mathscr{S}_X}   \int_{ \log( \ushort{r} ) - c_X }^{\bar{u} + 3 c_X }  \left[  \phi_2 ( u , x )  -  \phi_1 (u , x )   \right]^2 \, d F_{U} ( u )  d F_X(x) \ \leq \ \delta_k^2 \, .
\end{equation*}
Thus, the desired result follows by construction of the sub-subsequence $( L_{m_k} )_{ k \in \mathbb{N} }$.
Finally, Condition (\ref{eqa:bse2}) follows from Theorem 2.6.7 in \cite{vw23}, as well as Lemmas 2.6.16 and 2.6.20(viii) of this reference, since each function in $\mathscr{F}_R$ can be expressed as the composition of $\mathds{1}(0\leq \cdot)$ with a linear functional of $(1,U,X)$.

\paragraph{(b)}

Let $\check{\mathscr{F}}_{S}$ denote the class of functions $\check{\phi}_u$ on $\mathbb{R}^2$ of the form
$
\check{\phi}_{ u }  \left( \tilde{u} , \tilde{x} \right)  : =      \mathds{1} \left(\tilde{u} \leq u      \right)  +    f_U (u) \, \tilde{x} \, \tilde{u}
$
and indexed by $u \geq \log(\ushort{r} ) $.
In addition, define $\check{X}_{L,l} : = \bar{X}^\tau \hat{\Sigma}_X^{-1} \, ( X_l  - \bar{X} ) $ and note that we can write
\begin{equation*}
 \psi_l^\ast (  u  )   \, = \, \check{\phi}_{ u }  \left(  U_l^\ast  , \check{X}_{L,l}  \right)   - \E [  \check{\phi}_{ u }  \left(  U_l^\ast  , \check{X}_{L,l}  \right) \mid  \mathcal{S}_L  ] \, .
 \end{equation*}
Let $\check{P}_{UX}$ be the probability measure on $\mathbb{R}^2$ associated with the joint distribution of $\left( U , \, \mu_X^\tau {\Sigma}_X^{-1} \, ( X  - \mu_X ) \right) $ and also let $\check{\rho} ( \phi ) : =  ( \int \phi^2 \, d \check{P}_{UX} )^{1/2}$ denote the $\mathcal{L}_2 ( \check{P}_{UX} )$-norm on $\check{\mathscr{F}}_{S}$. Denote further
\begin{equation*}
\check{\mathscr{F}}_{ S , \delta } \ : = \  \left\{  (\phi_1 , \phi_2 ) \in \check{\mathscr{F}}_{S}^2 :  \,  \check{\rho} (\phi_2 -\phi_1  )  <  \delta \right\} .
\end{equation*}
Observe that $ ( \check{\mathscr{F}}_{S}   ,  \check{\rho} )$ constitutes a totally bounded metric space,  $\check{F}_{\mathrm{e}} ( \tilde{u}) := 1 + \| f_U \|_{\mathbb{R} , \infty }  \check{c}_X  |  \tilde{u} |$ can be taken as the envelope function, and that $\check{\mathscr{F}}_{S}$ is VC-type because it can be decomposed into the sum of two VC-subgraph classes.

By Theorem 2.11.1 and Lemma 2.11.6 in \cite{vw23}, as well as by the Lyndeberg-Feller central limit theorem \cite[Proposition 2.27]{vdVaart98}, the desired result follows by establishing the next statement: for any increasing sequence of positive integers $(L_m)_{m\in\mathbb{N}}$, there exists a subsequence $(L_{m_k})_{k\in\mathbb{N}}$ such that the following events hold w.p.1:
\begin{eqnarray}
& & \forall \, \eta > 0   , \  
\frac{1}{ L_{m_k} } \sum_{l=1}^{L_{m_k}} \E \left[  \check{F}_{\mathrm{e}} \left( U_l^\ast   \right) ^2 
\mathds{1} \left(    \check{F}_{\mathrm{e}} \left( U_l^\ast  \right) > \eta \sqrt{ L_{m_k} }      \right) \mid \mathcal{S}_{L_{m_k}} \right] \ \rightarrow \ 0 \, ;   \label{eqa:bc1} \\
& & \forall \delta_k \downarrow 0 , \,  \sup_{ (\phi_1 , \phi_2) \in \check{\mathscr{F}}_{ S , \delta_k }  }
\frac{1}{  L_{m_k} }  \sum_{l=1}^{ L_{m_k} } 
\E  \left\{  \left[  \phi_2 \left( U^\ast_l , \check{X}_{L_{m_k} ,l} \right)  -  \phi_1 \left(U^\ast_l , \check{X}_{L_{m_k} ,l} \right)   \right]^2  
 \, \middle| \,  \mathcal{S}_{L_{m_k}}   \right\}  \rightarrow   0  \, ;  \label{eqa:bc2} \\
& & \forall \, u , u^\prime \geq \log ( \ushort{r}  ) \,  , \  
\frac{1}{ L_{m_k} }
\sum_{l=1}^{L_{m_k}} \E \left[  \psi_l^\ast (  u  ) \psi_l^\ast (  u^\prime  )  \mid \mathcal{S}_{ L_{m_k} } \right]
\ \rightarrow \  \Psi_\nu  (  u  , u^\prime)    \, ;   \label{eqa:bc3}
\end{eqnarray}
where $\check{c}_X : = \sup_{x \in \mathscr{S}_X}  | \mu_X^\tau {\Sigma}_X^{-1} \, ( x  - \mu_X )   | +1$.

Before proceeding, I remark that the preceding statement implies that $L^{-1/2}\sum_l \psi_l^\ast$ is asymptotically $ \check{\rho} $-equicontinuous and that it converges weakly in $\ell^{\infty} \left(  [  \log(\ushort{r} )  , \infty ) \right)$ w.p.1, conditional on $\mathcal{S}_L$ and along the subsequence $(L_{m_k})_k$. Then, asymptotic tightness of $L^{-1/2}\sum_l \psi_l^\ast$ arises from Theorem 1.5.7 in \cite{vw23}, while tightness of its Gaussian limit follows by Lemma 1.3.8.(ii) in this reference.

Next I show in two steps that the preceding statement holds, i.e., given an increasing sequence $L_m \in \mathbb{N}$, the events in Eqs.\ (\ref{eqa:bc1})-(\ref{eqa:bc3}) hold w.p.1 along a certain subsequence $(L_{m_k})_{k\in\mathbb{N}}$.


\emph{Step 1.}
Let $\mathrm{TV} ( \phi , R)$ denote the total variation of a real function $\phi$ over a compact interval $R \subset \mathbb{R}$ and choose any $u_1 , u_2  \geq  \log(\ushort{r}) $ and $-\check{c}_X  \leq x  \leq \check{c}_X$. Note first that we can bound
\begin{eqnarray*}
\mathcal{T}_{ L}^{  \mathrm{s}1 } ( u_1 , u_2 , x) & : =  & \left| \,
\int_{ -\infty  }^{ \infty }  \left[  \check{\phi}_{u_2} ( \cdot , x )  -   \check{\phi}_{u_1} (\cdot ,x )   \right]^2 \, 
d \left( \hat{F}_{U,\lowmathcal{h}}   - F_U \right) 
\, \right|  \\
&  \leq  &  \left| \,		
  \left[  f_U (u_2) - f_U (u_1)    \right]^2 x^2 \int_{ -\infty  }^{ \log(\ushort{r})  } \tilde{u}^2 \, 
d \left( \hat{F}_{U,\lowmathcal{h}} (\tilde{u} )   - F_U (\tilde{u}) \right) 
\, \right| \\
& & \quad    +  \ 
\left| \hat{F}_{U,\lowmathcal{h}} [ \log(\ushort{r}) ]   - F_U [ \log(\ushort{r}) ] \right|  \left[  \check{\phi}_{u_2} ( \log(\ushort{r}) , x )  -   \check{\phi}_{u_1} ( \log(\ushort{r})  ,x )   \right]^2 \\
& & \quad    + \
\left\|  \hat{F}_{U,\lowmathcal{h}}  - F_U  \right\|_{\mathbb{R},\infty} \,
 \mathrm{TV} \left( \left[  \check{\phi}_{u_2} ( \cdot , x )  -   \check{\phi}_{u_1} (\cdot ,x )   \right]^2  ,  \,
 [ \log(\ushort{r}) , \,  \bar{u} + \lowmathcal{h} +  \check{c}_X \| \hat\beta - \beta\|_2   ]   \right) 
\end{eqnarray*}
and, as a consequence of part 3.(d) of this lemma, it follows that
\begin{equation*}
\mathcal{T}_{ L}^{  \mathrm{s}1 } \ : = \ \sup_{ ( u_1 , u_2 , x )} 
\left| \mathcal{T}_{ L}^{  \mathrm{s}1 } ( u_1 , u_2 , x)  \right| \, = \ o_p(1)
\end{equation*}
with the supremum being taken over $u_1 , u_2  \geq  \log(\ushort{r}) $ and $| x | \leq \check{c}_X$.
Second, write
\begin{eqnarray*}
& &  \frac{1}{L}  \sum_{l=1}^L  \int_{-\infty}^\infty  \left[  \check{\phi}_{u_2} ( u , \check{X}_{L,l} )  -   \check{\phi}_{u_1} (u , \check{X}_{L,l} )   \right]^2  \, d F_U (u)
\\ 
& & \quad  = \frac{1}{L}  \sum_{l=1}^L    
\int_{-\infty}^\infty  \left\{ \mathds{1} \left(u \leq u_2 \right)  - \mathds{1} \left( u\leq u_1 \right)  
+ [ f_U(u_2) - f_U (u_1) ]  \,  \check{X}_{L,l} u \right\}^2    d F_U(u)  \\
& & \quad  = \  \int_{-\infty}^\infty 
 \left[
\mathds{1} \left(u \leq u_2 \right)  - \mathds{1} \left( u\leq u_1 \right)  \right]^2    d F_U(u)
\,+ \, [ f_U(u_2) - f_U (u_1) ]^2  \E ( U^2) \, \frac{1}{L}  \sum_{l=1}^L \check{X}_{L,l}^2 \,  
\end{eqnarray*}
and observe that $L^{-1}  \sum_{l=1}^L \check{X}_{L,l}^2 =  \bar{X}^\tau \, \hat{\Sigma}_X^{-1} \,\bar{X} \overset{p}{\rightarrow}  \mu_{X}^\tau \, \Sigma_X^{-1} \,\mu_{X} $.
As a consequence, note that 
\begin{multline*}
\mathcal{T}_{ L}^{  \mathrm{s}2 } \  : =  \  
 \sup_{ u_1 , u_2  \geq  \log(\ushort{r}) } \,
\bigg| \,
 \frac{1}{L}  \sum_{l=1}^L  \int_{-\infty}^\infty  \left[  \check{\phi}_{u_2} ( u , \check{X}_{L,l} )  -   \check{\phi}_{u_1} (u , \check{X}_{L,l} )   \right]^2  \, d F_U (u) \\
 - \  \int_{-\infty}^\infty  \int_{-\infty}^\infty  \left[  \check{\phi}_{u_2} ( u , x )  -   \check{\phi}_{u_1} (u , x )   \right]^2  \, d F_U (u) d \check{F}_X(x)
\, \bigg| \, = \, o_p (1) ,
\end{multline*}
where $\check{F}_X(x)$ denotes the marginal distribution of the transformed covariate $\mu_X^\tau {\Sigma}_X^{-1} \, ( X  - \mu_X )$.

\emph{Step 2.} 
Using arguments similar to those used in the proof of part 5.(a), given an increasing sequence $L_m \in \mathbb{N}$, choose a sub-sequence $(L_{m_k})_{k}$ such that the following terms converge to zero a.s.\ along such a subsequence:
\begin{equation*}
\mathcal{T}_{ L}^{  \mathrm{s}0 } \  := \ \sup_{ c  \in \mathbb{R} } 
 \, \left|   \int_{-\infty}^c  u^2 \hat{f}_{U,\lowmathcal{h}} (u) \, du  - \E [ U^2\mathds{1} (   U\leq c ) ] \right|   \,  ,
\end{equation*}
$\mathcal{T}_{ L}^{  \mathrm{s}1 }$, $\mathcal{T}_{ L}^{  \mathrm{s}2 }$,
$\| \hat{F}_{U , \lowmathcal{h} }  - F_U  \|_{ \mathbb{R} , \infty}$,
$\| \mathrm{vec}(\hat{\Sigma}_X)  -  \mathrm{vec}(\Sigma_X ) \|_2$, $\| \bar{X} - \mu_X \|_2$, and $L^{2/5}\|  \hat\beta - \beta  \|_2 $.

Starting with the event in Eq.\ (\ref{eqa:bc1}), observe that such an event can be written as the intersection of a decreasing sequence of events; specifically,
\begin{multline*}
\left\{ \forall \, \eta > 0   , \  
\frac{1}{ L_{m_k} } \sum_{l=1}^{L_{m_k}} \E \left[  \check{F}_{\mathrm{e}} \left( U_l^\ast   \right) ^2 
\mathds{1} \left(    \check{F}_{\mathrm{e}} \left( U_l^\ast  \right) > \eta \sqrt{ L_{m_k} }      \right) \mid \mathcal{S}_{L_{m_k}} \right] \ \rightarrow \ 0 \right\}  \\
= \ \bigcap_{p \in \mathbb{N}}  \, \left\{ \frac{1}{ L_{m_k} }  \sum_{l=1}^{L_{m_k}} \E \left[  \check{F}_{\mathrm{e}} \left( U_l^\ast   \right) ^2 
\mathds{1} \left(    \check{F}_{\mathrm{e}} \left( U_l^\ast  \right) > \sqrt{ L_{m_k} }  / p    \right) \mid \mathcal{S}_{L_{m_k}} \right] \ \rightarrow \ 0 \right\} .
\end{multline*}
Then, it suffices to show that each event of the intersection on the right-hand side occurs w.p.1.
With this aim, fix $p \in \mathbb{N}$ and then observe that, for some finite constants $c_p , c^{\prime}_p >0$ and any $L$ sufficiently large, we can bound
\begin{eqnarray*}
& &  \frac{1}{ L }  \sum_{l=1}^{L} \E \left[  \check{F}_{\mathrm{e}} \left( U_l^\ast   \right) ^2 
\mathds{1} \left(    \check{F}_{\mathrm{e}} \left( U_l^\ast  \right) > \sqrt{ L }  / p    \right) \mid \mathcal{S}_{L} \right] \\
& & \quad   \leq \ 
   \int_{-\infty}^{  -c_p \, L^{1/2}  } \left( 1 + 2 \, \| f_U \|_{\mathbb{R} , \infty }   \check{c}_X | u | \right)
\hat{f}_{U , \lowmathcal{h} } (u ) \, du 
\, + \, \left( \| f_U \|_{\mathbb{R} , \infty }   \check{c}_X \right)^2 
\int_{-\infty}^{ -c_p \, L^{1/2}  } u^2  \hat{f}_{U , \lowmathcal{h} } (u ) \, du   \\
& & \quad   \leq \ 
  \| \hat{F}_{U , \lowmathcal{h} }  - F_U  \|_{ \mathbb{R} , \infty} \, + \,  F_U \left(   -c_p \, L^{1/2} \right)
\, + \,  c^{\prime}_p \, \left\{ \mathcal{T}_{  L}^{  \mathrm{s}0 }  +  \E [ U^2 \mathds{1}\left( U \leq  -c_p \, L^{1/2} \right)  ] \right\} .
\end{eqnarray*}
Considering now the event in Eq.\ (\ref{eqa:bc2}), write
\begin{equation*}
\E  \left\{  \left[ {\phi}_{2} \left( U^\ast_l , x \right)  -  {\phi}_{1} \left(U^\ast_l , x \right)   \right]^2  
 \, \middle| \,  \mathcal{S}_{L}   \right\} \, = \,   \int_{ -\infty  }^{ \infty }  \left[  {\phi}_{2} ( \cdot , x )  -   {\phi}_{1} (\cdot ,x )   \right]^2 \, d  \hat{F}_{U,\lowmathcal{h}} 
   \  \  \text{for} \  (\phi_1 , \phi_2) \in \check{\mathscr{F}}_{ S , \delta_k } 
\end{equation*}
and, by arguments similar to those used in the fourth step of the proof of part 5.(a), note that we can bound
\begin{equation*}
\sup_{ (\phi_1 , \phi_2) \in \check{\mathscr{F}}_{ S , \delta_k }  } \,
\frac{1}{  L_{m_k} }  \sum_{l=1}^{ L_{m_k} } 
\E  \left\{  \left[  \phi_2 \left( U^\ast_l , \check{X}_{L_{m_k} ,l} \right)  -  \phi_1 \left(U^\ast_l , \check{X}_{L_{m_k} ,l} \right)   \right]^2  
  \middle|   \mathcal{S}_{L_{m_k}}   \right\}
   \,  \leq \,  \mathcal{T}_{ L_{m_k} }^{  \mathrm{s}1 } + \mathcal{T}_{ L_{m_k} }^{  \mathrm{s}2 } +  \delta_k \, .
\end{equation*}
Finally, in view of the convergence result in Eq.\ (\ref{eqa:bc3}), note that $\sum_{l}  \bar{X}^\tau \hat{\Sigma}_X^{-1} \left( X_l - \bar{X} \right) =0 $ and write 
\begin{equation*}
\frac{1}{L} \sum_{l=1}^L \E \left[  \psi_l^\ast (  u  ) \psi_l^\ast (  u^\prime  ) \mid  \mathcal{S}_L  \right] 
\, = \,
\hat{F}_{U , \lowmathcal{h} } ( u  \wedge u^\prime )   -   \hat{F}_{U , \lowmathcal{h} } (u)   \hat{F}_{U , \lowmathcal{h} } ( u^\prime )    +   f_U (u) \left( \bar{X}^\tau \hat{\Sigma}_X^{-1} \bar{X} \right) f_U (u^\prime )   \hat{\sigma}_{U, \lowmathcal{h} }^2
\, ,
\end{equation*}
where $\hat{\sigma}_{U, \lowmathcal{h} }^2 : = \int u^2 d \hat{F}_{ U , \lowmathcal{h} } (u)  - \hat{\mu}_{  U , \lowmathcal{h}  }^2$.
Then, the desired result follows by construction of the subsequence $(L_{m_k})_{k \in \mathbb{N}}$.

\paragraph{(c)}
Here set $\mathbb{Y}_L^\ast  :=  \sqrt{L} \,  (\hat{F}_U^\ast -  \hat{F}_{U , \lowmathcal{h}} ) -  L^{-1/2} \sum_{l=1}^L \psi_l^\ast$.
Observe that part 5.(a) of this lemma and Eq.\ (\ref{eqa:characc}) imply $\mathbb{Y}_L^\ast \overset{\ast}{\leadsto} 0 $ in  $\ell^\infty ( [ \log ( \ushort{r} ) , \infty ) )$. To see this, choose any $\epsilon >0$ and note that we can bound 
\begin{equation*}
\hat{d}_{L} \left( \mathbb{Y}_L^\ast , 0  \right)  \ \leq \  \epsilon + 2 \,\Prob \left( \left\| \mathbb{Y}_L^\ast  \right\|_{  [ \log ( \ushort{r} ) , \infty ) ,\infty } > \epsilon  \mid \mathcal{S}_L \right) ,
\end{equation*}
which implies that
\begin{equation*}
\left\{ \lim_{m \rightarrow \infty}  \,  \hat{d}_{L_m} 
\left( \left\| \mathbb{Y}_{L_m}^\ast  \right\|_{  [ \log ( \ushort{r} ) , \infty ) ,\infty } , 0  \right) = 0 \right\}
 \subseteq 
\left\{ \limsup_{ m \rightarrow \infty }  \,  \hat{d}_{L_m} \left( \mathbb{Y}_{L_m}^\ast , 0  \right)  \leq \epsilon \right\} 
\end{equation*}
for any increasing sequence $L_m \in \mathbb{N}$, and therefore $ \hat{d}_{L} \left( \mathbb{Y}_{L}^\ast , 0  \right) = o_p (1) $.
The desired result then follows by Slutsky's lemma; see Example 1.4.7 of \cite{vw23}.
\qed

\subsubsection*{Proof of Lemma \ref{alem:approxGboot}}

For $r \in [\ushort{r}  , \bar{r} ]$,  define the transformation $T_r : \left( 0 , 1 \right) \rightarrow \mathbb{R}$ by 
\begin{equation*}
T_r  ( y )   \   : =  \  \lowmathcal{p}^{-1} ( y )^n   v_s 
\, + \,  n  \left[ 1 - \lowmathcal{p}^{-1} (y) \right] \lowmathcal{p}^{-1}  ( y)^{n-1} r 
\end{equation*}
and note that its derivative is given by
\begin{equation*}
T_r^\prime  ( y )  =       n \,  \lowmathcal{p}^{-1} (y)^{n-2} \, \frac{    (v_s - r) \, \lowmathcal{p}^{-1}(y)  \, +  \,  (n-1) r    \left[1 - \lowmathcal{p}^{-1}(y)  \right] }
{ \lowmathcal{p}^\prime [ \lowmathcal{p}^{-1}(y) ]  }  .
\end{equation*}
Moreover, letting $h = \tilde{H} - H$, note that integration by parts yields
\begin{equation*}
 \int_{r }^{\infty} w \, d (H + h) (w) -   \int_{r }^{\infty} w \, d H (w) \  = \ - r \, h (r )  -  \int_{r}^{\infty} h (w ) \,  dw \, 
\end{equation*}
and therefore
\begin{eqnarray*}
\Pi \left( \tilde{H} , r  \right) - \Pi \left( {H} , r  \right) & = & T_r   \left[   \tilde{H} (r )\right]  - T_r   \left[   H (r )\right] 
 -  r  \left[  \tilde{H} (r) - H (r) \right]  -   \int_{r}^{\infty} \left[  \tilde{H} (w) - H (w) \right] \,dw  \\ 
& = &   \Pi^\prime_H \left( \tilde{H} -  {H} , r  \right)  +  T_r   \left[   \tilde{H} (r )\right]  - T_r   \left[   H (r )\right]  - T_r^\prime   \left[   H (r )\right] \, \left[  \tilde{H} (r) - H (r) \right] \, .
\end{eqnarray*}
Then, the desired result follows from the inequality
\begin{equation*}
\sup_{r \in  [\ushort{r}  , \bar{r} ]} \,
\left|  
T_r   \left[   \tilde{H} (r )\right]  - T_r   \left[   H (r )\right]  - T_r^\prime   \left[   H (r )\right] \, \left[  \tilde{H} (r) - H (r) \right]
\right|
 \ \leq \ 
\underbrace{ \frac{1}{2} \sup_{r  , y  } \left|  T_r^{\prime\prime} (y)  \right| }_{ =: \, \bar{c}_\pi } \, \left\| \tilde{H}  - H \right\|_{  [ \ushort{r}  , \bar{r} ] , \infty }^2 \, ,
\end{equation*}
where the supremum on the right-hand side is taken over $(r , y ) \in [ \ushort{r}  , \bar{r} ]  \times  [ F_W( \ushort{r} )/2  , (1 +  F_W( \bar{r} ) )/2    ]$ and $ \bar{c}_\pi  <\infty$ follows by the fact that $\lowmathcal{p}^\prime \circ \lowmathcal{p}^{-1}$ is bounded away from zero on $[ F_W( \ushort{r} )/2  , (1 +  F_W( \bar{r} ) )/2    ]$.
\qed

\subsubsection*{Proof of Lemma \ref{alem:contG}}


To establish continuity of the c.d.f., pick $(c , \epsilon  ) \in \mathbb{R} \times ( 0 , 1)$ and consider grid points $\{ r_1 , \dots , r_{N (\epsilon)} \} \subseteq R$ such that $R$ can be covered by $N (\epsilon) \leq c_R \epsilon^{-m}$ hypercubes of diameter smaller than $\epsilon$,  each containing one grid point, where $c_R >0$ is a constant depending only on the diameter of $R$.
It follows from the triangle inequality that
\begin{multline*}
\Prob \left(  \left|  \|  \mathbb{G} \|_{R, \infty} - c  \right|  < \epsilon  \right) 
\\ 
\ \leq \ \Prob \left(  \left|  \|  \mathbb{G} \|_{R, \infty} - c  \right|  < \epsilon  ,  \,  
 \sup_{ r , r^\prime } \left|  \mathbb{G}(r^\prime)  -   \mathbb{G}(r)   \right|  < \epsilon^{1/4}  \right) 
\, + \, 
  \Prob \left( 
 \sup_{r^\prime , r} \left|  \mathbb{G}(r^\prime)  -   \mathbb{G}(r)   \right|  >  \epsilon^{1/4} 
 \right)  \\
 \ \leq \
   \Prob \left( 
 \left| \max_{j}  |  \mathbb{G}(r_j) |   - c \right| < \epsilon  + \epsilon^{1/4}
 \right) 
 \, + \,
   \Prob \left( 
 \sup_{r^\prime , r}\,  \left|  \mathbb{G}(r^\prime)  -   \mathbb{G}(r)   \right|  >  \epsilon^{1/4} 
 \right) \, , 
\end{multline*}
where the supremum is taken over $(  r , r^\prime  ) \in R^2$ such that $\|  r - r^\prime \|_\infty \leq \epsilon$ and the maximum across $j = 1, \dots , N (\epsilon)$.
The first part of the proof then follows by applying the Gaussian anti-concentration inequality in Comment 5 of \cite{cck15} to the finite-dimensional maximum, together with Theorem 2.3.7 of \cite{gn16} to the modulus of continuity;
specifically, both probabilities on the third line are $O(\epsilon^{1/4}\sqrt{\log(1/\epsilon)})$ as $\epsilon\to0^+$.

To establish strict monotonicity, let $\mathscr{H} (  \mathbb{G} )$ denote the reproducing kernel Hilbert space of $ \mathbb{G}$. Pick any nonnegative $y_1 < y_2$, define the set $\mathcal{C} = \{ \phi  \in  \mathscr{C}( R )  :  \| \phi \|_ { R , \infty} < (y_2 - y_1) / 4 \}$, and choose any $h \in  \mathscr{H}  (  \mathbb{G} )$ such that $\|  h \|_{R , \infty} =  (y_2 +  y_1) / 2 $, which is possible because $\mathscr{H}  (  \mathbb{G} )$ is a nontrivial linear subspace due to the covariance condition imposed on $\mathbb{G}$. Then, observe that 
\begin{equation*}
  \Prob \left( y_1 <    \| \mathbb{G}   \|_{R , \infty}  \leq y_2  \right) 
 \geq  \Prob \left(   \mathbb{G}  - h \in  \mathcal{C}  \right)   >0 ,
\end{equation*}
with the first inequality following by the triangle inequality, while the second follows from Theorem 2.4.5 and Corollary 2.6.18 in \cite{gn16}.
\qed

\subsubsection*{Proof of Lemma \ref{alem:supptG}}

Before proceeding with the proofs, three preliminary remarks are noteworthy. First, the Reproducing kernel Hilbert space (RKHS) of $\tilde{\mathbb{G}}$ is given by
\begin{equation*}
\mathscr{H} \left( \tilde{\mathbb{G}} \right)
 \, : = \, 
 \left\{  \phi \circ F_W (\cdot) +  c \, f_U \left( \log( \cdot ) \right)  \mathds{1} \left( \mu_X^\tau   \Sigma_X^{-1}  \mu_X \neq 0 \right)
 \in \mathbb{D}_0: \, 
\phi \in \mathcal{AC} , c \in  \mathbb{R} \right\} ,
\end{equation*}
where $\mathcal{AC}$ denotes the space of absolutely
continuous functions $\phi$ on $[F_W(\ushort{r}),1]$
satisfying $\phi(1)=0$ and
$\int_{F_W(\ushort{r})}^1|\phi^\prime(t)|^2\,dt<\infty$:
this result can be obtained by combining together Exercises 2.6.5 and 2.6.16 of \cite{gn16}.
Second, Corollary 2.6.7  in this reference and the fact that $\mathscr{H}(\tilde{\mathbb{G}})$ is dense
in $\mathbb{D}_0$ imply that $\tilde{\mathbb{G}}$ is supported on $\mathbb{D}_0$.
Third, since $\Pi^\prime_{F_W}$ is linear and continuous on $\mathbb{D}_0$, the aforementioned Exercise 2.6.5 also implies that the RKHS of $\Pi^\prime_{F_W} ( \tilde{\mathbb{G}} )$ is $\Pi^\prime_{F_W} ( \mathscr{H} \left( \tilde{\mathbb{G}} \right) )$.

Consider also the real function $\lowmathcal{b}$ defined by $\lowmathcal{b} (t ) : = \exp(-1/t)$ for $t>0$, and $\lowmathcal{b} (t ) : =0$ otherwise, and write $\phi_{r_1,r_2} ( \cdot ) : = \lowmathcal{b} (\cdot - r_1 ) \lowmathcal{b} ( r_2 - \cdot ) \in \mathbb{D}_0$ for $r_1 < r_2$, noting that $\phi_{r_1,r_2} \in \mathscr{S} (\tilde{\mathbb{G}})$ when $\ushort{r} \leq r_1 < r_2 \leq \bar{v}$ and that $\phi_{r_1,r_2} (t) = 0$ if and only if $t \notin (r_1 , r_2 )$.

\paragraph{1.}
Note that the subset $ \mathbb{D}_{A4.1} : =\{  \phi \in  \mathbb{D}_0  : \, \Pi^\prime_{F_W} (\phi , r ) =0    \}$ is a closed linear subspace of $ \mathbb{D}_0$ because $\Pi^\prime_{F_W} (\cdot , r )$ is a continuous linear functional on $ \mathbb{D}_0 $.
Hence, by the  0-1 Law \cite[Theorem 2.1.13]{gn16}, we must have either $\Prob \{   \Pi^\prime_{F_W} ( \tilde{\mathbb{G}}  ,  r )   = 0   \}  = 0$ or $\Prob \{   \Pi^\prime_{F_W} ( \tilde{\mathbb{G}}  ,  r )   = 0   \}  = 1$.
By contradiction, suppose that the latter holds and therefore $\mathscr{S} (\tilde{\mathbb{G}})  \subseteq \mathbb{D}_{A4.1} $.
However, $\mathscr{S} (\tilde{\mathbb{G}})  \subseteq \mathbb{D}_{A4.1} $ does not hold because $\phi_{r ,  \bar{v}}  \in \mathscr{S} (\tilde{\mathbb{G}})$ and $\phi_{r ,  \bar{v}} \notin   \mathbb{D}_{A4.1}$, since $ \Pi^\prime_{F_W} ( \phi_{r ,  \bar{v}}  , r )  = 0 -  \int_r^{\infty}  \phi_{r ,  \bar{v}}  < 0 $.

\paragraph{2.}
First, observe that the desired result holds for $c=0$. To see this, apply the same reasoning as in the preceding proof and, assuming without loss of generality that $r_1 < r_2$, note that the function $\phi_{r_1,r_2} \in \mathscr{S} (\tilde{\mathbb{G}})$ satisfies $ \Pi^\prime_{F_W} ( \phi_{r_1,r_2}   , r_2 ) -   \Pi^\prime_{F_W} ( \phi_{r_1,r_2}   , r_1 ) = \int_{r_1}^{r_2}   \phi_{r_1,r_2}  >0$. Second, the result for $c \neq 0$ follows from the fact that $ \Pi^\prime_{F_W} ( \tilde{\mathbb{G}}   , r_2 ) -   \Pi^\prime_{F_W} (  \tilde{\mathbb{G}}   , r_1 )$ is a centered normal random variable, as $ \Pi^\prime_{F_W} ( \cdot , r_2 ) -   \Pi^\prime_{F_W} (  \cdot   , r_1 )$ is a continuous linear functional on $ \mathbb{D}_0$, and by the Cameron-Martin formula \cite[Theorem 2.6.13]{gn16}.

\paragraph{3.} It follows from Fubini's theorem \cite[see, e.g.,][Lemmas 1.2.6 and 1.2.7]{vw23}. To see this, write
\begin{multline*}
\mathrm{cov} \left(  \Pi^\prime_{F_W} ( \tilde{\mathbb{G}}  ,  r_1 )   , \Pi^\prime_{F_W} ( \tilde{\mathbb{G}}  ,  r_2 )   \right) 
\\ \qquad = \ 
\varphi_{F_W} (r_1 )\varphi_{F_W}( r_2 )
\E\big[
\tilde{\mathbb{G}}(r_1)\tilde{\mathbb{G}}( r_2 )
\big] \, - \, 
\varphi_{F_W}(r_1)
\E\left[
\tilde{\mathbb{G}}(r_1)
\int_{r_2}^{\bar{v}} \tilde{\mathbb{G}}(w)\,dw
\right] \\
 - \
\varphi_{F_W} (r_2)\E\left[
\tilde{\mathbb{G}}(r_2)
\int_{r_1}^{\bar{v}} \tilde{\mathbb{G}}(v)\,dv
\right]
\, + \, 
\E\left[
\int_{r_2}^{\bar{v}} \int_{r_1}^{\bar{v}} \tilde{\mathbb{G}}(w) \tilde{\mathbb{G}}(v)\,dv
 \,dw
\right]
\end{multline*}
and note that Fubini's theorem applies because $| \E [ \tilde{\mathbb{G}}(\cdot_1 ) \tilde{\mathbb{G}}( \cdot_2 ) ] | =| \Psi_\omega ( \cdot_1  , \cdot_2 ) |$ is continuous and hence bounded in any compact subset of $[ \ushort{r} , \bar{v} ]^2$.

\paragraph{4.} Continuity follows by arguments similar to those used in the proof of Lemma \ref{alem:contG}. In particular, Lemmas \ref{alem:supptG}.1 and \ref{alem:supptG}.3 imply that the covariance function of $\Pi^\prime_{F_W} ( \tilde{\mathbb{G}} )$ indeed satisfies the conditions of Lemma \ref{alem:contG}. To establish strict monotonicity, first, I remark that $\Pi^\prime_{F_W} ( \mathscr{H} \left( \tilde{\mathbb{G}} \right) )$ contains the space of constant functions on $[ \ushort{r} , \bar{r} ]$. To see this, observe that $\phi_{F_W (\bar{r}) ,  1} \circ F_W \in \mathscr{H} \left( \tilde{\mathbb{G}} \right)$ and therefore 
\begin{equation*}
\Pi_{F_W}^\prime \left( \phi_{F_W (\bar{r}) , 1 } \circ F_W , r \right) 
\, = \,
 - \int_{ \bar{r} }^\infty  \phi_{F_W (\bar{r}) , 1 } \circ F_W  \, < \, 0 \quad \forall \ r \in [ \ushort{r} , \bar{r} ] \, ,
\end{equation*}
with the right-hand side not depending on $r$.
Second, pick any $y_1 < y_2$ and note that, by arguments similar to those used in the proof of Lemma \ref{alem:contG}, we can bound 
\begin{equation*}
\Prob \left( y_1 <   \sup_{r \in R} \, \Pi^\prime_{F_W} ( \tilde{\mathbb{G}} , r ) \leq y_2  \right) 
\ \geq  \  
\Prob \left(   \left\|  \Pi^\prime_{F_W} ( \tilde{\mathbb{G}} , \cdot ) -  \frac{y_1 + y_2}{2} \mathds{1} \left( \cdot \in [ \ushort{r} , \bar{r} ] \right)     \right\|_{R , \infty} < \frac{y_2 - y_1}{4} \right)   \  >  \   0 \, .
\end{equation*}

\paragraph{5.}  
To establish continuity, fix $c \geq 0$ and define $\check{\mathbb{G}} (r_1 , r_2) = \Pi^\prime_{F_W} ( \tilde{\mathbb{G}}  ,  r_2 ) - \Pi^\prime_{F_W} ( \tilde{\mathbb{G}}  ,  r_1 ) $ for $ r_1 , r_2  \in [ \ushort{r}  , \bar{r} ]$. Observe that $\mathrm{osc}_{R} \left( \Pi^\prime_{F_W} ( \tilde{\mathbb{G}}  ,  \cdot ) \right) = \| \check{\mathbb{G}} \|_{ R^2 , \infty } $ and that we can bound
\begin{equation*}
\Prob \left(   \| \check{\mathbb{G}} \|_{ R^2 , \infty }  = c  \right)  \ \leq \  \sum_{I , I^\prime}  \, \Prob \left(   \| \check{\mathbb{G}} \|_{   ( I \cap R) \times ( I^\prime \cap R) , \infty }  = c  \right)  ,
\end{equation*}
where the sum $\sum_{I , I^\prime} $ runs over all pairs of disjoint closed intervals  $I , I^\prime  \subset \mathbb{R}$ with rational endpoints such that both $I\cap R$ and $I^\prime \cap R$ are nonempty. Now pick an arbitrary pair $I , I^\prime$ of such intervals. By Lemma \ref{alem:contG}, the desired continuity result follows once we show that the Gaussian process $\check{\mathbb{G}}$, restricted to $\check{R}  : =  ( I \cap R) \times ( I^\prime \cap R)$, has (i) a Lipschitz continuous covariance function and (ii) a variance function that is bounded away from zero. To verify (i), just note that
\begin{equation*}
\mathrm{cov} \left(  \check{\mathbb{G}} (r) ,  \check{\mathbb{G}} (r^\prime) \right) 
=
\Psi_\pi  ( r_2 , r_2^\prime )  -  \Psi_\pi  ( r_2 , r_1^\prime ) - \Psi_\pi  ( r_1 , r_2^\prime ) + \Psi_\pi  ( r_1, r_1^\prime )
\end{equation*}
for $r : = (r_1,r_2), r^\prime : = (r_1^\prime , r_2^\prime ) \in \check{R}$. Property (ii) follows by contradiction: if the variance were zero at some $ (r_1,r_2) \in \check{R}$,  we would have $\Prob \left(   \Pi^\prime_{F_W} ( \tilde{\mathbb{G}}  ,  r_2 )   - \Pi^\prime_{F_W} ( \tilde{\mathbb{G}}  ,  r_1 )  = 0 \right) = 1$, contradicting the second part of this lemma.

To establish strict monotonicity, by arguments similar to those used in the previous part, it suffices to show that for any $c>0$ there exists $\phi\in\mathcal{AC}$ such that
\begin{equation*}
\mathrm{osc}_R\left(
\Pi'_{F_W}(\phi\circ F_W,\cdot)
\right)=c \, .
\end{equation*}
To this end, letting $r_1:=\min R < r_2 : = \max R$, define the function $\phi_0 : = \phi_{F_W(r_1) , F_W(r_2)}$.
Since $\phi_0$ admits a bounded continuous derivative on $[F_W(\ushort{r}),1]$ and
$\phi_0(1)=0$, it follows that
$\phi_0\in\mathcal{AC}$. Now consider the composition $h_0:=\phi_0\circ F_W$.
By construction, we have that $h_0(r_1)=h_0(r_2)=0$, whereas $h_0(w)>0$ for every $w\in(r_1,r_2)$, so
\begin{equation*}
\Pi'_{F_W}(h_0,r_2) -\Pi'_{F_W}(h_0,r_1) \ = \
\int_{r_1}^{r_2}h_0(w)\,dw  \ > \ 0  
\end{equation*}
and therefore $A_0 : = \mathrm{osc}_R( \Pi'_{F_W}(h_0) ) > 0$.
To complete the proof, define
$\check{\phi}:=c\phi_0/A_0\in\mathcal{AC}$
and note that
\begin{equation*}
\mathrm{osc}_R\left(
\Pi^\prime_{F_W}(\check{\phi}\circ F_W)
\right)
\, = \,
\frac{c}{A_0}
\mathrm{osc}_R\left(
\Pi^\prime_{F_W}(h_0)
\right)
\, = \, c.
\end{equation*}
\qed


\clearpage

\renewcommand{\thepage}{R.\arabic{page}}\setcounter{page}{1}

\pdfbookmark[1]{References}{References}

\bibliography{bib_inference_cdf-rev}	

\bibliographystyle{plainnat}

\end{document}